\documentclass[11pt,reqno]{amsart}

\usepackage[margin=1in]{geometry}
\usepackage{microtype}
\usepackage{amsmath,amssymb,amsthm,mathtools}
\usepackage{xcolor,booktabs,tabularx,array,etoolbox}
\usepackage[colorlinks=true,linkcolor=blue!45!black,citecolor=blue!45!black,urlcolor=blue!45!black]{hyperref}

\newcommand{\id}{\mathbf 1}
\newcommand{\idmap}{\mathrm{id}}
\newcommand{\R}{\mathbb R}
\newcommand{\C}{\mathbb C}
\newcommand{\cB}{\mathcal B}
\newcommand{\cH}{\mathcal H}
\newcommand{\cI}{\mathcal I}
\newcommand{\cL}{\mathcal L}

\newcommand{\Tr}{\operatorname{Tr}}
\newcommand{\supp}{\operatorname{supp}}
\newcommand{\diam}{\operatorname{diam}}
\newcommand{\dist}{\operatorname{dist}}
\newcommand{\ad}{\operatorname{ad}}

\newcommand{\e}{\mathrm e}
\newcommand{\norm}[1]{\left\lVert #1\right\rVert}
\newcommand{\abs}[1]{\left|#1\right|}
\newcommand{\ip}[2]{\left\langle #1,#2\right\rangle}
\newcommand{\ket}[1]{\lvert #1\rangle}

\theoremstyle{plain}
\newtheorem{thm}{Theorem}[section]
\newtheorem{lem}[thm]{Lemma}
\newtheorem{prop}[thm]{Proposition}
\newtheorem{cor}[thm]{Corollary}

\theoremstyle{definition}

\newtheorem{rem}[thm]{Remark}

\numberwithin{equation}{section}
\hypersetup{linktoc=all,bookmarksopen=true,bookmarksnumbered=true,
  pdftitle={Complete Modified Logarithmic Sobolev Inequalities for one-dimensional quantum Spin Chain},
  pdfauthor={Li Gao; Jingyu Guo; Sihao Wang},
  pdfsubject={Complete entropy contraction and worst-case mixing for one-dimensional quantum Gibbs samplers},
  pdfkeywords={quantum Gibbs states, complete modified logarithmic Sobolev inequality, entropy contraction, diamond norm, mixing times}
}

\newcommand{\cN}{\mathcal N}
\DeclareMathOperator{\Ad}{Ad}
\newcommand{\ii}{\mathrm i}

\makeatletter
\pretocmd{\@setaddresses}{\par\noindent\begin{minipage}{\textwidth}}{}{}
\apptocmd{\@setaddresses}{\end{minipage}}{}{}
\patchcmd{\@starttoc}{\addvspace{32\p@\@plus14\p@}}{}{}{}

\makeatother

\title[Complete MLSI for 1D Spin Chain]%
{Complete Modified Logarithmic Sobolev Inequalities for one-dimensional quantum Spin Chain}

\author{Li Gao}
\address{School of Mathematics and Statistics, Wuhan University, Wuhan 430072, China}
\address{Wuhan Institute of Quantum Technology, Wuhan 430075, China}
\email{gao.li@whu.edu.cn}

\author{Jingyu Guo}
\address{School of Mathematics and Statistics, Wuhan University, Wuhan 430072, China}
\email{guojingyu@whu.edu.cn}

\author{Sihao Wang}
\address{School of Mathematics and Statistics, Wuhan University, Wuhan 430072, China}
\email{wangsihao@whu.edu.cn}

\keywords{Quantum Gibbs samplers, spin chains, rapid mixing, relative entropy}
\date{October 20, 2026}

\begin{document}

\begin{abstract}
We prove a complete modified logarithmic Sobolev inequality with a system-size-independent constant for Chen--Kastoryano--Gily\'en Gibbs samplers on one-dimensional spin chain at every fixed positive temperature. The result holds for Hamiltonians with bounded finite-range interactions, including noncommuting and spatially nonuniform interactions. It yields trace and diamond mixing times of order $O(\log N+\log(1/\epsilon))$. The proof establishes complete spatial entropy factorization and absorbs quasi-local errors as conditional relative entropy at the original joint state and reference. The static estimate also yields a system-size-independent CMLSI for the single-site Petz heat-bath and regularised heat-bath dynamics.
\end{abstract}

\maketitle

\begingroup
\fontsize{8}{8.5}\selectfont
\tableofcontents
\endgroup

\section{Introduction}

The preparation of quantum Gibbs states connects the study of thermal
equilibrium with the design of quantum algorithms. A dissipative
approach prepares the target state by running a quantum Markov
evolution that leaves it invariant and converges to it from any initial
state. The convergence rate determines the evolution time required for
preparation. For a system of interacting spins, a basic problem is to
understand how this time depends on the number of sites and on the
temperature, particularly when the interaction terms do not commute.

The Gibbs samplers introduced by Chen, Kastoryano, and
Gily\'en in their breakthrough work \cite{CKG23} provide dynamics that satisfy exact
Kubo--Martin--Schwinger (KMS) detailed balance for noncommuting
Hamiltonians. Their mixing
properties have been studied in several regimes. At high temperature,
logarithmic trace-norm mixing is known for the Gaussian sampler on
qubit lattices~\cite{RSFA26}. In one dimension,
Bergamaschi and Chen~\cite{BC25} established a spectral gap independent
of the chain length at every fixed positive temperature. The direct
mixing bound obtained from this gap is
$O(N+\log(1/\epsilon))$, where $N$ is the number of sites and
$\epsilon$ is the target accuracy.

Recent results give substantially sharper bounds in one dimension.
Kim~\cite{Kim26} proved logarithmic trace-norm mixing for Petz
heat-bath dynamics with updates on blocks of a sufficiently large,
fixed length. Capel, Pérez-García, and Scandi~\cite{CPS26} established a modified
logarithmic Sobolev inequality (MLSI) with rate
$\Omega(1/\log N)$ for single-site and regularised heat-bath
dynamics. For the Gaussian CKG sampler on open chains,
Chen and Ding~\cite{CD26} obtained a trace-norm mixing time
$O(\log^4 N+\log(1/\epsilon))$ and an ordinary MLSI rate
$\Omega(1/\log^4 N)$. These results establish rapid mixing at all
fixed positive temperatures.

In this work, we provide a volume-independent
complete entropy decay rate for the CKG generators. More precisely, we establish a
complete modified logarithmic Sobolev inequality (CMLSI) with a
constant independent of system size at every fixed inverse
temperature $\beta>0$. The result applies to open chains and cycles with
uniformly bounded, finite-range
interactions, without commutativity or translation-invariance
assumptions. It implies the mixing time in both trace distance and diamond norm distance
\[
 t_{\mathrm{Tr}}(\epsilon)\leq t_\diamond(\epsilon)
 =O\!\left(\log N+\log\frac1\epsilon\right).
\]
The implicit constant depends only on the fixed model parameters but not on the size $N$.
This order is optimal in the sense that it is obtained by the independent single-site depolarizing dynamics for the vanishing Hamiltonian $H=0$.

The complete entropy decay inequality controls more than relaxation of the spins
alone. If they are initially entangled with a reference system on
which the dynamics does not act, it gives exponential decay of
relative entropy towards the product of the unchanged reference
marginal and the Gibbs state, with the same rate for every finite
reference dimension~\cite{GJL20,GR22}. At fixed model parameters,
this joint relative entropy is bounded by $O(N)$ for all inputs,
hence the uniform exponential decay yields the logarithmic
mixing bound in diamond norm. This extends the usual passage from
MLSI to trace-norm mixing~\cite{KT13} to inputs entangled with an
arbitrary quantum memory.

Our main technical input is a complete spatial factorization of
relative entropy. It bounds the global relative entropy by the sum
of its single-site conditional contributions, each defined as the
loss of relative entropy upon tracing out that site. Spatial
entropy factorization (also called approximate tensorization) has been an effective tool for deriving entropy
decay~\cite{CLP18,BCLPR21}, but noncommuting interactions complicate
the comparison of a chain with its subregions: a marginal of the
Gibbs state need not coincide with the Gibbs state of the restricted
Hamiltonian. We use one-dimensional Gibbs locality, building on
complex-time estimates~\cite{Araki69,PGPH23}, together with stability
of conditional relative entropy under local filtering. An interval
recursion and refinement of bounded regions into single sites then
give the factorization with no additive error and with constants
independent of both system size and reference dimension.

To apply this static estimate to CKG dynamics, we also need control
of conditional relative entropies on growing regions. We compare each single-site conditional
relative entropy with local entropy production, allowing errors
expressed in terms of the larger regions. These errors have summable
coefficients that can be made sufficiently small. Combining
this estimate with the global entropy factorization proves CMLSI.

\subsection{Model and notions}\label{main:sec-basic-notions}
Let $\Lambda$ be a 1D chain with size $N=|\Lambda|$ and $\cH_\Lambda=\bigotimes_{i\in \Lambda} \mathbb{C}^d$ be the total Hilbert space with local dimension $d$. A path is an open chain and a cycle has
periodic boundary conditions. We consider the finite-range, bounded-strength Hamiltonians on $\cH_\Lambda$
\begin{equation}\label{main:eq-model-bounds}
 \begin{gathered}
 H=\sum_{X\subseteq\Lambda}h_X,\qquad
 h_X=h_X^*,\\
 h_X=0\quad\text{if }\diam(X)>r_0,\qquad
 \sup_i\sum_{X\ni i}\norm{h_X}\leq J_0,
 \end{gathered}
\end{equation}
where each $h_X$ acts on $X$ and $\norm{\cdot}$ denotes the operator norm. The interaction terms $h_X$ can be non-commuting and
spatially non-uniform. The interaction diameter $r_0$ and incident strength $J_0$
are bounded independently of the chain size $N$. At inverse temperature $\beta>0$, the Gibbs state is
$\sigma=\e^{-\beta H}/\Tr\e^{-\beta H}.$

We consider either the Gaussian or the shifted-Metropolis generator $\cL$ introduced by Chen, Kastoryano, and Gily\'en \cite{CKG23}.
Its trace-adjoint $\cL_*$ generates a quantum Markov semigroup $T_{t,*}=\e^{t\cL_*}$ converging to the Gibbs state 
\[ \lim_{t\to \infty}\e^{t\cL_*}\rho = \sigma  \qquad \forall \rho \]
The detailed definitions of CKG generators are specified in \eqref{main:def-generator} and Section~\ref{main:not-proposals}.

For two quantum states $\rho,\sigma$, recall the relative entropy $D(\rho \Vert \sigma )=\Tr(\rho (\log \rho-\log \sigma))$ 
with natural logarithms.
For a quantum Markov semigroup $T_{t,*}=e^{\cL_* t}$ with generator $\cL$, the entropy production is defined as
\begin{equation}
 \begin{aligned}
 \cI_{\cL}(\rho)
 =-\left.\frac{\mathrm d}{\mathrm dt}D(T_{t,*}\rho\Vert \sigma)\right|_{t=0}
 =-\Tr\!\left[\cL_*(\rho)
           \bigl(\log\rho-\log\sigma\bigr)\right].
 \end{aligned}
\end{equation}
We say $T_{t,*}$ satisfies the modified log-Sobolev inequality (MLSI) for $\alpha>0$ if 
$$\cI_{\cL}(\rho)\geq2\alpha D(\rho\Vert\sigma) \qquad \text{for every faithful state }\rho, $$
which is equivalent to the exponential decay of relative entropy
\[ D(T_{t,*}\rho\Vert \sigma)\leq\e^{-2\alpha t}D(\rho\Vert \sigma).\]
The ordinary MLSI inequality requires this for any faithful input state $\rho$; the complete inequality requires the same constant for all
finite reference systems $R$ and faithful joint inputs on $\cH_R\otimes\cH_\Lambda$. 

For the mixing estimates, let $\mathcal E_*(X)=\sigma\Tr X$
be the replacement channel and $\norm X_1=\Tr|X|$ the trace norm. For a quantum Markov semigroup $T_{t,*}$ converging to the Gibbs state $\sigma$, define
the trace and diamond distances (\cite[Chapter~3]{Watrous18})
\[
 d_{\mathrm{Tr}}(t)=\frac12\sup_\rho\norm{T_{t,*}(\rho)-\sigma}_1,
 \qquad
 d_\diamond(t)=\frac12\norm{T_{t,*}-\mathcal E_*}_\diamond,
\]
where the first supremum is over input states. We define the corresponding mixing times
\[
 t_{\mathrm{Tr}}(\epsilon)=\inf\{t\geq0:d_{\mathrm{Tr}}(t)\leq\epsilon\},
 \qquad
 t_\diamond(\epsilon)=\inf\{t\geq0:d_\diamond(t)\leq\epsilon\}.
\]

\subsection{Main results}

\begin{thm}[Complete MLSI]
\label{main:thm-complete}
Let $H=\sum_{X\subset \Lambda}h_{X}$ be a $1D$ Hamiltonian on a chain or cycle with local dimension $d$, finite interaction range $r_0\geq 1$ and  
bounded interaction strength $J_0>0$. For every fixed inverse temperature $\beta>0$, there is a constant $\alpha=\alpha(\beta,d,r_0,J_0)>0$ such that for either CKG sampler $\cL$, every finite reference system $R$, and every faithful joint state
$\rho$ on $\cH_R\otimes\cH_\Lambda$,
\begin{equation}
2\alpha D(\rho\Vert\rho_R\otimes\sigma)\le \cI_{\cL}(\rho).
\end{equation}
Consequently, for every joint state $\rho$ and $t\geq0$,
\begin{equation}
 D\!\left((\idmap_R\otimes e^{\cL_* t})(\rho)\,\middle\Vert\,
          \rho_R\otimes\sigma\right)
 \leq\e^{-2\alpha t}D(\rho\Vert\rho_R\otimes\sigma).
\end{equation}
The MLSI constant $\alpha$ is independent of the chain length $N=|\Lambda|$ and the reference dimension $|\cH_R|$.
\end{thm}

Theorem~\ref{main:thm-complete} gives a size-independent entropy decay rate for any joint input $\rho$ with
an arbitrary retained reference system. It applies to the CKG generators at every fixed
inverse temperature $\beta>0$, while the decay rate $\alpha$ depends on $\beta$. Compared with the high-temperature CKG mixing result~\cite{RSFA26},
the all-temperature spectral-gap estimate~\cite{BC25}, and the
subsequent rapid-mixing and ordinary MLSI bounds~\cite{CD26},
our result establishes a size-independent complete entropy-decay
inequality that is uniform over arbitrary quantum reference systems.

\begin{cor}[Mixing time]\label{main:thm-optimal}
Under the assumptions of Theorem~\ref{main:thm-complete},
\begin{equation}
 \frac12\norm{T_{t,*}-\mathcal E_*}_\diamond
 \leq\sqrt{N(\log d+\beta J_0)}\,\e^{-\alpha t}.
\end{equation}
In particular, for $0<\epsilon<1$,
\begin{equation}
 t_{\mathrm{Tr}}(\epsilon)\leq t_\diamond(\epsilon)
 =O_{\beta,d,r_0,J_0}\!\left(\log N+\log\frac1\epsilon\right).
\end{equation}
\end{cor} 
Corollary~\ref{main:thm-optimal} exploits the above entropy estimate to give
a channel-distance bound, uniformly over entangled inputs. The
$\log N+\log(1/\epsilon)$ order improves the $(\log N)^4+\log(1/\epsilon)$ bound in ~\cite{CD26} and is optimal over the class of admissible Hamiltonians. Indeed, for \(H=0\), both samplers reduce to independent single-site depolarizing dynamics, for which a product-state test gives the trace-norm mixing lower bound \(\Omega(\log N+\log(1/\epsilon))\). This establishes the worst-case optimality of our trace- and diamond-norm mixing bounds.

Table~\ref{main:tab-comparison} records selected mixing bounds at fixed model
parameters.
The CKG theorems of~\cite{BC25,CD26} are stated for open chains
with $d=2^q$ and single-site Pauli proposals. The unscaled Pauli
normalization in~\cite{CD26} differs from ours by a factor $d^2$ in
the generator, which does not change the displayed orders at fixed $d$. Kim's main theorem is formulated on cycles, with open boundaries
treated in~\cite[Appendix~D.2]{Kim26}.

\begin{table}[!t]
\centering
\small
\setlength{\tabcolsep}{4pt}
\renewcommand{\arraystretch}{1.15}
\begin{tabularx}{\textwidth}{@{}
>{\raggedright\arraybackslash}p{0.20\textwidth}
>{\raggedright\arraybackslash}p{0.20\textwidth}
>{\raggedright\arraybackslash}p{0.09\textwidth}
>{\raggedright\arraybackslash}p{0.20\textwidth}
>{\raggedright\arraybackslash}X@{}}
\toprule
Reference & Geometry / interaction & Temp. & Dynamics & Mixing time \\
\midrule
Kochanowski et al.~\cite{KACR24}
& 1D / commuting & All & Davies
& $O(\log N+\log(1/\epsilon))$ \\
Rouz\'e et al.~\cite{RSFA26}
& Lattices / general & High & Gaussian CKG
& $O(\log N+\log(1/\epsilon))$ \\
Bergamaschi--Chen~\cite{BC25}
& 1D / general & All & CKG
& $O(N+\log(1/\epsilon))$ \\
Kim~\cite{Kim26}
& 1D / general & All & Fixed-block Petz heat-bath
& $O(\log N+\log(1/\epsilon))$ \\
Capel et al.~\cite{CPS26}
& 1D / general & All & Single-site and regularised HB
& $O(\log N[\log N+\log(1/\epsilon)])$ \\
Chen--Ding~\cite{CD26}
& 1D / nearest-neighbour & All & Gaussian CKG
& $O(\log^4 N+\log(1/\epsilon))$ \\
This work
& 1D / general & All & Gaussian and Metropolis CKG
& $O(\log N+\log(1/\epsilon))$ \\
\bottomrule
\end{tabularx}
\caption{Selected mixing-time bounds.
``All'' means every finite positive temperature; ``General'' allows
noncommuting, spatially nonuniform interactions. The displayed bounds concern trace-norm mixing times, up to
model-dependent constants. Our result also controls diamond-norm
mixing through a volume-independent CMLSI.}
\label{main:tab-comparison}
\end{table}

Our argument has two main ingredients. The first is a
size-independent spatial entropy factorization
(Theorem~\ref{main:thm-static}), which bounds the global relative
entropy and sums of regional conditional relative entropies in
terms of single-site conditional relative entropies. The second
is a local CKG entropy-production estimate
(Theorem~\ref{main:prop-dynamic}), which controls single-site
conditional relative entropies by entropy production up to
regional error terms. These errors are absorbed using the
regional factorization bound.

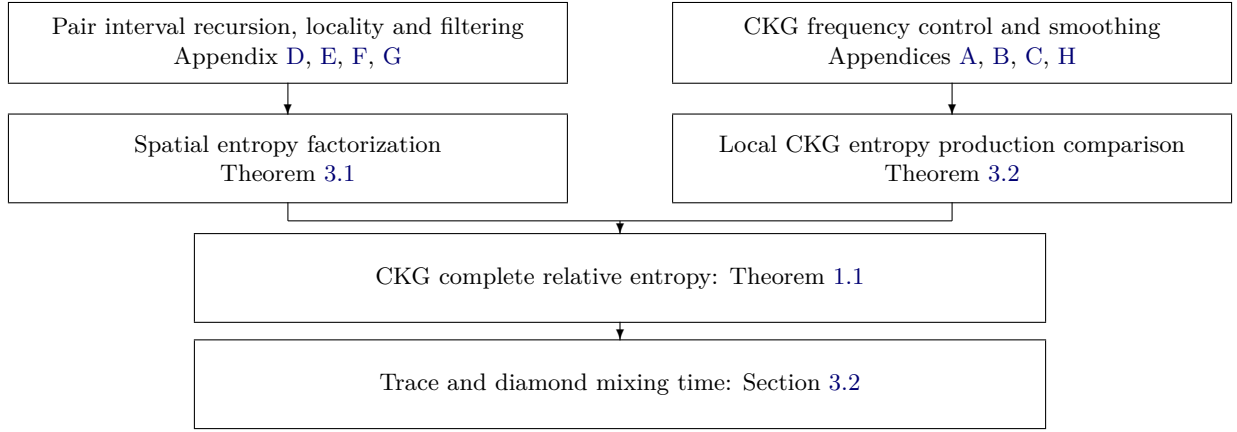
\begin{figure}[!htbp]
\centering
\setlength{\unitlength}{1pt}
\footnotesize
\begin{picture}(460,186)
\put(0,156){\framebox(210,30){\shortstack{
Pair interval recursion, locality and filtering \\
Appendix~\ref{sec:tech-ch_filter_transport}, \ref{sec:tech-ch_block_refinement}, \ref{er:section}, \ref{sec:tech-ch_spatial_factorization} 
}}}
\put(250,156){\framebox(210,30){\shortstack{
CKG frequency control and smoothing\\
Appendices~\ref{noise:section}, \ref{sec:tech-ch_static_calculus}, \ref{gl:section}, \ref{ar:introduction}
}}}
\put(105,156){\vector(0,-1){12}}
\put(355,156){\vector(0,-1){12}}
\put(0,111){\framebox(210,33){\shortstack{
Spatial entropy factorization\\ Theorem~\ref{main:thm-static}
}}}
\put(250,111){\framebox(210,33){\shortstack{
Local CKG entropy production comparison\\
Theorem~\ref{main:prop-dynamic}
}}}
\put(105,111){\line(0,-1){7}}
\put(355,111){\line(0,-1){7}}
\put(105,104){\line(1,0){250}}
\put(230,104){\vector(0,-1){5}}
\put(70,66){\framebox(320,33){\shortstack{
CKG complete relative entropy: Theorem~\ref{main:thm-complete}
}}}
\put(230,66){\vector(0,-1){7}}
\put(70,26){\framebox(320,33){\shortstack{
Trace and diamond mixing time:
Section~\ref{main:sec-contraction-mixing}
}}}
\end{picture}
\caption{Spatial factorization and local CKG entropy-production
estimates yield a complete modified logarithmic Sobolev
inequality with a constant independent of chain size.
Entropy contraction then gives mixing-time bounds.}
\label{main:fig-proof-logic}
\end{figure}

The rest of the paper is organized as follows.
Section~\ref{main:sec-preliminaries} recalls the spin-chain notation,
the generators, and the conditional quantities needed in Section~\ref{main:sec-proof}, where
the two main ingredients are stated and combined to prove Theorem~\ref{main:thm-complete}. The passage from entropy contraction to mixing bounds is discussed in
Section~\ref{main:sec-contraction-mixing}. Figure~\ref{main:fig-proof-logic} illustrates the logical structure of the proof, from
the spatial and dynamical inputs to complete relative entropy decay
and mixing of the CKG samplers. Section~\ref{main:sec-discussion} provides a summary and further discussion.

The appendices contain the detailed proofs of
Theorems~\ref{main:thm-static} and~\ref{main:prop-dynamic}.
Appendix~\ref{noise:section} develops conditional relative entropy stability
and continuity under local noise.
Appendix~\ref{sec:tech-ch_static_calculus} develops tools for second variations of conditional relative entropy and local filtering. Appendix~\ref{gl:section} supplies the Gibbs
locality and physical collar estimates, and
Appendix~\ref{sec:tech-ch_filter_transport} treats filtering across a
boundary. Appendix~\ref{br:section} refines bounded physical blocks into singleton
entropy deficits. Appendix~\ref{er:section} establishes an interval recursion
at the original state and reference on both paths and cycles, and
Appendix~\ref{sf:preliminaries} sums its spatial errors to obtain fixed
window covers and regional charges, thereby proving Theorem~\ref{main:thm-static}. Appendix~\ref{ar:introduction}
establishes arithmetic-energy bounds, paired-noise stability, and local
perturbation estimates, followed by CKG frequency comparisons and physical
tail bounds, thereby proving Theorem~\ref{main:prop-dynamic}.

\begingroup
\makeatletter
\let\addcontentsline\@gobblethree
\makeatother
\noindent\textbf{AI Statement.} The authors acknowledge assistance from
ChatGPT (GPT-6.1 Sol) during exploratory discussions and manuscript preparation. All mathematical arguments and proofs in
this manuscript were verified and reformulated by the authors, who take full responsibility for all content.

\noindent\textbf{Acknowledgement.} Li Gao, Jingyu Guo, and Sihao Wang are
supported in part by the National Natural Science Foundation of China
(grant no.~12401163) and the Department of Science and Technology of
Hubei Province (project nos.~2025EHA041 and 2025AFA044).
\par
\endgroup

\section{Preliminaries}\label{main:sec-preliminaries}
We describe the quantum spin systems and Chen-Kastoryano-Gily\'en dynamics considered in this paper,
then introduce conditional relative entropy and the modified log-Sobolev inequality
used to control mixing times.

\medskip
\noindent{\bf Notation.}
For a finite-dimensional Hilbert space $\cH$, let $\cB(\cH)$ denote its
matrix algebra. We write $\norm{\cdot}$, $\norm{\cdot}_1$, and
$\norm{\cdot}_2$ for the operator, trace, and Hilbert--Schmidt norms,
respectively. Traces are standard unnormalized matrix traces, and
logarithms are natural. A state is described as a positive matrix of trace one;
it is faithful if it is strictly positive on the space under consideration.
Operators on a subsystem are extended by the identity on its complement.

\subsection{Quantum spin chains}\label{main:sec-spin-chains}
Let $\Lambda$ be a one-dimensional labeled chain or cycle with $|\Lambda|=N$ sites,
each site with a local Hilbert space $\C^d$. Thus $N$ is the system size, $d$ is the local dimension, and the full Hilbert space has dimension $d^N$. A chain has two boundary sites
and a cycle is periodic. Distances and diameters are measured in the graph metric with unit edge lengths. For a subset $U\subseteq\Lambda$, write
\[
 \cH_U=\bigotimes_{i\in U}\C^d,\qquad
 \dim\cH_U=d^{|U|},\qquad U^c=\Lambda\setminus U.
\]

We consider Hamiltonians on a one-dimensional chain or cycle 
\[
 \begin{gathered}
 H=\sum_{X\subseteq\Lambda}h_X,\qquad h_X=h_X^*
 \end{gathered}
\]
where $h_X$ is supported on $X$. We assume $H$ has finite interaction range $r_0$ and bounded 
interaction strength $J_0$, 
\[
 \begin{gathered}
 h_X=0\quad\text{if }\diam(X)>r_0,\qquad
 \sup_{i\in\Lambda}\sum_{X\ni i}\norm{h_X}\leq J_0,
 \end{gathered}
\]
We omit the scalar interaction on the empty set.
The range $r_0$ bounds the spatial diameter of an interaction support, and the strength bound $J_0$
controls the sum of the norms of all terms touching a given site. The strength bound gives
\[
 \norm H\leq\sum_{X\ne\emptyset}\norm{h_X}
 \leq\sum_{i\in\Lambda}\sum_{X\ni i}\norm{h_X}\leq NJ_0.
\]
The $h_X$ terms may involve several sites and not commute, and they need not be translation invariant.
For example, nearest-neighbour interactions have range $r_0=1$, and $J_0$ bounds the sum of the norms of the onsite term and the two adjacent bond terms at each site. The Gibbs state at inverse temperature $\beta>0$ is 
\begin{equation}
 \sigma=\frac{\e^{-\beta H}}Z,\qquad Z=\Tr\e^{-\beta H}.
\end{equation}
where $Z$ is the partition function. Throughout the paper, we fix the parameters $\beta,d,r_0,J_0$. These parameters remain fixed as $N$ varies.

For an operator $X\in \cB(\cH_\Lambda)$, let $X_U=\Tr_{U^c}X$ be its marginal on $U$.
We distinguish the marginal $\sigma_U$ of the global Gibbs state $\sigma$ from
the Gibbs state $\sigma^U$ of the restricted Hamiltonian $H_U$:
\begin{equation}
 H_U=\sum_{X\subseteq U}h_X,\qquad
 \sigma^U=\frac{\e^{-\beta H_U}}{Z_U},\qquad
 Z_U=\Tr_U\e^{-\beta H_U}.
\end{equation}
In particular,
$\sigma^\Lambda=\sigma$, whereas $\sigma_U$ and $\sigma^U$ generally
differ because the marginal retains the effect of interactions across
the boundary of $U$. This subscript and superscript convention follows
\cite{CPS26}.

For a site $i\in\Lambda$ and an integer $\ell\geq0$, define the interval
\begin{equation}\label{main:eq-physical-balls}
 S_{i,\ell}=\{x\in\Lambda:\dist(i,x)\leq r_0\ell\}.
\end{equation}
The parameter $\ell$ varies as a localization scale, with physical
radius $r_0\ell$. These intervals are nested, with $S_{i,0}=\{i\}$ a singleton, and satisfy
$|S_{i,\ell}|\leq\min\{N,1+2r_0\ell\}$. They eventually cover the chain.

\subsection{Chen-Kastoryano-Gily\'en generators}\label{main:not-proposals}
We use the Gibbs samplers introduced by Chen, Kastoryano, and Gily\'en~\cite{CKG23}. Let $H=\sum_{X\subseteq\Lambda}h_X$ be a Hamiltonian on a spin chain or cycle described as above. 
For each site $i$, choose a Hermitian basis
$\{A_{i,a}\}_{a=0}^{d^2-1}$ with
\[
 A_{i,0}=\frac{1}{d}\id_i,\qquad
 \Tr_i(A_{i,a}A_{i,b})=\frac{1}{d}\delta_{ab},
\]
The index $i$ labels the proposal site and $a$ labels the onsite basis
element. As $\{\sqrt d A_{i,a}\}_{a}$ gives a Hilbert--Schmidt orthonormal
basis, we have, for every matrix $X$,
\begin{equation}\label{main:eq-proposal-completeness}
 \sum_{a=0}^{d^2-1}A_{i,a}^*XA_{i,a}
 =\frac{\id_i}{d}\otimes\Tr_iX,\qquad
 \sum_{a=0}^{d^2-1}A_{i,a}^*A_{i,a}=\id.
\end{equation}

In the Heisenberg picture, the CKG generator has the form 
\begin{equation}\label{main:def-generator}
 \begin{aligned}
 \cL&=\sum_{i\in\Lambda}\cL_i,\\
 \cL_i(Y)&=\mathrm i\Bigl[\sum_{a}B_{i,a},Y\Bigr]+\sum_{a}\int_{-\infty}^{\infty}\gamma(\omega)
 \left(J_{i,a}(\omega)^*YJ_{i,a}(\omega)
       -\frac12\{J_{i,a}(\omega)^*J_{i,a}(\omega),Y\}\right)
       \,\mathrm d\omega.
 \end{aligned}
\end{equation}
Here $\{X,Y\}=XY+YX$ denotes the anti-commutator. For each site $i$ and a basis element
$a=1,\ldots,d^2-1$,  the jumps $J_{i,a}(\omega), \omega \in \mathbb{R}$ are obtained by filtering Hamiltonian evolution:
\begin{equation}
 J_{i,a}(\omega)=\frac1{\sqrt{2\pi}}
 \int_\R f_\beta(t)\e^{-\mathrm i\omega t}
        \e^{\mathrm itH}A_{i,a}\e^{-\mathrm itH}\,\mathrm dt,
\end{equation}
with Gaussian filter function
\[
 f_\beta(t)=\beta^{-1/2}\left(\frac2\pi\right)^{1/4}
             \e^{-t^2/\beta^2},\qquad
 \int_\R|f_\beta(t)|^2\,\mathrm dt=1.
\]
With this filter function, we consider the Gaussian and Metropolis rates, respectively:
\begin{equation}
 \gamma_G(\omega)=\e^{-(\beta\omega+1)^2/2},\qquad
 \gamma_M(\omega)=\e^{-\max\{\beta\omega+\frac12,0\}}.
\end{equation}
The coherent term is determined as
\begin{align}
 R_{i,a}&=\int_\R\gamma(\omega)J_{i,a}(\omega)^*J_{i,a}(\omega)
                   \,\mathrm d\omega,\notag\\
 B_{i,a}&=\frac{\mathrm i}{2}
          \tanh\!\left(\frac\beta4\ad_H\right)(R_{i,a}),
 \qquad \ad_H(Y)=[H,Y].
 \label{main:eq-original-correction}
\end{align}
The spectral calculus in an energy basis $H|m\rangle=E_m|m\rangle$ is given by:
\[
 (B_{i,a})_{mn}=\frac{\mathrm i}{2}
       \tanh\!\left(\frac\beta4(E_m-E_n)\right)(R_{i,a})_{mn}.
\]
The generator in \eqref{main:def-generator} has GKLS
form~\cite{GKS76,Lindblad76}, hence it generates a quantum Markov semigroup $T_t=e^{\cL t}$ in the Heisenberg picture that is completely positive and unital. The coherent term ensures KMS detailed balance with
respect to $\sigma$~\cite{CKG23}. More precisely, the
Kubo--Martin--Schwinger (KMS) inner product is
\[
 \langle X,Y\rangle_\sigma
 =\Tr[\sigma^{1/2}X^*\sigma^{1/2}Y],
\]
and each site generator satisfies
\[
 \langle X,\cL_i(Y)\rangle_\sigma
 =\langle\cL_i(X),Y\rangle_\sigma,\qquad
 \cL_i(\id)=0,\qquad \cL_{i,*}(\sigma)=0.
\]
Hence each site semigroup and the full semigroup preserve the same
global Gibbs state $\sigma$. Appendix~\ref{sec:tech-ch_actual_rows} verifies the CKG construction with the
normalization used here.

\subsection{Modified log-Sobolev inequality and mixing}
\label{main:sec-conditional}
For two states $\rho,\sigma$, the quantum relative entropy~\cite{Umegaki62} is
\[
 D(\rho\Vert \sigma)=\Tr[\rho(\log\rho-\log  \sigma)]
\]
if $\supp\rho\subseteq\supp\sigma$, and is $+\infty$ otherwise. The relative entropy is nonnegative and non-increasing under a completely positive trace-preserving map $T_*$~\cite{Lindblad75}
\begin{align}\label{eq:dpi}D(T_*\rho\Vert T_*\sigma)\le  D(\rho\Vert \sigma)\ .\end{align}

{\bf \noindent Entropy production.} Let $\sigma=\e^{-\beta H}/Z$ be the Gibbs state and $T_{t,*}=e^{\cL_*t}$ be the Schr\"odinger picture quantum Markov semigroup above that preserves $\sigma=T_{t,*}\sigma$. Each $T_{t,*}$ is a quantum channel. Let $\rho_t=T_{t,*}(\rho)$ be the time evolution of an initial state $\rho$. The entropy
production is
\begin{equation}\label{main:eq-production-definition}
 \begin{aligned}
 \cI_{\cL}(\rho)
 =-\left.\frac{\mathrm d}{\mathrm dt}D(T_{t,*}\rho\Vert \sigma)\right|_{t=0}
 =-\Tr\!\left[\cL_*(\rho)
           \bigl(\log\rho-\log\sigma\bigr)\right].
 \end{aligned}
\end{equation}
$\cI_{\cL}(\rho)$ is always non-negative by differentiating
relative entropy data processing \eqref{eq:dpi} for the semigroup $T_{t,*}$.
Replacing $\cL_*$ with $\cL_{i,*}$, the one site entropy production is \[\cI_i=-\Tr\!\left[\cL_{i,*}(\rho)
           \bigl(\log\rho-\log\sigma\bigr)\right],\] 
  which is the dissipation of the global relative entropy under the
update centered at $i$. As the entropy production is linear over the generator, we have
\[
 \cI_{\cL}(\rho)=\sum_{i\in\Lambda}\cI_i(\rho),\qquad
 \cI_i(\rho)\geq0.
\]

\vspace{10pt}
{\bf \noindent Modified log-Sobolev inequalities.} We say that $\cL$ satisfies a \emph{modified logarithmic Sobolev
inequality} (MLSI) with constant $\alpha>0$ if
\begin{equation}
 2\alpha D(\rho\Vert \sigma)\le \cI_{\cL}(\rho)
\end{equation}
for every faithful state $\rho$. The constant $\alpha$ describes exponential entropy decay rate along the evolution $T_{t,*}$. Indeed, 
the MLSI implies
\[
 \frac{\mathrm d}{\mathrm dt}D(T_{t,*}\rho\Vert \sigma)
 =-\cI_{\cL}(T_{t,*}\rho)\leq-2\alpha D(T_{t,*}\rho\Vert \sigma)\qquad \Longleftrightarrow
 \qquad D(T_{t,*}\rho\Vert \sigma)\leq\e^{-2\alpha t}D(\rho\Vert \sigma).
\]
The entropy contraction extends to singular states by approximation. 

To formulate complete inequalities, let $R$ be a finite reference
system and let $\rho$ be a state on $\cH_R\otimes\cH_\Lambda$, with
$\rho_R=\Tr_\Lambda\rho$ being the marginal on the reference system. The state $\rho$ may be
entangled between the chain $\Lambda$ and the reference $R$, where $R$ may have arbitrary dimension. We consider the amplified  semigroup $\idmap_R\otimes T_{t,*}$ that leaves $R$ untouched. The product state $\rho_R\otimes\sigma$ is invariant under this semigroup. Writing $\mathcal E_*(X)=\sigma\Tr X$ for the replacer channel to the Gibbs state, we have
\[
 (\idmap_R\otimes\mathcal E_*)(\rho)=\rho_R\otimes\sigma.
\] 

We say that $\cL$ satisfies a \emph{complete modified logarithmic Sobolev
inequality} (CMLSI) with constant $\alpha>0$ if, for every finite reference system $R$ and every faithful joint state $\rho\in\cB(\cH_R\otimes\cH_\Lambda)$,  
\begin{equation}
 2\alpha D(\rho\Vert\rho_R\otimes\sigma)\leq\cI_{\cL}^{R}(\rho):=-\Tr\!\left[ (\idmap_R\otimes\cL_*)(\rho)
           \bigl(\log\rho-\log(\rho_R\otimes\sigma)\bigr)\right],
\end{equation}
where $\rho_R\otimes\sigma=(\idmap_R\otimes\mathcal E_*)(\rho)$ is the invariant product state associated with $\rho$.
This is equivalent to exponential relative entropy decay for amplified semigroups $\idmap_R\otimes T_{t,*}$  uniformly over all finite reference systems $R$,
\begin{equation}\label{main:eq-cmlsi-convention2}
 D((\idmap_R\otimes T_{t,*})(\rho)\Vert \rho_R\otimes \sigma )\leq \e^{-2\alpha t}D(\rho\Vert \rho_R\otimes \sigma)
\end{equation}

\vspace{10pt}
{\bf \noindent Mixing times.}
By quantum Pinsker's inequality~\cite{HOT81}, CMLSI \eqref{main:eq-cmlsi-convention2} therefore gives
\[
 \frac12\norm{(\idmap_R\otimes T_{t,*})(\rho)-\rho_R\otimes \sigma}_1
 \leq\e^{-\alpha t} \sqrt{\frac{D(\rho\Vert \rho_R\otimes \sigma)}2}.
\]
To bound the initial entropy uniformly, write
$S(\rho)=-\Tr(\rho\log\rho)$ for the von Neumann entropy. 
By $\norm H\leq NJ_0$ and
$\log \Tr(e^{-\beta H})\leq N\log d+\beta NJ_0$, this yields
\[
 \begin{aligned}
 D(\rho\Vert \rho_R\otimes \sigma)&=-S(\rho)+S(\rho_R)
       +\beta\Tr[\rho(\id_R\otimes H)]+\log Z\leq2N(\log d+\beta J_0),
 \end{aligned}
\]
where we used the conditional entropy bound $S(\rho_R)-S(\rho)\leq N\log d$. 

Taking the supremum over joint input states gives the diamond-norm
bound, by the state-input characterization for a difference of
channels~\cite[Theorem~3.51]{Watrous18}:
\[
 \frac12\norm{T_{t,*}-\mathcal E_*}_\diamond
 \leq\sqrt{N(\log d+\beta J_0)}\,\e^{-\alpha t}.
\]
Consequently, for $0<\epsilon<1$, the mixing times defined in
Section~\ref{main:sec-basic-notions} satisfy
\[
 t_{\mathrm{Tr}}(\epsilon)\leq t_\diamond(\epsilon)
 \leq\frac1{2\alpha}
       \log\!\left(\frac{N(\log d+\beta J_0)}{\epsilon^2}\right).
\]
A size-independent CMLSI constant therefore gives
$O(\log N+\log(1/\epsilon))$ mixing at fixed model parameters.
The same calculation gives the trace-mixing bound from
ordinary MLSI.

\section{Entropy factorization and mixing}\label{main:sec-proof}
\subsection{Conditional entropy and spatial factorization}\label{main:sec-static}
Let $\cH_\Lambda=\bigotimes_{i\in\Lambda}\mathbb{C}^d$ be the Hilbert space
of a quantum spin system on a chain $\Lambda$, and let $H$ be a Hamiltonian
on $\cH_\Lambda$. For states $\rho\geq0$ and $\omega>0$ on $\cH_\Lambda$, define
\begin{equation}
 \Delta_V(\rho\mid \omega)=D(\rho\Vert \omega)-D(\rho_{V^c}\Vert \omega_{V^c}).
\end{equation}
where $\rho_{V^c}=\Tr_{V}\rho$ is the marginal state on $V^c$.
The quantity $\Delta_V$ describes the relative
entropy loss when the subregion $V$ is traced out. We write
$\Delta_i=\Delta_{\{i\}}$ for the conditional relative entropy for a single site $\{i\}$.
The data-processing inequality for partial trace gives, for $U\subseteq V$,
\begin{equation}\Delta_U(\rho\mid \omega)\leq\Delta_V(\rho\mid \omega),
\end{equation}
 by $\rho_{V^c}=\Tr_{V\setminus U}\rho_{U^c}$ and similarly for $\omega $. Fix a Gibbs state $\sigma=e^{-\beta H}/Z$ with inverse temperature $\beta>0$. We use the short notation 
\[ \Delta_V(\rho):=\Delta_V(\rho\mid \sigma)=D(\rho \Vert \sigma)
      -D(\rho_{V^c}\Vert \sigma_{V^c})\ . \]
 
The same definition extends to the joint states with the reference system $\cH_R \otimes \cH_{\Lambda}$. For a joint state $\rho \in \cB(\cH_R \otimes \cH_{\Lambda})$ and a faithful state $\phi\in \cB(\cH_{\Lambda})$ on the spin system, write
\begin{align}
 \Delta_V(\rho;\phi):=\Delta_V(\rho \mid \rho_R\otimes \phi), \qquad \Delta_V(\rho):= \Delta_V(\rho ;\sigma).
\end{align}
Note that for any faithful state $\eta$ on the reference system $R$,
\[
 D(\rho\Vert\eta\otimes\sigma)
 -D(\rho_{V^c}\Vert\eta\otimes\sigma_{V^c})=\Delta_V(\rho),
\]
because the two $\log\eta$ terms cancel under partial trace.
Thus the conditional difference is independent of the auxiliary
choice $\eta$ over $R$. 

The first main ingredient in our proof is the following spatial entropy factorization for the global relative entropy and the physical balls in \eqref{main:eq-physical-balls}.
\begin{thm}[Spatial entropy factorization]
\label{main:thm-static}
 Let $H=\sum_{X\subset \Lambda}h_{X}\in \cB(\bigotimes_{i\in\Lambda}\mathbb{C}^d)$ be a Hamiltonian on a $1D$ chain or cycle $\Lambda$ with local dimension $d$ and $\sigma=e^{-\beta H}/\Tr e^{-\beta H}$ be the Gibbs state with inverse temperature $\beta>0$. 
Assume that $H$ has finite interaction range $r_0$ and strength $J_0$, 
\[
 \begin{gathered}
 h_X=0\quad\text{if }\diam(X)>r_0,\qquad
 \sup_{i\in\Lambda}\sum_{X\ni i}\norm{h_X}\leq J_0,
 \end{gathered}
\]
There exist constants $C_s,C_r$ depending only on $\beta,d,r_0,J_0$ such that for every state $\rho$,
\begin{equation}\label{main:eq-static-singletons}
  D(\rho\Vert \sigma)
  \leq C_s\sum_{i\in \Lambda}\Delta_i(\rho).
\end{equation}
and for every integer $\ell\geq0$, 
\begin{equation}
  \sum_{i\in \Lambda}\Delta_{S_{i,\ell}}(\rho )
  \leq C_r(1+\ell)\sum_{i\in \Lambda}\Delta_i(\rho ).
\end{equation}
The same estimates apply to joint states $\rho$ on $\cH_R\otimes\cH_\Lambda$ with an arbitrary reference system $\cH_R$.
The constants $C_s,C_r$ are independent of the length $|\Lambda|$ and the reference dimension $|\cH_R|$.
\end{thm}
The proof is given in Appendix~\ref{sf:static-proof}.

The above statement is a property of the Gibbs state and does not
involve a choice of sampler. The singleton estimate turns local dissipation bounds into a
global inequality. The regional estimate also controls growing
neighborhoods by a coefficient linear in their radius, with refinement
costs confined to fixed-size physical windows. In comparison,
\cite[Theorem~3.1]{CPS26} uses a weak block factorization with an additive
error and logarithmic block scale. Our bounds here have no additive terms and are uniform over the retained quantum reference.

\subsection{Entropy contraction and mixing}
\label{main:sec-contraction-mixing}
Recall that for $i\in \Lambda$, the single-site entropy production is \[\cI_i=-\Tr\!\left[\cL_{i,*}(\rho)
           \bigl(\log\rho-\log\sigma\bigr)\right],\] 
           where $\cL_i$ is the single-site CKG Lindbladian \eqref{main:def-generator} with either the Gaussian rates $\gamma_G$ or the Metropolis rates $\gamma_M$.
\begin{thm}[Local entropy-production estimate]
\label{main:prop-dynamic}
Under the assumptions of Theorem~\ref{main:thm-static}, for every $\kappa>0$, there are nonnegative coefficients $(z_\ell)_{\ell\geq0}$ and a constant $A>0$, depending only on $\beta,d,r_0,J_0,\kappa$, such that
\begin{equation}\label{main:eq-dynamic-error}
  \sum_{\ell\geq0}z_\ell\e^{r_0\ell}\leq\kappa
\end{equation}
and, for every site $i$,
\begin{equation}\label{main:eq-dynamic-bridge}
  \Delta_i(\rho)
  \leq A\cI_i(\rho)
      +\sum_{\ell\geq0}z_\ell
          \Delta_{S_{i,\ell}}(\rho).
\end{equation}
The estimate holds for all faithful joint states $\rho$ with a finite reference system $R$, for both Gaussian and Metropolis samplers on chains and cycles.
\end{thm}
The proof is given in Appendix~\ref{ar:dynamic-proof}.

Theorem~\ref{main:prop-dynamic} connects the spatial entropy
factorization of Theorem~\ref{main:thm-static} to the CKG dynamics
by bounding each single-site conditional relative entropy $\Delta_i$ in terms
of its entropy production $\cI_i$ and regional error terms $\Delta_{S_{i,\ell}}$ with tunably
small coefficients $z_\ell$. After summing over sites, the regional
factorization allows these errors to be absorbed, while the
singleton factorization yields the size-independent CMLSI in
Theorem~\ref{main:thm-complete}.

We now present the proof of Theorem~\ref{main:thm-complete} based on Theorem~\ref{main:thm-static} and Theorem~\ref{main:prop-dynamic}. Corollary~\ref{main:thm-optimal} then follows from the standard discussion from Section~\ref{main:sec-conditional}.
\begin{proof}[Proof of Theorem~\ref{main:thm-complete}]
Fix $C_s,C_r$ from Theorem~\ref{main:thm-static}. Apply Theorem~\ref{main:prop-dynamic} with
$\kappa=\frac1{2C_r}$. The regional estimate gives
\[
 \sum_i\sum_{\ell\geq0}z_\ell\Delta_{S_{i,\ell}}(\rho) \leq C_r\sum_{\ell\geq0}z_\ell(1+\ell) \sum_i\Delta_i(\rho)\leq\frac12\sum_i\Delta_i(\rho).
\]
where the second inequality uses $1+\ell\leq\e^{r_0\ell}$. Summing \eqref{main:eq-dynamic-bridge} over the sites $i\in \Lambda$ and using
$\cI_{\cL}=\sum_i\cI_i$ gives
\[
  \sum_i\Delta_i(\rho)\leq A\cI_{\cL}(\rho)+\frac12\sum_i\Delta_i(\rho)
\]
which yields
\[
  \sum_i\Delta_i(\rho)\leq 2A\cI_{\cL}(\rho).
\]
Together with \eqref{main:eq-static-singletons}, we obtain 
\[
  D(\rho\Vert\sigma)\leq2C_sA\cI_{\cL}(\rho).
\]
with MLSI constant $\alpha=\frac1{4C_sA}>0$. 
The choices of constants $C_s, C_r,A$ depend only on the fixed model parameters $\beta,d,r_0,J_0$. The same estimates extend to joint states with an arbitrary reference system.
\end{proof}

\subsection{Heat-bath dynamics}

The single-site heat-bath updates of~\cite{CPS26} reconstruct an erased
site by the Petz recovery map~\cite{Petz86} for the global Gibbs reference.
For $i^c=\Lambda\setminus\{i\}$ and
$\sigma_{i^c}=\Tr_i\sigma$, define
\[
 P_{i,*}(X)=\sigma^{1/2}
 \bigl(\id_i\otimes\sigma_{i^c}^{-1/2}(\Tr_iX)
                         \sigma_{i^c}^{-1/2}\bigr)\sigma^{1/2}.
\]
The Petz channels $P_{i,*}$ preserve $\sigma$,
and their powers $\displaystyle E_{i,*}=\lim_{m\to\infty}P_{i,*}^{\,m}$ converge to the trace-adjoint of the
$\sigma$-preserving conditional expectation $E_{i}$ onto the fixed
algebra~\cite[Proposition~2.1]{CPS26}.
Put 
\begin{equation}
 \cL_{\mathrm{HB},*}=\sum_i(P_{i,*}-\idmap),\qquad
 \cL_{\mathrm{RHB},*}=\sum_i(E_{i,*}-\idmap),
\end{equation}
where each site has weight one. 

Write
$Q_{\mathrm{HB},i,*}=P_{i,*}$ and
$Q_{\mathrm{RHB},i,*}=E_{i,*}$. For $a\in\{\mathrm{HB},\mathrm{RHB}\}$,  define $\cI_{a,i}$ by
\eqref{main:eq-production-definition} with generator
$Q_{a,i,*}-\idmap$, and put $\cI_a=\sum_i\cI_{a,i}$.

\begin{cor}[Heat-bath entropy decay]
Let $H=\sum_{X\subset \Lambda}h_{X}\in \cB(\bigotimes_{i\in\Lambda}\mathbb{C}^d)$ be a $1D$ Hamiltonian with bounded interaction range and strength as in \eqref{main:eq-model-bounds}. 
For $a\in\{\mathrm{HB},\mathrm{RHB}\}$ and every joint state $\rho\in \cB(\cH_R\otimes\cH_\Lambda)$ with arbitrary reference system $R$,
\begin{equation}
 D\!\left((\idmap_R\otimes\e^{t\cL_{a,*}})(\rho)
       \,\middle\Vert\,\rho_R\otimes\sigma\right)
 \leq \e^{-t/C_s}D(\rho\Vert\rho_R\otimes\sigma), \qquad \forall \ t \ge 0
\end{equation}
Their trace and diamond mixing times are
$O_{\beta,d,r_0,J_0}(\log N+\log(1/\epsilon))$.
\end{cor}

\begin{proof}
For faithful $\rho$ and either update
$Q_*\in\{\idmap_R\otimes P_{i,*},\idmap_R\otimes E_{i,*}\}$,
expanding the relative entropies gives
\[
 \begin{split}
 &-\Tr\!\left[(Q_*(\rho)-\rho)
             \bigl(\log\rho-\log(\rho_R\otimes\sigma)\bigr)\right]=D(\rho\|\rho_R\otimes \sigma)-D(Q_*(\rho)\Vert\rho_R\otimes\sigma)
                      +D(Q_*(\rho)\Vert\rho).
 \end{split}
\]
The update $P_{i,*}$ factors through $\Tr_i$ followed by the Petz recovery channel,
which maps $\sigma_{i^c}$ to $\sigma$. The data-processing inequality gives
\[
 D(Q_*(\rho)\Vert\rho_R\otimes\sigma)
 \leq D(\Tr_i\rho\Vert\rho_R\otimes\sigma_{i^c})
\]
for $Q_*=\idmap_R\otimes P_{i,*}$, and
the same inequality passes to the power limit defining $E_{i,*}$.
Dropping the last term gives the complete update inequality
of~\cite[Proposition~2.2, Eq.~(2.17)]{CPS26}:
\begin{equation}\label{main:eq-hb-local-deficit}
 \cI_{a,i}(\rho)\geq\Delta_i(\rho),
 \qquad a\in\{\mathrm{HB},\mathrm{RHB}\}.
\end{equation}
Summation and Theorem~\ref{main:thm-static} give
\[
 \cI_a(\rho)\geq\sum_i\Delta_i\geq C_s^{-1}D(\rho\|\rho_R\otimes \sigma).
\]
The CMLSI constant is therefore at least $1/(2C_s)$. The conclusions on mixing times follow from
the argument in
Section~\ref{main:sec-contraction-mixing}, with $\alpha=1/(2C_s)$.
\end{proof}

\begin{rem}[Comparison with previous heat-bath bounds]
The update inequality \eqref{main:eq-hb-local-deficit} is from
\cite{CPS26}. The ordinary
MLSI rate in~\cite{CPS26} is $\Omega(1/\log N)$, with trace mixing
$O(\log N(\log N+\log(1/\epsilon)))$ on chains with $d=2^q$. The improvement here follows from the volume-independent
entropy factorization in Theorem~\ref{main:thm-static}.
Our argument uses conditional relative entropy, rather than local fixed-algebra
CMLSI~\cite[Remarks~2.3--2.4]{CPS26}. 
\end{rem}

\section{Discussion}\label{main:sec-discussion}

We establish a system-size-independent complete modified logarithmic
Sobolev inequality for the Gaussian and Metropolis CKG samplers on
one-dimensional chains and cycles at every fixed positive temperature.
The proof combines complete spatial entropy factorization for Gibbs
states with a local comparison of entropy production.
The spatial factorization is independent of the sampler and also yields
complete entropy contraction for the heat-bath and regularised heat-bath
dynamics.
These results give logarithmic mixing times, uniformly over auxiliary
quantum systems and allowing initial entanglement, strengthening the
connection between equilibrium locality and dynamical relaxation in
noncommuting spin chains.

Two quantitative questions remain.
First, our estimates are uniform in system size but do not give sharp
dependence on temperature.
Tracking this dependence would clarify the low-temperature cost of
entropy contraction and mixing.
Second, it is natural to ask whether analogous complete entropy
factorization holds in higher dimensions under suitable assumptions
on Gibbs locality.
Extending the present argument would require estimates that control
boundary contributions and accumulated errors across overlapping regions.
Such estimates could provide a route to complete entropy contraction
beyond one dimension.

\appendix
\section{Conditional relative entropy under local depolarizing noise}
\label{noise:section}
Let $V\subseteq\Lambda$ be a subset with $V^c=\Lambda\setminus V$ being its complement.
Let $\rho\geq0$ and $\sigma>0$ be states on $\cH_\Lambda=\cH_V\otimes \cH_{V^c}$. Write
\[
 \rho_{V^c}=\Tr_V\rho,\qquad \sigma_{V^c}=\Tr_V\sigma,
\]
as the marginal densities. Recall the conditional relative entropy defined in Section~\ref{main:sec-static}
\[
 \Delta_V(\rho\mid\sigma)
 =D(\rho\Vert\sigma)-D(\rho_{V^c}\Vert\sigma_{V^c}).
\]

For a matrix $A$, define the left and right multiplication operators
by $L_A(X)=AX$ and $R_A(X)=XA$, respectively. For $u>0$, set
\[
 M_u=L_\rho+uR_\sigma,\qquad
 M_{u,V^c}=L_{\rho_{V^c}}+uR_{\sigma_{V^c}}.
\]
Since $\sigma>0$, both operators are strictly positive on their respective Hilbert--Schmidt spaces. Define
\[
 Z_u=M_u^{-1}(\rho-\sigma),\qquad
 Z_{u,V^c}=M_{u,V^c}^{-1}(\rho_{V^c}-\sigma_{V^c}).
\]
For a strictly positive operator $M$ on matrix space, define the weighted $L_2$-norm
\[
 \norm X_M^2=\Tr[X^*M(X)].
\]
In particular,
\[
 \norm X_{M_u}^2=\Tr(\rho XX^*)+u\Tr(\sigma X^*X).
\]
Set
\[
 \mathcal Q_u(\rho\mid\sigma)=\norm{\rho-\sigma}_{M_u^{-1}}^2=\norm{Z_u}_{M_u}^2.
\]
By \cite[Lemma~2.3]{GG26Entropy}, we have
\begin{equation}\label{noise:eq-full-integral}
 D(\rho\Vert\sigma)=\int_0^\infty
 \frac{u}{(1+u)^2}\mathcal Q_u(\rho\mid\sigma)\,du.
\end{equation}
The representation extends to $\rho\geq0$ by approximation.
The following lemma follows from the resolvent identity of
Carlen and Vershynina~\cite[Lemma~2.1]{CV19}, under the parameterization of
the integral representation~\eqref{noise:eq-full-integral}. For convenience, we present a proof here in our notation.
\begin{lem}\label{noise:canonical}
Let $\rho\geq0$ and $\sigma>0$ be states on $\cH_\Lambda=\cH_V\otimes \cH_{V^c}$.
Then,
\begin{equation}\label{noise:eq-conditional-integral}
 \Delta_V(\rho\mid\sigma)=\int_0^\infty\frac{u}{(1+u)^2}
 \norm{Z_u-\id_V\otimes Z_{u,V^c}}_{M_u}^2\,du.
\end{equation}
\end{lem}

\begin{proof} The defining equation for $Z_{u,V^c}$ and partial trace duality give, for every matrix $Y$ on
$\cH_{V^c}$,
\begin{align}\label{eq:partial}
 \Tr\!\left[(\id_V\otimes Y)^*M_u
       (Z_u-\id_V\otimes Z_{u,V^c})\right]
=\Tr\!\left[Y^*(\rho_{V^c}-\sigma_{V^c}
       -\rho_{V^c}Z_{u,V^c}-uZ_{u,V^c}\sigma_{V^c})\right]=0.
\end{align}
Thus $\id_V\otimes Z_{u,V^c}$ is the orthogonal projection of $Z_u$
onto the subspace $\{\id_V\otimes Y:Y\in\cB(\cH_{V^c})\}$
with respect to the $M_u$-weighted inner product. Moreover,
\[
 \norm{\id_V\otimes Z_{u,V^c}}_{M_u}^2
 =\norm{Z_{u,V^c}}_{M_{u,V^c}}^2
 =\mathcal Q_u(\rho_{V^c}\mid\sigma_{V^c}).
\]
Since $\norm{Z_u}_{M_u}^2=\mathcal Q_u(\rho\mid\sigma)$, the Pythagorean identity yields
\begin{equation}
 \mathcal Q_u(\rho\mid\sigma)-\mathcal Q_u(\rho_{V^c}\mid\sigma_{V^c})
 =\norm{Z_u-\id_V\otimes Z_{u,V^c}}_{M_u}^2.
\end{equation}
Therefore, \eqref{noise:eq-conditional-integral} follows from the integral representation
 \eqref{noise:eq-full-integral}.
\end{proof}

Completing the square gives
\begin{equation}
 \mathcal Q_u(\rho\mid\sigma)=\max_Y
 \left\{2\Re\Tr[(\rho-\sigma)Y]-\Tr(\rho YY^*)-u\Tr(\sigma Y^*Y)\right\},
\end{equation}
with unique maximizer $Z_u$. As a supremum of affine functions, $\mathcal Q_u$ is jointly convex. Indeed, it was proved in \cite{LR99} that $\mathcal Q_u$ satisfies the data processing inequality: for every quantum channel $T$ (with inverses taken on $\supp T\sigma$),
\begin{equation}
 \mathcal Q_u(T\rho\mid T\sigma)\leq\mathcal Q_u(\rho\mid\sigma).
\end{equation}
The preceding identities extend to positive matrices $\rho\geq0$ and $\sigma>0$, with the relative entropy $D$ defined by the same trace formula. Namely,
\begin{equation}\label{noise:eq-homogeneous-full-integral}
 \int_0^\infty\frac{u}{(1+u)^2}\mathcal Q_u(\rho\mid\sigma)\,du
 =D(\rho\Vert\sigma)-\Tr\rho+\Tr\sigma<\infty.
\end{equation}
In particular
\[
 \mathcal Q_u(a\rho\mid a\sigma)=a\mathcal Q_u(\rho\mid\sigma),\qquad a>0.
\]

Define the local conditional expectation
\[
 \Omega_V(X)=\frac{\id_V}{d^{|V|}}\otimes\Tr_V X,
\]
which is the orthogonal projection onto the local observables $\id_V\otimes \cB(\cH_{V^c})$ in the Hilbert-Schmidt norm. The complete index of the partial trace conditional expectation $\Omega_V$ gives the standard matrix order inequality
\begin{equation}\label{noise:eq-physical-ceiling}
 X\leq d^{2|V|}\Omega_V(X),\qquad X\geq0.
\end{equation}

For $0\leq r\leq1$, define the local depolarizing map
\[
 \mathcal D_{V,r}=r\idmap+(1-r)\Omega_V,\qquad
 \rho_r=\mathcal D_{V,r}(\rho)=r\rho+(1-r)\Omega_V(\rho),
\]
and similarly for $\sigma_r$.

\begin{lem}\label{noise:conditional-measure}
Let $\rho\geq0$ and $\sigma>0$ be states, and set $K=[0,d^{2|V|}]$.
There is a finite positive Borel measure $\mu$ on $K$, with $\mu(\{0\})=0$, such that
\begin{equation}\label{noise:eq-stieltjes}
 \Delta_V(\rho_r\mid\sigma_r)=r^2\int_K\frac{\mu(d\lambda)}{1-r+r\lambda},\qquad 0<r\leq1,
\end{equation}
and
\begin{equation}
 \mu(K)\leq d^{2|V|}\Delta_V(\rho\mid\sigma).
\end{equation}
Set $G(x)=\Tr(x\Omega_V(\rho)-\rho)_+$, where $A_+$ denotes the positive part.
If $\sigma\geq m\Omega_V(\sigma)$ for $0<m\leq1$, then
\begin{equation}\label{noise:eq-endpoint-density}
 \mu(d\lambda)=G(\lambda)\,d\lambda,\qquad 0<\lambda<m.
\end{equation}
\end{lem}

\begin{proof}
Write $\Delta_r=\Delta_V(\rho_r\mid\sigma_r)$.
The marginals are fixed along the path $\rho_{r,V^c}=\rho_{V^c}$, so $Z_{u,V^c}$ is independent of $r$.
Set
\[
 M_{r,u}=L_{\rho_r}+uR_{\sigma_r},\qquad S_u=\rho-\sigma-M_u(\id_V\otimes Z_{u,V^c}).
\]
Equation~\eqref{eq:partial} gives $\Omega_V(S_u)=0$.
By bimodularity,
\[
 \rho_r-\sigma_r-M_{r,u}(\id_V\otimes Z_{u,V^c})=rS_u.
\]
Lemma~\ref{noise:canonical} therefore yields
\begin{equation}\label{noise:eq-paired-source}
 \Delta_r=r^2\int_0^\infty\frac{u}{(1+u)^2}\norm{S_u}_{M_{r,u}^{-1}}^2\,du,\qquad 0<r\leq1.
\end{equation}

Define
\[
 T_u=M_{0,u}^{-1/2}M_{1,u}M_{0,u}^{-1/2},\qquad v_u=M_{0,u}^{-1/2}S_u.
\]
Since $0<T_u\leq d^{2|V|}\idmap$ and $M_{r,u}=(1-r)M_{0,u}+rM_{1,u}$,
\[
 \norm{S_u}_{M_{r,u}^{-1}}^2=\langle v_u,((1-r)\idmap+rT_u)^{-1}v_u\rangle_{\mathrm{HS}}.
\]
For each $u>0$, $B\mapsto\norm{\id_B(T_u)v_u}_2^2$ is a finite positive spectral measure on $K$.
Its value at each Borel set $B$ is measurable in $u$, since $T_u$ and $v_u$ are continuous
and finite-dimensional Borel functional calculus is measurable.
Define the measure
\[
 \mu(B)=\int_0^\infty\frac{u}{(1+u)^2}\norm{\id_B(T_u)v_u}_2^2\,du,\qquad B\subseteq K\text{ Borel}.
\]
Note that $T_u>0$ gives $\mu(\{0\})=0$.
For every nonnegative Borel function $f$,
\[
 \int_K f(\lambda)\,\mu(d\lambda)=\int_0^\infty\frac{u}{(1+u)^2}\langle v_u,f(T_u)v_u\rangle_{\mathrm{HS}}\,du.
\]
Taking $f(\lambda)=(1-r+r\lambda)^{-1}$, with its value at $0$ immaterial when $r=1$,
and using~\eqref{noise:eq-paired-source} proves~\eqref{noise:eq-stieltjes}.
At $r=1$, this gives
\[
 \mu(K)\leq d^{2|V|}\int_K\frac{\mu(d\lambda)}{\lambda}=d^{2|V|}\Delta_1<\infty.
\]

Work on $\cH_V\otimes\supp\rho_{V^c}$, on which $\rho_0=\Omega_V(\rho)>0$.
The positive-part integral formula for relative entropy~\cite[Theorem~6]{Frenkel23},
written in hockey-stick form, gives after substituting $t=1-r+rx$
\[
 \frac{D(\rho_r\Vert\rho_0)}{r^2}=\int_0^1\frac{\Tr(x\rho_0-\rho)_+}{1-r+rx}\,dx+\int_1^{d^{2|V|}}\frac{\Tr(\rho-x\rho_0)_+}{1-r+rx}\,dx,\qquad 0<r<1.
\]
The upper limit follows from $\rho\leq d^{2|V|}\rho_0$.
These integrals define a finite positive measure $\mu_{\mathrm{in}}$ on $K$ satisfying
\[
D(\rho_r\Vert\rho_0)=r^2\int_K\frac{\mu_{\mathrm{in}}(d\lambda)}{1-r+r\lambda},\qquad 0<r<1,
\]
and
\begin{equation}\label{noise:eq-input-density}
 \mu_{\mathrm{in}}(d\lambda)=G(\lambda)\,d\lambda,\qquad 0<\lambda<1.
\end{equation}

Now assume $\sigma\geq m\sigma_0$, where $\sigma_0=\Omega_V(\sigma)$.
Since the marginals are fixed,
\[
 H(r):=\Delta_r-D(\rho_r\Vert\rho_0)=-\Tr\!\left[\rho_r(\log\sigma_r-\log\sigma_0)\right].
\]
For $s\geq0$, define the two-sided map
\[
 \mathcal R_s(X)=(\sigma_0+s\id)^{-1/2}X(\sigma_0+s\id)^{-1/2}.
\]
It is self-adjoint for the Hilbert--Schmidt inner product and preserves the local algebra.
Since $\Omega_V(\sigma_0-\sigma)=0$, this gives
\[
 \Tr\!\left[\mathcal R_s(\rho_0)\mathcal R_s(\sigma_0-\sigma)\right]=\Tr\!\left[\mathcal R_s^2(\rho_0)(\sigma_0-\sigma)\right]=0.
\]
The reference bounds also give
\[
 m\id\leq\id-\mathcal R_s(\sigma_0-\sigma)=\mathcal R_s(\sigma+s\id)\leq d^{2|V|}\id.
\]
The resolvent formula for the logarithm yields
\[
 H(r)=r\int_0^\infty\Tr\!\left[\mathcal R_s(\rho_r)\mathcal R_s(\sigma_0-\sigma)\bigl(\id-r\mathcal R_s(\sigma_0-\sigma)\bigr)^{-1}\right]ds.
\]
Using $\rho_r=\rho_0+r(\rho-\rho_0)$, the preceding cancellation removes the term of order $r$.
Since $\sigma_0>0$, $\mathcal R_s$ is bounded near $s=0$ and is $O(s^{-1})$ as $s\to\infty$.
In particular,
\[
 \int_0^\infty\!\left(\norm{\mathcal R_s(\rho_0)}_1\norm{\mathcal R_s(\sigma_0-\sigma)}^2+\norm{\mathcal R_s(\rho-\rho_0)}_1\norm{\mathcal R_s(\sigma_0-\sigma)}\right)ds<\infty.
\]
For $0<r<(2d^{2|V|})^{-1}$, the geometric series for
$\bigl(\id-r\mathcal R_s(\sigma_0-\sigma)\bigr)^{-1}$ and $(1-r+r\lambda)^{-1}$
are uniformly dominated by $\sum_{n\geq0}2^{-n}$.
The preceding integrable bound and the finiteness of $\mu$ and $\mu_{\mathrm{in}}$
justify termwise integration after removing the vanishing linear term.
Comparing coefficients of $r^{n+2}$ gives, for $n\geq0$,
\[
 \int_K(1-\lambda)^n(\mu-\mu_{\mathrm{in}})(d\lambda)=\int_0^\infty\Tr\!\left[\mathcal R_s(\rho_0)\mathcal R_s(\sigma_0-\sigma)^{n+2}+\mathcal R_s(\rho-\rho_0)\mathcal R_s(\sigma_0-\sigma)^{n+1}\right]ds.
\]
Since the functions $(1-\lambda)^n$ span the polynomials, linear combinations give the identity
for every real polynomial $p$.
On the right, the matrix inside the trace at $n=0$ is multiplied on the right by
$p\bigl(\id-\mathcal R_s(\sigma_0-\sigma)\bigr)$.

Let $f\in C(K)$ be real and compactly supported in $(0,m)$,
and choose real polynomials $p_j\to f$ uniformly on $K$.
The spectral bound implies
\[
 \norm{p_j\bigl(\id-\mathcal R_s(\sigma_0-\sigma)\bigr)}\leq\sup_{\lambda\in[m,d^{2|V|}]}|p_j(\lambda)|\longrightarrow0,
\]
uniformly in $s$.
The integrable bound above and the finiteness of $\mu$ and $\mu_{\mathrm{in}}$
allow passage to the limit in the polynomial identities, giving
$\int_K f(\lambda)\,(\mu-\mu_{\mathrm{in}})(d\lambda)=0$.
Thus $\mu=\mu_{\mathrm{in}}$ on $(0,m)$,
and~\eqref{noise:eq-input-density} proves~\eqref{noise:eq-endpoint-density}.
\end{proof}
The function $G$ is convex and satisfies $G(0)=0$.
Hence $G(x)/x$ is nondecreasing for $x>0$.

\begin{lem}\label{noise:entropy-retention}
Let $V\subseteq\Lambda$, and let $\rho\geq0$ and $\sigma>0$ be states on $\cH_\Lambda$.
Set\[
 \rho_r=r\rho+(1-r)\Omega_V(\rho),\qquad \sigma_r=r\sigma+(1-r)\Omega_V(\sigma).
\]
For $0<r\leq1$,
\begin{equation}
 \Delta_V(\rho_r\mid\sigma_r)\leq r\Delta_V(\rho\mid\sigma).
\end{equation}
If $\sigma\geq m\Omega_V(\sigma)$ for $0<m\leq1$, then
\begin{equation}\label{noise:eq-retention}
 b_m(r)\Delta_V(\rho\mid\sigma)\leq\Delta_V(\rho_r\mid\sigma_r),
\end{equation}
where
\[
 b_m(r)=r-\frac{1-r}{m}\log\left(1+\frac{rm}{1-r}\right)>0,
\]
with $b_m(1)=1$ defined by continuity.
For $0<r<1$, the coefficient satisfies
\begin{equation}\label{noise:eq-retention-loss}
 1-b_m(r)=(1-r)+\frac{1-r}{m}\log\left(1+\frac{rm}{1-r}\right)
 \leq\frac{1-r}{m}\left[1+\log\left(1+\frac1{1-r}\right)\right],
\end{equation}
so $b_m(r)\to1$ as $r\to 1^-$.
\end{lem}

\begin{proof}
Write $\Delta_r=\Delta_V(\rho_r\mid\sigma_r)$ and $\Delta_1=\Delta_V(\rho\mid\sigma)$. The first inequality $\Delta_r\leq r\Delta_1$ follows from joint convexity of relative entropy and the invariance of the marginals $\rho_{r,V^c}=\rho_{V^c}$.

Now assume $\sigma\geq m\Omega_V(\sigma)$ and $0<r<1$.
Let $\mu$ be the measure on $K=[0,d^{2|V|}]$ in Lemma~\ref{noise:conditional-measure}.
The scalar bounds
\[
 \frac{t}{1+t}\leq\log(1+t)<t,\qquad t>0,
\]
give $b_m(r)>0$ and, for $\lambda\in K\cap [m,\infty)$, by the monotonicity of $\lambda\mapsto \frac{\lambda}{1-r+r\lambda}$
\[
 \frac{r^2\lambda}{1-r+r\lambda}\geq\frac{r^2m}{1-r+rm}\geq
 \frac{r^2}{m}\int_0^m\frac{\lambda}{1-r+r\lambda}\,d\lambda=b_m(r)
\]
Hence, on this region,
$\frac{r^2}{1-r+r\lambda}\geq\frac{b_m(r)}{\lambda}$.
Thus the contribution to $\Delta_r$ is at least $b_m(r)$ times
the corresponding contribution to $\Delta_1$.
For $\lambda\in (0,m)$, the integral contributions from Lemma~\ref{noise:conditional-measure} to $\Delta_r$ and $\Delta_1$ are
\[
 r^2\int_0^m\frac{G(\lambda)}{1-r+r\lambda}\,d\lambda,\qquad
 \int_0^m\frac{G(\lambda)}\lambda\,d\lambda
\]
respectively.
Since $G(\lambda)/\lambda$ and $\lambda/(1-r+r\lambda)$ are nondecreasing, the integral Chebyshev inequality yields
\[
\begin{aligned}
 r^2\int_0^m\frac{G(\lambda)}\lambda\frac{\lambda}{1-r+r\lambda}\,d\lambda
 \geq\frac{r^2}{m}
 \left(\int_0^m\frac{\lambda}{1-r+r\lambda}\,d\lambda\right)
 \left(\int_0^m\frac{G(\lambda)}\lambda\,d\lambda\right)=b_m(r)\int_0^m\frac{G(\lambda)}\lambda\,d\lambda.
\end{aligned}
\]
Combining the estimates on the two regions proves~\eqref{noise:eq-retention}. Finally, $m\leq1$ and $rm\leq1$ give
\[
 1-b_m(r)=1-r+\frac{1-r}{m}\log\left(1+\frac{rm}{1-r}\right)
 \leq\frac{1-r}{m}\left[1+\log\left(1+\frac1{1-r}\right)\right].
\]
The right side tends to zero as $r\to 1$, proving~\eqref{noise:eq-retention-loss} and the limit.
\end{proof}

For $0<m\leq1$ and $0<r<1$, set
\[
 \omega_m(r)=1-b_m(1-r)
 =r+\frac{r}{m}\log\left(1+\frac{(1-r)m}{r}\right).
\]
By~\eqref{noise:eq-retention-loss},
\[
 \omega_m(r)=O_m\!\left(r(1+\log(1/r))\right)\longrightarrow0 \text{ as } r\to 0.
\]

\begin{cor}\label{noise:union-continuity}
Let $\rho\geq0$ and $\sigma>0$ be states on $\cH_\Lambda$, and let $I,J\subseteq\Lambda$ be two subsets.
Assume
\[
 \sigma\geq m\Omega_J(\sigma),\qquad 0<m\leq1.
\]
For $0<r<1$, set
\[
 \rho_r=r\rho+(1-r)\Omega_J(\rho),\qquad \sigma_r=r\sigma+(1-r)\Omega_J(\sigma).
\]
Then
\begin{equation}\label{noise:eq-union-continuity}
 \abs{\Delta_I(\rho_r\mid\sigma_r)-\Delta_I(\rho\mid\sigma)}
 \leq\omega_m(1-r)\Delta_{I\cup J}(\rho\mid\sigma).
\end{equation}
\end{cor}

\begin{proof}
Put $J'=J\setminus I$.
Define the full and reduced entropy losses
\[
\begin{aligned}
 L_{\mathrm{full}}&=D(\rho\Vert\sigma)-D(\rho_r\Vert\sigma_r),\\
 L_{\mathrm{red}}&=D(\rho_{I^c}\Vert\sigma_{I^c})
 -D((\rho_r)_{I^c}\Vert(\sigma_r)_{I^c}).
\end{aligned}
\]
Then
\begin{align}\label{eq:diff}
 \Delta_I(\rho_r\mid\sigma_r)-\Delta_I(\rho\mid\sigma)=L_{\mathrm{red}}-L_{\mathrm{full}}.
\end{align}
Note that the partial traces satisfy
\[
 \Tr_I\mathcal D_{J,r}=\mathcal D_{J',r}\Tr_I,\qquad
 \sigma_{I^c}\geq m\Omega_{J'}(\sigma_{I^c}).
\]
Since the $J^c$-marginals are unchanged,
\[
 L_{\mathrm{full}}=\Delta_J(\rho\mid\sigma)-\Delta_J(\rho_r\mid\sigma_r),
\]
and the same identity holds for the reduced pair $\rho_{I^c},\sigma_{I^c}$ with $J'$.
Applying Lemma~\ref{noise:entropy-retention} to both pairs yields
\[
 0\leq L_{\mathrm{full}}\leq\omega_m(1-r)\Delta_J(\rho\mid\sigma),\qquad
 0\leq L_{\mathrm{red}}\leq\omega_m(1-r)\Delta_{J'}(\rho_{I^c}\mid\sigma_{I^c}).
\]
Data processing implies that $\Delta_J(\rho\mid\sigma)\leq\Delta_{I\cup J}(\rho\mid\sigma)$, while the chain rule gives
\[
 \Delta_{I\cup J}(\rho\mid\sigma)
 =\Delta_I(\rho\mid\sigma)+\Delta_{J'}(\rho_{I^c}\mid\sigma_{I^c}).
\]
Thus both losses lie in
$[0,\omega_m(1-r)\Delta_{I\cup J}(\rho\mid\sigma)]$,
which proves~\eqref{noise:eq-union-continuity}
by~\eqref{eq:diff}.
\end{proof}

\begin{rem}
All the estimates in this section remain valid in the presence of
an arbitrary finite-dimensional reference system $R$.
Indeed, the proofs apply with $\cH_\Lambda$ replaced by
$\cH_\Lambda\otimes\cH_R$ and $\cH_{V^c}$ replaced by
$\cH_{V^c}\otimes\cH_R$, with each local map acting trivially on $R$.
\end{rem}
\section{Conditional relative entropy calculus}\label{sec:tech-ch_static_calculus}
\label{sc:section}

\subsection{Conditional second variation}

For a state $\rho>0$ on $\cH_\Lambda$, define the
Bogoliubov--Kubo--Mori (BKM) operator~\cite{Mori65,PT93} by
\[
 \Gamma_\rho(X)=\int_0^1\rho^sX\rho^{1-s}\,ds.
\]
This is a strictly positive operator on Hilbert--Schmidt matrix space.
Its inverse is the derivative of the matrix logarithm:
\begin{equation}
 D(\log)_\rho[X]=\Gamma_\rho^{-1}(X)
 =\int_0^\infty(\rho+s\id)^{-1}X(\rho+s\id)^{-1}\,ds.
\end{equation}

Let $A=A^*$ satisfy $\Tr A=0$, and set $\rho_t=\rho+tA$
on an interval where $\rho_t>0$.
For any fixed state $\sigma>0$ on the same space,
\[
 \begin{aligned}
 \frac{d}{dt}D(\rho_t\Vert\sigma)
 =\Tr[A(\log\rho_t-\log\sigma)], \qquad
 \frac{d^2}{dt^2}D(\rho_t\Vert\sigma)
 =\Tr[A\Gamma_{\rho_t}^{-1}(A)]
 =\norm{A}_{\Gamma_{\rho_t}^{-1}}^2.
 \end{aligned}
\]
In particular, the second derivative is independent of $\sigma$.

Set $M_u=L_\rho+uR_\rho$ for $u>0$.
Applying the second derivative to the relative entropy integral representation
\eqref{noise:eq-full-integral} and using dominated convergence, we obtain
\begin{equation}\label{sc:bkm-quadratic-integral}
 \norm{A}_{\Gamma_\rho^{-1}}^2
 =2\int_0^\infty\frac{u}{(1+u)^2}
 \norm{A}_{M_u^{-1}}^2\,du.
\end{equation}
 
The BKM quadratic form satisfies the order
comparison~\cite[Lemma~2.1]{GR22}:
if $\rho,\sigma>0$ are states with $a\rho\leq\sigma\leq b\rho$,
where $a,b>0$, then for every $A=A^*$ with $\Tr A=0$,
\begin{equation}\label{sc:bkm-quadratic-order}
 b^{-1}\norm{A}_{\Gamma_\rho^{-1}}^2
 \leq\norm{A}_{\Gamma_\sigma^{-1}}^2
 \leq a^{-1}\norm{A}_{\Gamma_\rho^{-1}}^2.
\end{equation}
This follows from inverse-order monotonicity and the integral
representation~\eqref{sc:bkm-quadratic-integral}.

For a state $\rho>0$, a Hermitian trace-zero matrix $A$,
and an erased subset $V\subseteq\Lambda$, define the conditional second variation by
\[
 \mathcal C_V(\rho;A):=\left.\frac{d^2}{dt^2}
 \Delta_V(\rho+tA\mid\sigma)\right|_{t=0}
 =\norm{A}_{\Gamma_\rho^{-1}}^2
 -\norm{A_{V^c}}_{\Gamma_{\rho_{V^c}}^{-1}}^2,
\]
where the above identity holds for any fixed state $\sigma>0$.

Recall that the range
$\operatorname{Ran}\Omega_V
=\{\id_V\otimes Y:Y\in\mathcal B(\cH_{V^c})\}$
consists of local observables acting trivially on $V$.
For $u>0$ and $Z\in \cB(\cH_\Lambda)$, we define the weighted $L_2$-norm distance
\[
 \operatorname{dist}_{M_u}(Z,\operatorname{Ran}\Omega_V)
 =\inf_{Y\in\operatorname{Ran}\Omega_V}\norm{Z-Y}_{M_u}.
\]
The following is the second-order derivation of Lemma \ref{noise:canonical}. 
\begin{lem}\label{sc:conditional-second-variation}
Let $\rho>0$ be a state on $\cH_\Lambda$,
let $A=A^*$ satisfy $\Tr A=0$, and let $V\subseteq\Lambda$.
For $M_u=L_\rho+uR_\rho$ and $Z_u=M_u^{-1}(A)$,
\begin{equation}\label{sc:conditional-quadratic-quotient}
 \mathcal C_V(\rho;A)
 =2\int_0^\infty\frac{u}{(1+u)^2}
 \operatorname{dist}_{M_u}(Z_u,\operatorname{Ran}\Omega_V)^2\,du
 \geq0.
\end{equation}
In particular, for fixed $\sigma>0$, the conditional relative entropy $\rho\mapsto \Delta_V(\rho\mid\sigma)$
is convex on the state space.
\end{lem}

\begin{proof}
Set
\[
 M_{u,V^c}=L_{\rho_{V^c}}+uR_{\rho_{V^c}},
 \qquad
 Z_{u,V^c}=M_{u,V^c}^{-1}(A_{V^c}).
\]
As in the proof of Lemma~\ref{noise:canonical}, the partial-trace duality gives
\[
 \begin{aligned}
 \Tr\!\left[
 (\id_V\otimes Y)^*
 M_u(Z_u-\id_V\otimes Z_{u,V^c})
 \right]=
 \Tr\!\left[
 Y^*(A_{V^c}-M_{u,V^c}(Z_{u,V^c}))
 \right]=0
 \end{aligned}
\]
for every matrix $Y$ on $\cH_{V^c}$.
Thus $\id_V\otimes Z_{u,V^c}$ is the $M_u$-weighted orthogonal projection
of $Z_u$ onto $\operatorname{Ran}\Omega_V$.
Moreover,
\[
 \norm{Z_u}_{M_u}^2=\norm{A}_{M_u^{-1}}^2,
 \qquad
 \norm{\id_V\otimes Z_{u,V^c}}_{M_u}^2
 =\norm{A_{V^c}}_{M_{u,V^c}^{-1}}^2.
\]
The Pythagorean identity therefore gives
\[
 \begin{aligned}
 \norm{A}_{M_u^{-1}}^2
 -\norm{A_{V^c}}_{M_{u,V^c}^{-1}}^2
 =\norm{Z_u-\id_V\otimes Z_{u,V^c}}_{M_u}^2
 =\operatorname{dist}_{M_u}(Z_u,\operatorname{Ran}\Omega_V)^2.
 \end{aligned}
\]
Integrating with weight $2u/(1+u)^2$ and applying
\eqref{sc:bkm-quadratic-integral} to the full and marginal quadratic forms
proves~\eqref{sc:conditional-quadratic-quotient}.
\end{proof}

The conditional quadratic form  $A\mapsto\mathcal C_V(\rho;A)$ is therefore positive semi-definite,
although it may be degenerate. Let $\rho,\sigma>0$ be states on $\cH_\Lambda$, and set
\[
 \rho_t=\sigma+t(\rho-\sigma),\qquad 0\leq t\leq1.
\]
Integrating the second derivative formula twice 
gives (see \cite[Lemma 2.3]{GR22})
\[
 D(\rho\Vert\sigma)
 =\int_0^1(1-t)\norm{\rho-\sigma}_{\Gamma_{\rho_t}^{-1}}^2\,dt.
\]
Subtracting the same formula for the marginal pair $(\rho_{V^c},\sigma_{V^c})$ yields
\begin{equation}
 \Delta_V(\rho\mid\sigma)
 =\int_0^1(1-t)\mathcal C_V(\rho_t;\rho-\sigma)\,dt.
\end{equation}

\subsection{Local filter comparison}
For local filtering, we use the extensions of $D$ and $\Delta_V$ for positive operators
from Appendix~\ref{noise:section}:
for $\rho\geq0,\sigma>0$ and $a,b>0$,
\[
 D(a\rho\Vert b\sigma)
 =aD(\rho\Vert\sigma)+a\Tr\rho\log(a/b), \qquad \Delta_V(\rho\mid\sigma)
 =D(\rho\Vert\sigma)-D(\rho_{V^c}\Vert\sigma_{V^c})
\]
Moreover, 
\begin{equation}\label{sc:independent-scaling}
 \Delta_V(a\rho\mid b\sigma)=a\Delta_V(\rho\mid\sigma),
\end{equation}
where the additional terms cancel in the conditional difference since partial trace preserves the trace.
For $\rho\ne0$, independent normalization and partial-trace data
processing~\cite{Lindblad75} give
\[
 \Delta_V(\rho\mid\sigma)
 =(\Tr\rho)\,
 \Delta_V\!\left(
 \frac{\rho}{\Tr\rho}\,\middle|\,
 \frac{\sigma}{\Tr\sigma}
 \right)\geq0. 
\]

\begin{lem}\label{sc:filters}
Let $\rho\geq0$ and $\sigma>0$ be states on $\cH_\Lambda$,
let $V\subseteq\Lambda$, and let
$K=K_V\otimes\id_{V^c}$ with $K_V$ invertible.
Set
\begin{equation}
 \rho^K=\frac{K\rho K^*}{\Tr(K\rho K^*)},
 \qquad
 \sigma^K=\frac{K\sigma K^*}{\Tr(K\sigma K^*)}.
\end{equation}
Then
\begin{equation}\label{sc:normalized-filter-detailed}
 \norm{K^{-1}}^{-2}\Delta_V(\rho\mid\sigma)
 \leq
 \Tr(K\rho K^*)\,\Delta_V(\rho^K\mid\sigma^K)
 \leq
 \norm{K}^2\Delta_V(\rho\mid\sigma).
\end{equation}
\end{lem}

\begin{proof}
We first work with positive matrices $\rho\geq0,\sigma>0$.
Set
\[
 C_0=\frac{K_V}{\norm{K_V}},
 \qquad
 C_1=(\id_V-C_0^*C_0)^{1/2},
\]
and define
\[
 \rho_j=(C_j\otimes\id_{V^c})\rho(C_j^*\otimes\id_{V^c}),
 \qquad
 \sigma_j=(C_j\otimes\id_{V^c})\sigma(C_j^*\otimes\id_{V^c}).
\]
Both are considered on
$\operatorname{Ran}C_j\otimes\cH_{V^c}$, where $\sigma_j>0$.

Since $C_0^*C_0+C_1^*C_1=\id_V$, the map
$\rho\mapsto\rho_0 \oplus\rho_1$ is completely positive and trace preserving.
Data processing~\cite{Lindblad75}, extended to positive matrices
by independent normalization, and direct-sum additivity give
\[
 D(\rho_0\Vert\sigma_0)+D(\rho_1\Vert\sigma_1)\leq D(\rho\Vert\sigma).
\]
Note that after taking partial trace on $V$, $V^c$-marginals satisfy
\[
 (\rho_0)_{V^c}+(\rho_1)_{V^c}=\Tr_V( (C_0\otimes\id_{V^c})\rho(C_0^*\otimes\id_{V^c}) +(C_1\otimes\id_{V^c})\rho(C_1^*\otimes\id_{V^c}) )=\Tr_V\rho =\rho_{V^c},
\]
and similarly  $(\sigma_0)_{V^c}+(\sigma_1)_{V^c}=\sigma_{V^c}$.
Applying data processing for $\omega_0\oplus \omega_1\mapsto \omega_0+\omega_1$ therefore gives
\[
D((\rho_0)_{V^c}\Vert(\sigma_0)_{V^c})+D((\rho_1)_{V^c}\Vert(\sigma_1)_{V^c})
 \geq D(\rho_{V^c}\Vert\sigma_{V^c}).
\]
Subtracting yields
\begin{equation}
 \Delta_V(\rho_0\mid\sigma_0)+\Delta_V(\rho_1\mid\sigma_1)
 \leq\Delta_V(\rho\mid\sigma).
\end{equation}
Each summand is nonnegative, so
\[
 \Delta_V(\rho_0\mid\sigma_0)\leq\Delta_V(\rho\mid\sigma).
\]
For $j=0$, the scaling gives
\[
 \Delta_V(K\rho K^*\mid K\sigma K^*)
 =\norm{K}^2\Delta_V(\rho_0\mid\sigma_0)
 \leq\norm{K}^2\Delta_V(\rho\mid\sigma).
\]
Applying this upper bound to $K^{-1}$ gives the reverse bound:
\[
 \Delta_V(\rho\mid\sigma)
 \leq\norm{K^{-1}}^2\Delta_V(K\rho K^*\mid K\sigma K^*).
\]

For the state $\rho^K=\frac{K\rho K^*}{\Tr(K\rho K^*)}$, the scaling gives
\[
 \Tr(K\rho K^*)\,\Delta_V(\rho^K\mid\sigma^K)
 =\Delta_V(K\rho K^*\mid K\sigma K^*),
\]
which proves the claim by the two bounds above.
\end{proof}

For $K$ and $\rho^K$ as in Lemma~\ref{sc:filters},
testing against positive matrices on $V^c$ and using
$\norm{K^{-1}}^{-2}\id\leq K^*K\leq\norm{K}^2\id$ gives
\begin{equation}
 \norm{K^{-1}}^{-2}\rho_{V^c}
 \leq
 \Tr(K\rho K^*)\,(\rho^K)_{V^c}
 \leq
 \norm{K}^2\rho_{V^c}.
\end{equation}
Hence $\rho_{V^c}$ and $(\rho^K)_{V^c}$,
as well as their corresponding further marginals, have the same support.

All preceding statements remain valid with an arbitrary finite reference system $R$,
with all maps acting trivially on $R$ and all physical partial traces retaining it.
In this setting, for a state $\rho\geq0$ on $R\Lambda$, and a state $\sigma>0$ on $\Lambda$,
the conditional relative entropy $\Delta_V(\rho|\sigma)$ is replaced by $\Delta_V(\rho;\sigma)$
\begin{equation}\label{sc:reference-cancellation}
 \Delta_V(\rho;\sigma)=D(\rho\Vert\eta\otimes\sigma)
 -D(\rho_{V^c}\Vert\eta\otimes\sigma_{V^c}),
\end{equation} which holds for any $\eta>0$ on $R$, 
as $\log\eta$ terms cancel by partial trace.
Thus the conditional formulas above, the continuity estimate in Corollary~\ref{noise:union-continuity}, and convexity from Lemma~\ref{sc:conditional-second-variation} also apply to conditional relative entropy $\Delta_V(\rho;\sigma)$ with references, including inputs with different reference marginals.

\section{Locality and filtering}
\label{gl:section}

\subsection{Onsite evolution and local comparison bounds}

Let $H=\sum_{X\subseteq\Lambda}h_X$ be a Hamiltonian on a one-dimensional
spin chain or cycle with Hilbert space
$\mathcal H_\Lambda=\bigotimes_{i\in\Lambda}\mathbb C^d$, interaction
range $r_0$, and strength $J_0$, as in
Section~\ref{main:sec-spin-chains}.
We take the range bound $r_0\geq1$ to be an integer.
We study locality and filtering properties of the Gibbs state
$\sigma=\e^{-\beta H}/\Tr\e^{-\beta H}$ at inverse temperature $\beta>0$.
We omit the scalar interaction on the empty set.
Unless otherwise stated, constants depend only on $\beta,d,r_0,J_0$
and are independent of $N=|\Lambda|$.

Write $\tau_H^z(Y)=\e^{\ii zH}Y\e^{-\ii zH}$.
For an onsite operator $X$ at $i$ with $\norm{X}\leq1$, let
$S_{i,\ell}=\{j\in\Lambda:\operatorname{dist}(j,i)\leq r_0\ell\}$
and define
\[
 X_0(z)=\tau_{H_{S_{i,0}}}^z(X),\qquad
 X_\ell(z)=\tau_{H_{S_{i,\ell}}}^z(X)
 -\tau_{H_{S_{i,\ell-1}}}^z(X),\quad \ell\geq1.
\]
Each shell is supported on $S_{i,\ell}$ and vanishes for $\ell\geq1$
whenever $S_{i,\ell-1}=\Lambda$.

The following lemma gives an explicit version of Araki's locality
estimates~\cite{Araki69}; see also~\cite[Proposition~3.2(i)]{BCP22}
and~\cite[Theorem~2.2]{PGPH23}.
We include a self-contained proof to obtain explicit constants valid
uniformly for chains, cycles, and restricted Hamiltonians.

\begin{lem}
\label{gl:onsite}
For every integer $\ell\geq0$, $t\in\R$ and $z\in\mathbb C$,
\begin{equation}
 \norm{X_\ell(t)}\leq\frac{(v_0|t|)^\ell}{\ell!},\qquad
 \norm{\tau_H^z(X)}\leq c_{|z|},
\end{equation}
where
\[
 v_0=4r_0(1+J_0),\qquad
 c_s=\exp\!\bigl[2J_0s+v_0s\e^{v_0s}\bigr],\quad s\geq0.
\]
For $i\in U\subseteq\Lambda$, the same bounds hold for $H_U$,
with balls $S_{i,\ell}\cap U$.
\end{lem}

\begin{proof}
For $\ell\geq1$, put
$\delta H_\ell=H_{S_{i,\ell}}-H_{S_{i,\ell-1}}$.
The added layer has at most $2r_0$ sites, including when a cycle
ball closes.
Every term in $\delta H_\ell$ contains an added site, so
\[
 \norm{\delta H_\ell}\leq2r_0J_0\leq v_0/2,\qquad
 2\norm{H_{S_{i,\ell}}}\leq2J_0+v_0\ell.
\]
For $\ell\geq2$, finite range $r_0$ makes $\delta H_\ell$ commute
with $\tau_{H_{S_{i,\ell-2}}}^s(X)$.
Duhamel's formula therefore gives, for every $\ell\geq1$,
\[
 X_\ell(z)=\ii\int_0^z\tau_{H_{S_{i,\ell}}}^{z-s}
 \bigl([\delta H_\ell,X_{\ell-1}(s)]\bigr)\,ds.
\]
Parametrizing by $s=uz$ and using
$\norm{\tau_H^w}\leq\e^{2\norm H|\operatorname{Im}w|}$ gives
\[
 \norm{X_\ell(z)}\leq v_0|z|\int_0^1
 \e^{(2J_0+v_0\ell)(1-u)|\operatorname{Im}z|}
 \norm{X_{\ell-1}(uz)}\,du.
\]
Starting from $\norm{X_0(z)}\leq\e^{2J_0|\operatorname{Im}z|}$,
induction yields
\begin{equation}
 \norm{X_\ell(z)}
 \leq\e^{(2J_0+v_0\ell)|\operatorname{Im}z|}
 \frac{(v_0|z|)^\ell}{\ell!}.
\end{equation}
Setting $z=t$ proves the real-time bound.
The finite shell sum telescopes to $\tau_H^z(X)$, and
$|\operatorname{Im}z|\leq|z|$ gives
\[
 \norm{\tau_H^z(X)}
 \leq\e^{2J_0|z|}\sum_{\ell\geq0}
 \frac{(v_0|z|\e^{v_0|z|})^\ell}{\ell!}=c_{|z|}.
\]
Restriction preserves the interaction bounds, so the same proof
applies with balls intersected with $U$.
\end{proof}

Let $V\subseteq U\subseteq\Lambda$ and
$\sigma_U=\Tr_{U^c}\sigma$ be the marginal on $U$.
The reset map on $V$ acts within $U$ as
\[
 \Omega_V(\sigma_U)=\frac{\id_V}{d^{|V|}}
 \otimes\sigma_{U\setminus V}.
\]

\begin{lem}
\label{gl:conditional-index}
For every $V\subseteq U\subseteq\Lambda$,
\begin{equation}
 c_{\beta/2}^{-2|V|}\Omega_V(\sigma_U)
 \leq\sigma_U
 \leq c_{\beta/2}^{2|V|}\Omega_V(\sigma_U).
\end{equation}
\end{lem}

\begin{proof}
The Gibbs normalization constant cancels in conjugation, so
Lemma~\ref{gl:onsite} at $z=\pm\ii\beta/2$ gives
\begin{equation}
 \norm{\sigma^{\pm1/2}X\sigma^{\mp1/2}}\leq c_{\beta/2}
\end{equation}
for every one-site unitary $X$.
Taking positive squares and using unitarity yields
\[
 c_{\beta/2}^{-2}\sigma
 \leq X\sigma X^*
 \leq c_{\beta/2}^2\sigma.
\]
Average over the one-site Weyl unitaries $X_{pq}$ at $i$, using
the depolarizing identity~\cite[Eq.~(4.109)]{Watrous18}:
\[
 \frac1{d^2}\sum_{p,q=0}^{d-1}
 X_{pq}\sigma X_{pq}^*=\Omega_i(\sigma).
\]
Rearranging gives the one-site comparison.
The resets commute and satisfy
$\Omega_V=\prod_{i\in V}\Omega_i$.
Applying their positivity and the one-site comparisons successively
gives the factors $c_{\beta/2}^{\pm2|V|}$.
Taking the partial trace over $\Lambda\setminus U$ proves the claim.
\end{proof}

\subsection{Gibbs ratios across a buffer}

Partition the entire chain into consecutive intervals $U,V,W$,
with $V\neq\emptyset$, and put $b=|V|$.
For its Gibbs state $\sigma$, define
\[
 R_{U,W}=\sigma_{UW}(\sigma_U\otimes\sigma_W)^{-1}-\id.
\]
We use the convention $\sigma_\emptyset=1$.
For $n\geq1$, let $J_n$ be the union of the two boundary strips
adjacent to $V$: the last $\min(n,|U|)$ sites of $U$ and the first
$\min(n,|W|)$ sites of $W$.
Thus $J_n\subseteq U\cup W$; it contains no sites of the buffer $V$.

\begin{figure}[htbp]
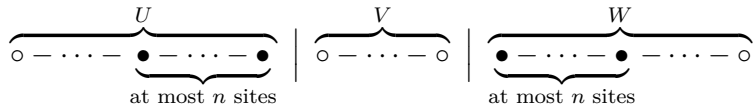

\centering
\[
 \overbrace{\circ-\cdots-
 \underbrace{\bullet-\cdots-\bullet}_{\text{at most }n\text{ sites}}}^{U}
 \ \Big|\
 \overbrace{\circ-\cdots-\circ}^{V}
 \ \Big|\
 \overbrace{
 \underbrace{\bullet-\cdots-\bullet}_{\text{at most }n\text{ sites}}
 -\cdots-\circ}^{W}
\]
\caption{The filled sites form $J_n$, consisting of two boundary
strips outside the buffer.}
\end{figure}

\noindent\textbf{Orbit regrouping for cycles.}
On a cycle, fix a nonempty proper interval $I=\{p,\ldots,q\}$,
and let $\pi$ map sites to the orbits of the reflection
$x\mapsto p+q-x$, in cyclic coordinates.
These orbits form a path with interaction range at most $r_0$,
strength at most $2J_0$, and local dimension $d^2$, after padding
singleton orbits by independent maximally mixed $d$-dimensional
factors with zero Hamiltonian; no reflection symmetry is required.
A partition $U,V,W$ is called reflection-saturated if its pieces
are the preimages of consecutive intervals partitioning this path.
For such partitions, set $b=|\pi(V)|$ and define $J_n$ by lifting
the two boundary strips from the orbit path.
Consequently,
\[
 |J_n|\leq2n\quad\text{on chains},\qquad
 |J_n|\leq4n\quad\text{on cycles},
\]
where $|J_n|$ counts original physical sites, while $b,n$ count
orbit groups on cycles.
Padding cancels from the Gibbs ratios.

\medskip
\noindent\textbf{Cut estimates.}
Extend the finite path interaction by zero to $\mathbb Z$.
For adjacent finite intervals $U,V$, define
\[
 E_{U,V}=\e^{-\beta H_{UV}}\e^{\beta(H_U+H_V)}.
\]
Applying~\cite[Corollary~3.4 and Remark~3.5]{BCP22} to
$\{\beta h_X\}$ gives a model-only constant $c_{\mathrm{cut}}>1$
such that
\begin{align}
 \norm{E_{U,V}},\ \norm{E_{U,V}^{-1}}\leq c_{\mathrm{cut}}.
\end{align}
Choose $c_{\mathrm{cut}}$ uniformly for the original and
regrouped models.

For an integer $n\geq1$, let $U_n$ consist of the $\min(n,|U|)$
sites of $U$ closest to the cut, and let $V_n$ consist of the
$\min(n,|V|)$ sites of $V$ closest to the cut.
A shorter interval is retained in full.
On a regrouped cycle, these sites are orbit groups.

\begin{figure}[htbp]
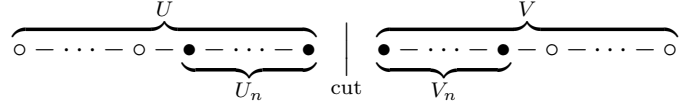

\centering
\[
 \overbrace{\circ-\cdots-\circ-
 \underbrace{\bullet-\cdots-\bullet}_{U_n}}^{U}
 \ \underset{\mathrm{cut}}{\Big|}
 \ \overbrace{\underbrace{\bullet-\cdots-\bullet}_{V_n}
 -\circ-\cdots-\circ}^{V}
\]
\caption{A two-sided cut restriction.
Filled sites are retained and open sites are discarded.
At most $n$ sites are retained on each side,
giving at most $2n$ sites in total.}
\end{figure}

The one-sided restrictions are $E_{U_n,V}$ and $E_{U,V_n}$:
only the indicated interval is shortened.
The two-sided restriction is
\[
 E_{U_n,V_n}
 =\e^{-\beta H_{U_nV_n}}\e^{\beta(H_{U_n}+H_{V_n})}.
\]
All restricted operators are extended by the identity outside
their retained intervals.
Restriction preserves the interaction bounds, so these operators
and their inverses have norm at most $c_{\mathrm{cut}}$.

With the same choice of $c_{\mathrm{cut}}$, set
\[
 f_{\mathrm{cut}}(n)
 =\frac{c_{\mathrm{cut}}^n}{(\lfloor n/r_0\rfloor+1)!},
 \qquad n\geq1.
\]
The one-sided truncation errors satisfy
\[
 \norm{E_{U,V}-E_{U_n,V}}\leq f_{\mathrm{cut}}(n),\qquad
 \norm{E_{U,V}-E_{U,V_n}}\leq f_{\mathrm{cut}}(n).
\]
Applying these bounds successively gives
\begin{align}\label{eq:cut}
 \norm{E_{U,V}-E_{U_n,V_n}}\leq2f_{\mathrm{cut}}(n).
\end{align}
The same truncation bounds hold for inverses and adjoints.

\medskip
\noindent\textbf{Clustering and local indistinguishability.}
By~\cite[Theorem~1 and the following remarks]{KK25}, there are
model-only constants $K\geq1$ and $c>0$ such that, for every finite
interval $I$ and operators $X,Y$ supported in $I$,
\[
 \abs{\Tr(\sigma^I XY)-\Tr(\sigma^I X)\Tr(\sigma^I Y)}
 \leq K\e^{-c\sqrt{\dist(\supp X,\supp Y)}}\norm{X}\norm{Y}.
\]
The constants are independent of $I$ and the support sizes.

To match the positive-sign Gibbs convention in~\cite{KK25},
use the auxiliary interaction
$\widetilde h_X=\norm{h_X}\id_X-h_X\geq0$.
Its sum over $X\subseteq I$ is $-H_I$ plus a scalar, so its
positive-sign Gibbs state is $\sigma^I$.
Moreover, $\norm{\widetilde h_X}\leq2\norm{h_X}$, and each term
involves at most $r_0+1$ sites.
The incident and pair norm sums are at most $2J_0$, and the pair
sum vanishes beyond distance $r_0$.
Thus the assumptions of~\cite{KK25} hold with exponential profile
$\overline J(x)=\e^{-x}$ and strength parameter
$2(1+J_0)\e^{r_0}$; the tail bounds follow from geometric series.
Restriction and zero extension preserve these bounds.

For adjacent finite intervals $U,V,W$, an integer $n\geq1$ with
$|V|\geq2n$, and $X$ supported on $U$,
\cite[Proposition~7.1]{BCP22} gives
\[
 \abs{\Tr(\sigma^{UVW}X)-\Tr(\sigma^{UV}X)}
 \leq K\e^{-c\sqrt n}\norm{X}.
\]
Here $K$ is chosen large enough to absorb the fixed prefactors
and the factorially decaying truncation term in that proposition.
The corresponding right-end estimate compares $\sigma^{UVW}$
and $\sigma^{VW}$ on operators supported on $W$.

The first estimate below adapts the proof of
\cite[Proposition~8.1]{BCP22} using the clustering bound
of~\cite{KK25}.
We construct the support-restricted approximants needed below.

\begin{lem}
\label{gl:relative-mixing}
Let $\sigma$ be the Gibbs state on a chain $\Lambda$ partitioned into
consecutive intervals $U,V,W$ with $b=|V|\geq1$. Write
\[
 R=R_{U,W}:=\sigma_{UW}(\sigma_U\otimes\sigma_W)^{-1}-\id.
\]
There are constants $K\geq1$ and $c>0$, depending only on the model,
and operators
$R_n$ supported on $J_n$, such that
\begin{equation}
 \norm{R}\leq K\e^{-c\sqrt b},
 \label{gl:ordinary-ratio}
\end{equation}
and
\begin{equation}
 \norm{R-R_n}\leq
 \frac{K^{n+1}}{(\lfloor n/r_0\rfloor+1)!},\qquad n\geq1.
 \label{gl:ratio-collars}
\end{equation}
For each fixed finite partition, $R_n=R$ for all sufficiently
large $n$.
The same conclusions hold for reflection-saturated cycle
partitions, with $b,n$ counted in orbit groups.
\end{lem}

\begin{proof}
Use the representation in
\cite[Section~8, Step~1, Eqs.~(37)--(38)]{BCP22}:
\begin{equation}
 \id+R^*=\lambda X_{\mathrm L}^{-1}X_{\mathrm R}^{-1}Y,
\end{equation}
where
\[
 X_{\mathrm L}=\Tr_{VW}(\sigma^{VW}E_{U,VW}^*),\qquad
 X_{\mathrm R}=\Tr_{UV}(\sigma^{UV}E_{UV,W}^*),
\]
\[
 Y=\Tr_V(\sigma^VE_{U,V}^*E_{UV,W}^*),\qquad
 \lambda=\frac{Z_{UVW}Z_V}{Z_{UV}Z_{VW}},
\]
and $Z_U=\Tr_U\e^{-\beta H_U}$, with the same convention for
other regions.
The cut estimates and~\cite[Corollary~4.4]{BCP22} uniformly bound
these factors, $X_{\mathrm L}^{-1},X_{\mathrm R}^{-1},\lambda$,
and $R$ by model-only constants.
Throughout the proof, $K\geq1$ may increase and $c>0$ may decrease.

Replace each expansional by its two-sided $n$-site cut restriction,
keeping the Gibbs states unchanged, and denote the resulting factors by
$X_{\mathrm L,n},X_{\mathrm R,n},Y_n$.
Choose an integer $n_0\geq1$ depending only on the model such that
\[
 2f_{\mathrm{cut}}(n)
 \max_{a\in\{\mathrm L,\mathrm R\}}\norm{X_a^{-1}}
 \leq\tfrac12,\qquad n\geq n_0.
\]
The uniform inverse bounds ensure that this choice is independent
of the partition.
For $n\geq n_0$, the Neumann series bounds
$X_{\mathrm L,n}^{-1},X_{\mathrm R,n}^{-1}$ uniformly.
Define
\[
 R_n=\bigl(\lambda X_{\mathrm L,n}^{-1}
 X_{\mathrm R,n}^{-1}Y_n-\id\bigr)^*.
\]
The cut estimates~\eqref{eq:cut} and the inverse-difference identity
$A^{-1}-B^{-1}=A^{-1}(B-A)B^{-1}$ give
\[
 \norm{R-R_n}\leq Kf_{\mathrm{cut}}(n),\qquad n\geq n_0.
\]
Partial trace removes the buffer $V$-support, and inversion preserves
each outer boundary algebra, so $R_n$ is supported on $J_n$.
This construction is exact once all intervals are retained.
For $n<n_0$, set $R_n=0$.
Increasing $K$ absorbs the fixed prefactor and the base
$c_{\mathrm{cut}}$ in $f_{\mathrm{cut}}$, and covers these
finitely many indices, proving~\eqref{gl:ratio-collars}.

For the smallness estimate \eqref{gl:ordinary-ratio}, follow the proof of
\cite[Proposition~8.1, Steps~2--3]{BCP22}, using the clustering
input from~\cite[Theorem~1 and the following remarks]{KK25}
and the local indistinguishability bound stated above.
Write $\epsilon(s)=K\e^{-c\sqrt{s}}$.
In the proofs of~\cite[Propositions~8.5 and~8.7]{BCP22},
the factorial cut estimates make the shell sums with their
Schmidt dimension factors uniformly finite.
The same arguments therefore apply with this choice of $\epsilon$.

For $b\geq6$, take $m=\lfloor b/3\rfloor$.
The intervening buffers have length at least $b/3$, and the
boundary comparisons permit radius $m\geq b/6$.
The cited estimates give
\[
 |\lambda-1|+\norm{Y-X_{\mathrm L}X_{\mathrm R}}
 \leq K\bigl[f_{\mathrm{cut}}(m)+\epsilon(b/6)\bigr].
\]
The ratio identity and the uniform inverse bounds imply
\[
 \norm{R}\leq K\bigl[f_{\mathrm{cut}}(m)+\epsilon(b/6)\bigr].
\]
Factorial decay and the definition of $\epsilon$ yield
\eqref{gl:ordinary-ratio} after increasing $K$ and decreasing $c$.
The uniform bound on $R$ covers the small buffer lengths.
The cycle case follows from the regrouping with parameters
$(\beta,d^2,r_0,2J_0)$.

\end{proof}

\subsection{Localized product-reference filters}

For the partitions in Lemma~\ref{gl:relative-mixing}, write
$R=R_{U,W}$ and define
\[
 C_{\mathrm{prod}}
 =\bigl(\sigma_{UW}(\sigma_U\otimes\sigma_W)^{-1}\bigr)^{1/2},
\]
using the principal square root.
Since $\id+R$ is similar to the positive matrix
$(\sigma_U\otimes\sigma_W)^{-1/2}
\sigma_{UW}(\sigma_U\otimes\sigma_W)^{-1/2}$,
similarity covariance gives
\begin{equation}
 C_{\mathrm{prod}}(\sigma_U\otimes\sigma_W)
 C_{\mathrm{prod}}^*=\sigma_{UW}.
 \label{gl:exact-productization}
\end{equation}

All filters on $UW$ are extended by the identity on $V$.
For the original full Gibbs state $\sigma$, define
\begin{equation}
 \cN_\sigma(X)=\max\bigl\{\norm{X},
 \norm{\sigma^{-1/2}X\sigma^{1/2}},
 \norm{\sigma^{1/2}X\sigma^{-1/2}}\bigr\}.
\end{equation}
This norm is submultiplicative and invariant under adjoints.

If $X$ is supported on a physical set $J$, it commutes with
$\Omega_J(\sigma)$.
The two-sided comparison in Lemma~\ref{gl:conditional-index}
therefore gives
\[
 X\sigma X^*\leq c_{\beta/2}^{2|J|}\norm{X}^2\Omega_J(\sigma)
 \leq c_{\beta/2}^{4|J|}\norm{X}^2\sigma.
\]
Conjugating by $\sigma^{-1/2}$ and applying the same argument
to $X^*$ yields
\begin{equation}
 \cN_\sigma(X)\leq c_{\beta/2}^{2|J|}\norm{X}.
\end{equation}
Here $|J|$ counts original physical sites, also on
reflection-saturated cycles.

\begin{thm}
\label{gl:full-reference}
Let $\sigma$ be the global Gibbs state on a chain
or a cycle with partition $U,V,W$ as in
Lemma~\ref{gl:relative-mixing}, and let $R_n$ be the approximants
constructed there.
There are constants $c_{\mathrm{loc}}\geq1$,
$c_{\mathrm{full}}>0$ depending only on the model and integer thresholds
$b_{\mathrm{full}},n_{\mathrm{full}}\geq1$ such that, for
$b\geq b_{\mathrm{full}}$:
\begin{enumerate}
\item
The exact filter satisfies
\begin{equation}
 \cN_\sigma(C_{\mathrm{prod}}-\id)
 +\cN_\sigma(C_{\mathrm{prod}}^{-1}-\id)
 \leq c_{\mathrm{loc}}\e^{-c_{\mathrm{full}}\sqrt b}\leq1/16.
\end{equation}
\item
For every $n\geq n_{\mathrm{full}}$, the principal square root
$C_n=(\id+R_n)^{1/2}$ is invertible, and $C_n$ is
supported on $J_n$.
Moreover,
\begin{equation}
 \cN_\sigma(C_{\mathrm{prod}}-C_n)
 +\cN_\sigma(C_{\mathrm{prod}}^{-1}-C_n^{-1})
 \leq f_{\mathrm{full}}(n)\leq1/8,
\end{equation}
where
$f_{\mathrm{full}}(n)
=c_{\mathrm{loc}}^{\,n+1}/(\lfloor n/r_0\rfloor+1)!$.
For each fixed finite partition, $C_n=C_{\mathrm{prod}}$
for all sufficiently large $n$.
\end{enumerate}
\end{thm}

\begin{proof}
Keep the decay rate $c$ from Lemma~\ref{gl:relative-mixing} fixed,
and allow $K$ to increase.

\noindent\textbf{Step 1.}
By Lemma~\ref{gl:relative-mixing}, the identity
$R-R_n=\sum_{k\geq n}(R_{k+1}-R_k)$ has only finitely many
nonzero terms.
Each increment is supported on $J_{k+1}$, with
$|J_{k+1}|\leq4(k+1)$ counted in physical sites.
The support estimate therefore gives
\[
 \cN_\sigma(R-R_n)
 \leq\sum_{k\geq n}c_{\beta/2}^{8(k+1)}
 \norm{R_{k+1}-R_k}\leq g(n),\qquad n\geq1,
\]
where
$g(n)=c_{\mathrm{tail}}^{\,n+1}/(\lfloor n/r_0\rfloor+1)!$
for a model-only $c_{\mathrm{tail}}\geq1$.
The last inequality follows by bounding each increment by the
two adjacent errors in Lemma~\ref{gl:relative-mixing} and
absorbing the exponential weight and tail sum into the
factorial bound.

Choose
\[
 c_*=\min\left\{1,\frac{c}{16(1+\log c_{\beta/2})}\right\},
 \qquad m=\lfloor c_*\sqrt b\rfloor.
\]
Then $8c_*\log c_{\beta/2}\leq c/2$.
For $b\geq4/c_*^2$, we also have $m\geq c_*\sqrt b/2\geq1$.
Applying the support estimate to $R_m$ yields
\[
 \cN_\sigma(R)
 \leq c_{\beta/2}^{8m}\bigl[\norm{R}+g(m)\bigr]+g(m).
\]
Lemma~\ref{gl:relative-mixing} and the choice of $c_*$ bound
the term containing $\norm{R}$ by $K\e^{-c\sqrt b/2}$.
Since $g$ has finite exponential moments of every order,
$(1+c_{\beta/2}^{8m})g(m)$ is bounded by a model-only multiple
of $\e^{-cm/c_*}$.
Increasing $K$ gives
\[
 \cN_\sigma(R)\leq K\e^{-c\sqrt b/2},
 \qquad b\geq4/c_*^2.
\]

\noindent\textbf{Step 2.}
Set
\[
 c_{\mathrm{loc}}=4\max\{1,c_{\mathrm{tail}},K\},\qquad
 c_{\mathrm{full}}=c/2.
\]
Then $4g(n)\leq f_{\mathrm{full}}(n)$ for every $n\geq1$.
Choose
\[
 b_{\mathrm{full}}
 =\left\lceil\max\left\{\frac4{c_*^2},
 \left(\frac{\log(16c_{\mathrm{loc}})}
 {c_{\mathrm{full}}}\right)^2\right\}\right\rceil,
\]
and choose an integer $n_{\mathrm{full}}\geq1$ such that
$f_{\mathrm{full}}(n)\leq1/8$ for every $n\geq n_{\mathrm{full}}$.
Such a choice exists by factorial decay.
Throughout these ranges,
\[
 c_{\mathrm{loc}}\e^{-c_{\mathrm{full}}\sqrt b}\leq1/16,
 \qquad \cN_\sigma(R)\leq1/64,\qquad
 \cN_\sigma(R_n)\leq1/64+1/32<1/8.
\]

For $\cN_\sigma(X),\cN_\sigma(Y)\leq1/8$, the ordered identity
$X^k-Y^k=\sum_{j=0}^{k-1}X^{k-1-j}(X-Y)Y^j$ gives
\[
 \cN_\sigma(X^k-Y^k)
 \leq k\,8^{1-k}\cN_\sigma(X-Y).
\]
The coefficients of both binomial series have modulus at most one,
and $\sum_{k\geq1}k8^{1-k}<2$.
Consequently,
\[
 \cN_\sigma\bigl((\id+X)^{\pm1/2}-(\id+Y)^{\pm1/2}\bigr)
 \leq2\cN_\sigma(X-Y).
\]
These convergent series define the principal powers.

Taking $(X,Y)=(R,0)$ proves the first assertion:
\[
 \cN_\sigma(C_{\mathrm{prod}}-\id)
 +\cN_\sigma(C_{\mathrm{prod}}^{-1}-\id)
 \leq4\cN_\sigma(R)
 \leq c_{\mathrm{loc}}\e^{-c_{\mathrm{full}}\sqrt b}\leq1/16.
\]
Taking $(X,Y)=(R,R_n)$ proves the second estimate:
\[
 \cN_\sigma(C_{\mathrm{prod}}-C_n)
 +\cN_\sigma(C_{\mathrm{prod}}^{-1}-C_n^{-1})
 \leq4g(n)\leq f_{\mathrm{full}}(n)\leq1/8.
\]
The series for the two powers of $\id+R_n$ are mutually inverse
and remain in the algebra on $J_n$, proving invertibility
and support.
Finally, $R_n=R$ for all sufficiently large $n$, so
$C_n=C_{\mathrm{prod}}$ eventually.
\end{proof}

For $b\geq b_{\mathrm{full}}$ and $n\geq n_{\mathrm{full}}$,
the triangle inequality and Theorem~\ref{gl:full-reference} give
\[
 \max\{\cN_\sigma(C_n),\cN_\sigma(C_n^{-1})\}
 \leq1+\frac1{16}+\frac18<2.
\]

The comparison orders and filter identities extend by
tensoring with a state $\omega$ on any finite reference system.
Moreover,
\[
 \cN_{\omega\otimes\sigma}(\id\otimes X)=\cN_\sigma(X),
\]
with conjugations taken on $\supp\omega$.
Thus the localization estimates hold with the same constants
after adjoining a reference system $R$, independently of its
dimension $|\cH_R|$ and state $\omega$.

\section{Conditional relative entropy under filters across boundaries}\label{sec:tech-ch_filter_transport}
\label{ft:section}

\subsection{Stability under local filtering}
We follow the conventions of matrices, marginals, and local-depolarizing noise as in  Appendix~\ref{noise:section}. Recall that each physical site has local dimension $d$, so $\dim\cH_V=d^{|V|}$ and the partial trace conditional expectation $\Omega_V(X)=d^{-|V|}\id_V\otimes\Tr_VX$. For an operator $K$, write $$K^{-*}=(K^*)^{-1} \qquad \kappa(K)=\norm K\norm{K^{-1}},$$
where $\kappa(K)$ is the condition number of $K$.
In this appendix, let $\rho$ denote the input state and $\sigma$ the faithful reference state on $\cH_\Lambda$.

Let $I\subset \Lambda$ be a subset and \begin{equation}
 \Delta_I(\rho\mid \sigma )=D(\rho\Vert \sigma)-D(\rho_{I^c}\Vert \sigma_{I^c}).
\end{equation}
be the conditional relative entropy.
 For $0<r<1$, recall the $I$-local depolarizing map $$\rho_r=\mathcal D_{I,r}(\rho)=r\rho+(1-r)\Omega_I(\rho)$$ and similarly for $\sigma_r$. We will use the coefficient $$\omega_m(1-r)=1-b_m(r)
 =1-r+\frac{1-r}{m}\log\left(1+\frac{r m}{1-r}\right)$$ from Corollary~\ref{noise:union-continuity}, which controls the change of erased relative entropy deficit under $\mathcal D_{I,r}$ by the union conditional relative entropy. The following theorem extends this comparison to an invertible filter whose support may extend beyond the erasure subset $I$. There is no containment assumption between $I$ and $J$. 

\begin{thm}
\label{ft:union-transport}
Let $\sigma>0$ be a faithful state on $\Lambda$. Let $I,J\subseteq\Lambda$ be subsets, and let $K$ be an invertible operator supported on $J$. Assume
\[
 \epsilon=\norm{K-\id}\leq\frac18,\qquad
 \sigma\geq m\Omega_J(\sigma),\qquad 0<m\leq1.
\]
For every state $\rho$, the normalized states
\[
 \rho^K=\frac{K\rho K^*}{\Tr K\rho K^*},\qquad
 \sigma^K=\frac{K\sigma K^*}{\Tr K\sigma K^*},
\]
satisfy
\begin{equation}
 \abs{\Delta_I(\rho^K\mid \sigma^K)-\Delta_I(\rho\mid \sigma)}
 \leq\frac{1000d^{|J|/2}}{m}\sqrt{\epsilon}\,
 \Delta_{I\cup J}(\rho\mid \sigma).
 \label{ft:union-bound}
\end{equation}
The numerical constant $1000$ is universal, and the same estimate applies uniformly
to joint states with a reference system $R$ of arbitrary finite dimension.
\end{thm}

\subsection{Resolvents and local filter comparison}
We follow the resolvent integral formulas from Appendix~\ref{noise:section} with the extension to positive matrices. Let $\rho\geq0$ and $\sigma>0$ be positive operators. For $u>0$, recall the multiplication operators $M_u$ and score vectors $Z_u$
\begin{align}
 &M_u(\rho,\sigma)=L_\rho+uR_\sigma,\qquad
 Z_u(\rho,\sigma)=M_u(\rho,\sigma)^{-1}(\rho-\sigma),\qquad
\nonumber\\
&\mathcal Q_u(\rho\mid\sigma)=\langle \rho-\sigma, M_u^{-1}(\rho-\sigma)\rangle=\norm{Z_u}_{M_u}^2
\end{align}
For $V\subseteq\Lambda$, use the local algebra $\mathcal M_{V^c}=\operatorname{Ran}\Omega_V=\{\id_V\otimes Y\} $ and  $\operatorname{dist}_{M_u}$ for the distance in the $M_u$-weighted $L_2$-norm. As shown in Appendix \ref{noise:section},  $\id_V\otimes Z_{u,V^c}$ is the $M_u$-orthogonal projection of $Z_u$ onto $\mathcal{M}_{V^c}$.
Define the orthogonal vector
\[
 W_{u,V}=Z_u-\id_V\otimes Z_{u,V^c}.
\]
Then the conditional resolvent quantity is
\begin{align}
\mathcal Q_{u,V}(\rho\mid\sigma)
 :&=\mathcal Q_u(\rho\mid\sigma)-\mathcal Q_u(\rho_{V^c}\mid\sigma_{V^c})\nonumber\\
 &=\norm{Z_u}_{M_u}^2-\norm{Z_{u,V^c}}_{M_{u,V^c}}^2\nonumber\\
 &=\norm{W_{u,V}}_{M_u}^2
 =\operatorname{dist}_{M_u}(Z_u,\mathcal{M}_{V^c})^2\geq0 \label{eq:square}
\end{align}
Apply the unnormalized entropy integral \eqref{noise:eq-homogeneous-full-integral} separately to the full and reduced pairs:
\begin{align*}
 D(\rho\Vert \sigma)
 &=\Tr \rho-\Tr \sigma+\int_0^\infty \frac{u}{(1+u)^2}\mathcal Q_u(\rho\mid\sigma)\,du,\\
 D(\rho_{V^c}\Vert \sigma_{V^c})
 &=\Tr \rho_{V^c}-\Tr \sigma_{V^c}
   +\int_0^\infty \frac{u}{(1+u)^2}\mathcal Q_u(\rho_{V^c}\mid\sigma_{V^c})\,du.
\end{align*}
Subtracting the two identities therefore cancels the linear trace terms and gives
\begin{align}
 \Delta_V(\rho\mid \sigma)
 &=D(\rho\Vert \sigma)-D(\rho_{V^c}\Vert \sigma_{V^c})\notag\\
 &=\int_0^\infty \frac{u}{(1+u)^2}\norm{W_{u,V}}_{M_u}^2\,du=\int_0^\infty \frac{u}{(1+u)^2}\dist_{M_u}(Z_u,\mathcal M_{V^c})^2\,du.
 \label{ft:homogeneous-conditional-integral}
\end{align}
which shows that Lemma \ref{noise:canonical} holds for positive $\rho,\sigma$.

For states $\rho,\sigma$ with $\sigma>0$ and an invertible matrix $K$, define the conditional quantity
\begin{equation}
 \Gamma_I(K;\rho,\sigma)
 =\int_0^\infty \frac{u}{(1+u)^2}
 \dist_{M_u}(Z_u,K^*\mathcal M_{I^c}K^{-*})^2\,du.
 \label{ft:similarity-resource}
\end{equation}
Set the constant
\[
 \eta_{K,\sigma}
 =\norm{K^*K-\id}+\norm{\sigma^{-1/2}(K^*K-\id)\sigma^{1/2}}.
\]
The following proposition compares the square roots of $\Delta_I(\rho^K\mid \sigma^K)$ and $\Gamma_I(K;\rho,\sigma)$.
\begin{prop}
\label{ft:similarity}
Let $\rho,\sigma>0$ be states on $\Lambda$. Let $K$ be invertible and supported on $J$, with $\eta_{K,\sigma}\leq1/8$. For every subset $I$, the quantity \eqref{ft:similarity-resource} is finite and nonnegative, and
\begin{equation}
 \abs{\sqrt{\Delta_I(\rho^K\mid \sigma^K)}-\sqrt{\Gamma_I(K;\rho,\sigma)}}
 \leq3\eta_{K,\sigma}\sqrt{\Gamma_I(K;\rho,\sigma)}
 +2\eta_{K,\sigma}\sqrt{\Delta_J(\rho\mid \sigma)}.
 \label{ft:similarity-root}
\end{equation}
\end{prop}
We need the following lemma of metric comparison under filtering.
\begin{lem}
\label{ft:metric-comparison-lemma}
Let $\rho,\sigma>0$ be states on $\Lambda$, and let $K$ be invertible with $\eta_{K,\sigma}\leq1/8$. Set 
\[
 a_K=\norm{K^*K-\id},\qquad
 b_{K,\sigma}=\norm{\sigma^{-1/2}(K^*K-\id)\sigma^{1/2}},
 \qquad a_K+b_{K,\sigma}=\eta_{K,\sigma}.
\]
Define
\[
 \alpha=\min\left\{1-a_K,\frac{(1-b_{K,\sigma})^2}{1+a_K}\right\},
 \qquad
 \beta=\max\left\{1+a_K,\frac{(1+b_{K,\sigma})^2}{1-a_K}\right\}.
\]
For every $u>0$ and every matrix $Y$,
\begin{equation}
 \alpha\norm Y_{M_u(\rho,\sigma)}^2
 \leq\norm{K^{-*}YK^*}_{M_u(K\rho K^*,K \sigma K^*)}^2
 \leq\beta\norm Y_{M_u(\rho,\sigma)}^2,
 \label{ft:metric-comparison}
\end{equation}
where $\alpha\geq1-2\eta_{K,\sigma}$ and $\beta\leq1+3\eta_{K,\sigma}$.
\end{lem}

\begin{proof}
Write $A=K^*K$.
Cyclicity gives
\begin{align}
 \norm{K^{-*}YK^*}_{M_u(K\rho K^*,K \sigma K^*)}^2
 &=
 \Tr \rho YA Y^*
 +u\Tr \sigma A Y^*A^{-1}YA\notag\\
 &=
 \norm{\rho^{1/2}YA^{1/2}}_{\mathrm{HS}}^2
 +u\norm{A^{-1/2}YA\sigma^{1/2}}_{\mathrm{HS}}^2.
 \label{ft:similarity-metric}
\end{align}
The eigenvalues of $A^{1/2}$ lie in $[\sqrt{1-a_K},\sqrt{1+a_K}]$.
With $D_\sigma=\sigma^{-1/2}A\sigma^{1/2}$, the second norm is
$
 \norm{A^{-1/2}(Y\sigma^{1/2})D_\sigma}_{\mathrm{HS}}.
$
Since $\norm{D_\sigma-\id}=b_{K,\sigma}$, its singular values lie in $[1-b_{K,\sigma},1+b_{K,\sigma}]$.
Consequently,
\begin{align*}
 (1-a_K)\norm{\rho^{1/2}Y}_{\mathrm{HS}}^2
 &\leq\norm{\rho^{1/2}YA^{1/2}}_{\mathrm{HS}}^2
 \leq(1+a_K)\norm{\rho^{1/2}Y}_{\mathrm{HS}}^2,\\
 \frac{(1-b_{K,\sigma})^2}{1+a_K}\norm{Y\sigma^{1/2}}_{\mathrm{HS}}^2
 &\leq\norm{A^{-1/2}(Y\sigma^{1/2})D_\sigma}_{\mathrm{HS}}^2
 \leq\frac{(1+b_{K,\sigma})^2}{1-a_K}\norm{Y\sigma^{1/2}}_{\mathrm{HS}}^2.
\end{align*}
Applying the above estimate to \eqref{ft:similarity-metric} proves~\eqref{ft:metric-comparison}.
Finally,
\[
 \frac{(1-b_{K,\sigma})^2}{1+a_K}\geq1-a_K-2b_{K,\sigma}
 \geq1-2\eta_{K,\sigma},
\]
and, since $a_K+b_{K,\sigma}=\eta_{K,\sigma}\leq1/8$,
\[
 \frac{(1+b_{K,\sigma})^2}{1-a_K}-1
 =\frac{a_K+2b_{K,\sigma}+b_{K,\sigma}^2}{1-a_K}\leq\frac{2\eta_{K,\sigma}+\eta_{K,\sigma}^2}{1-\eta_{K,\sigma}}\le \frac{2+\frac{1}{8}}{1-\frac{1}{8}}\eta_{K,\sigma}
 \leq3\eta_{K,\sigma}.
\]
Together with $1-a_K\geq1-2\eta_{K,\sigma}$ and $1+a_K\leq1+3\eta_{K,\sigma}$, these give the stated bounds on $\alpha,\beta$.
\end{proof}

\begin{proof}[Proof of Proposition~\ref{ft:similarity}]
Put
\[
 \rho'=K\rho K^*,\qquad \sigma'=K\sigma K^*.
\]
Then $\rho^K=\rho'/\Tr \rho'$ and $\sigma^K=\sigma'/\Tr \sigma'$. Use $A,a_K,b_{K,\sigma},\alpha,\beta$ from Lemma~\ref{ft:metric-comparison-lemma}, and write $M_u'=M_u(\rho',\sigma')$. Every $X\in\mathcal M_{(I\cup J)^c}$ commutes with $K^*$, so $K^*XK^{-*}=X$. Thus $\mathcal M_{(I\cup J)^c}\subseteq K^*\mathcal M_{I^c}K^{-*}$, and
\[
 0\leq\Gamma_I(K;\rho,\sigma)\leq\Delta_{I\cup J}(\rho\mid \sigma)<\infty.
\]
Let $Z=Z_u(\rho,\sigma)$ and $Z'_u=Z_u(\rho',\sigma')$. Direct multiplication gives
\[
 \rho'K^{-*}ZK^*=K\rho ZK^*,
 \qquad
 uK^{-*}ZK^*\sigma'=uK^{-*}ZA\sigma K^*.
\]
Subtract the congruence of $\rho Z+uZ\sigma=\rho-\sigma$. We obtain
\[
 M_u'(K^{-*}ZK^*)=(\rho'-\sigma')+F_u,
 \qquad
 F_u=uK^{-*}[Z,A]\sigma K^*.
\]
Consequently,
\[
 Z_u'-K^{-*}Z_uK^*=-(M_u')^{-1}F_u.
\]

For the original pair, write
\[
 W_{u,J}=Z-\id_J\otimes Z_{u,J^c}.
\]
Since $A-\id$ is supported on $J$,
\[
 [Z_u,A]=[W_{u,J},A-\id].
\]
Let $V_u=K^{-*}[Z,A]K^{-1}$, so that $F_u=uV_u\sigma'$. Since $M_u'\geq uR_{\sigma'}$, inverse order gives
\[
 \norm{Z'_u-K^{-*}ZK^*}_{M_u'}^2
 =\ip{F_u}{(M_u')^{-1}F_u}_{\mathrm{HS}}
 \leq\frac1u\Tr F_u^*F_u(\sigma')^{-1}
 =u\Tr \sigma'V_u^*V_u.
\]
The factorization $K\sigma^{1/2}$ of $\sigma'$ differs from $(\sigma')^{1/2}$ by a right unitary. Hence
\[
 \norm{V_u(\sigma')^{1/2}}_{\mathrm{HS}}
 =\norm{K^{-*}[Z_u,A]\sigma^{1/2}}_{\mathrm{HS}}
 \leq\norm{K^{-1}}\norm{[W_{u,J},A-\id]\sigma^{1/2}}_{\mathrm{HS}}.
\]
For every $W$,
\[
 \norm{W(A-\id)\sigma^{1/2}}_{\mathrm{HS}}
 \leq b_{K,\sigma}\norm{W\sigma^{1/2}}_{\mathrm{HS}},
\]
while left multiplication by $A-\id$ costs $a_K$. Therefore by \eqref{eq:square}
\[
 \norm{Z'_u-K^{-*}ZK^*}_{M_u'}\leq\sqrt u\norm{K^{-1}}\eta_{K,\sigma} \sqrt{\Tr W_{u,J}^*W_{u,J}\sigma}
 \leq\norm{K^{-1}}\eta_{K,\sigma}\sqrt{\mathcal Q_{u,J}(\rho\mid\sigma)}.
\]
Square the preceding inequality and integrate against $\frac{u}{(1+u)^2}\,du$. By \eqref{noise:eq-conditional-integral}, taking square roots gives
\begin{equation}
 \left(\int_0^\infty \frac{u}{(1+u)^2}\norm{Z'_u-K^{-*}Z_uK^*}_{M_u'}^2\,du\right)^{1/2}
 \leq\norm{K^{-1}}\eta_{K,\sigma}\sqrt{\Delta_J(\rho\mid \sigma)}.
\end{equation}
The conditional relative entropy on the right is for the original pair $(\rho,\sigma)$.

Set \[
 \mathcal H_I=
 \int_0^\infty \frac{u}{(1+u)^2}
 \dist_{M_u'}(K^{-*}Z_uK^*,\mathcal M_{I^c})^2\,du.
\]
and recall the conditional integral formula 
\[ \Delta_I(\rho'\mid \sigma')=\int_0^\infty \frac{u}{(1+u)^2}
 \dist_{M_u'}(Z_u',\mathcal M_{I^c})^2\,du.\] 
By triangle inequality,
\[
 \abs{\dist(x,\mathcal U)-\dist(y,\mathcal U)}\leq\norm{x-y}.
\]
Applying this to $Z'_u$ and $K^{-*}Z_uK^*$ with the above estimate in the $M_u'$ norm,
\[
 \abs{\sqrt{\Delta_I(\rho'\mid \sigma')}-\sqrt{\mathcal H_I}}
 \leq\norm{K^{-1}}\eta_{K,\sigma}\sqrt{\Delta_J(\rho\mid \sigma)},
\]

In \eqref{ft:metric-comparison}, replace $Y$ by $Z_u-X$ and take the infimum over
$X\in K^*\mathcal M_{I^c}K^{-*}$. Integrating gives
\[
 \alpha\Gamma_I\leq\mathcal H_I\leq\beta\Gamma_I.
\]

By \eqref{sc:independent-scaling},  $\Delta_I(\rho^K\mid \sigma^K)=\Delta_I(\rho'\mid \sigma')/\Tr K\rho K^*$. Since
\[
 1-a_K\leq \Tr K\rho K^* \leq1+a_K,
 \qquad
 \norm{K^{-1}}\leq(1-a_K)^{-1/2},
\]
we have,
\begin{align*}
 \sqrt{\frac{\alpha}{\Tr K\rho K^*}}\sqrt{\Gamma_I}-&(1-a_K)^{-1}\eta_{K,\sigma}\sqrt{\Delta_J(\rho\mid \sigma)}
 \leq\sqrt{\Delta_I(\rho^K\mid \sigma^K)} \\
 &\leq\sqrt{\frac{\beta}{\Tr K\rho K^*}}\sqrt{\Gamma_I}+(1-a_K)^{-1}\eta_{K,\sigma}\sqrt{\Delta_J(\rho\mid \sigma)}.
\end{align*}
Moreover,
\[
 \frac{\alpha}{\Tr K\rho K^*}\geq1-3\eta_{K,\sigma},
 \qquad
\frac{\beta}{\Tr K\rho K^*}\leq1+5\eta_{K,\sigma}.
\]
Each square-root coefficient differs from one by at most $3\eta_{K,\sigma}$, and
$(1-a_K)^{-1}\leq8/7<2$. This proves
\eqref{ft:similarity-root}.\end{proof}

\subsection{Proof of Theorem~\ref{ft:union-transport}}

First assume $\rho>0$. The case $\epsilon=0$ is immediate,
so assume $\epsilon>0$. For
\[
 0<1-r\leq m,\qquad r\geq1/2,\qquad
 \delta=\frac{d^{|J|}\epsilon}{\sqrt{1-r}}\leq\frac1{48},
\]
we first prove an estimate depending on the noise parameter $r$, then choose $r$. Write
\[
 \rho_r=\mathcal D_{J,r}\rho,\qquad \sigma_r=\mathcal D_{J,r}\sigma.
\]
Recall that $\epsilon=\norm{\id-K}\le \frac{1}{8}$. Then 
\begin{equation}\label{eq:order}
 (1-\epsilon)^2\id\leq K^*K,\ KK^*\leq (1+\epsilon)^2\id,
 \qquad
 \kappa(K)^2=\norm{K^{-1}}^2\norm{K}^2\leq \frac{(1+\epsilon)^2}{(1-\epsilon)^2}\leq2.
\end{equation}
Set 
$$\rho_r^K=\frac{K\rho_rK^*}{\Tr K\rho_rK^*}\qquad \sigma_r^K=\frac{K\sigma_rK^*}{\Tr K\sigma_rK^*}.$$
We compare the conditional relative entropy along the sequence
\[
 (\rho,\sigma),\quad
 (\rho_r,\sigma_r),\quad
 (\rho_r^K,\sigma_r^K),\quad
 (\mathcal D_{J,r}\rho^K,\mathcal D_{J,r}\sigma^K),\quad
 (\rho^K,\sigma^K).
\]
For every $X\geq0$, the bounds \eqref{eq:order} give
\[
 (1-\epsilon)^2\,\Omega_J(X)\leq\Omega_J(KXK^*)\leq (1+\epsilon)^2\,\Omega_J(X).
\]
\medskip
\noindent\textbf{Step 1. Noise costs.}
Corollary~\ref{noise:union-continuity} gives
\begin{equation}
 \abs{\Delta_I(\rho_r\mid \sigma_r)-\Delta_I(\rho\mid \sigma)}
 \leq\omega_m(1-r)\Delta_{I\cup J}(\rho\mid \sigma).
 \label{ft:initial-noise-cost}
\end{equation}

For the filtered pair $(\rho^K,\sigma^K)$, since $\Omega_J(\sigma)$ commutes with the $J$-supported matrix $K$, by above estimate \eqref{eq:order} of $K$,
\[
 K\sigma K^*\geq mK\Omega_J(\sigma)K^*\geq m(1-\epsilon)^2\Omega_J(\sigma)\geq \frac{m (1-\epsilon)^2}{(1+\epsilon)^2}\Omega_J(K\sigma K^*),
\]
After normalization,
\[
 \sigma^K\geq\frac{m (1-\epsilon)^2}{(1+\epsilon)^2}\Omega_J(\sigma^K)
 \geq\frac m2\Omega_J(\sigma^K).
\]
Lemma~\ref{sc:filters} applied with $I\cup J$ gives
\begin{equation}
 \Delta_{I\cup J}(\rho^K\mid \sigma^K)
 \leq\frac{\norm K^2}{\Tr K\rho K^*}\Delta_{I\cup J}(\rho\mid \sigma)
 \leq\kappa(K)^2\Delta_{I\cup J}(\rho\mid \sigma),
 \label{ft:integrated-local-filter}
\end{equation}
where the last inequality uses $\Tr K\rho K^*\geq\norm{K^{-1}}^{-2}$.
Applying \eqref{ft:initial-noise-cost} to $\rho^K,\sigma^K$ with constant $m/2$ therefore yields
\begin{equation}
 \abs{\Delta_I(\mathcal D_{J,r}\rho^K\mid \mathcal D_{J,r}\sigma^K)-\Delta_I(\rho^K\mid \sigma^K)}
 \leq
 \kappa(K)^2\omega_{m/2}(1-r)\Delta_{I\cup J}(\rho\mid \sigma).
 \label{ft:final-noise-cost}
\end{equation}

\medskip
\noindent\textbf{Step 2. Filtering costs of noisy pairs.}
By \eqref{noise:eq-physical-ceiling},
\begin{equation}
 (1-r)\Omega_J(\rho)\leq \rho_r\leq d^{2|J|}\Omega_J(\rho),
 \qquad
  m\Omega_J(\sigma)\leq \sigma_r\leq d^{2|J|}\Omega_J(\sigma).
 \label{ft:regularized-orders}
\end{equation}
Since the $J$-supported matrix $K$ commutes with $\Omega_J(\rho)$ and $\Omega_J(\sigma)$, the bounds in
\eqref{ft:regularized-orders} imply
\begin{align}
 \norm{\rho_r^{1/2}(K-\id)\rho_r^{-1/2}},
 \ \norm{\rho_r^{1/2}(K-\id)^*\rho_r^{-1/2}}&\leq\delta,\notag\\
 \norm{\sigma_r^{-1/2}(K-\id)\sigma_r^{1/2}},
 \ \norm{\sigma_r^{-1/2}(K-\id)^*\sigma_r^{1/2}}&\leq\delta.
 \label{ft:regularized-multipliers}
\end{align}
For example,
\[
 (K-\id)^*\rho_r(K-\id)\leq d^{2|J|}\epsilon^2\Omega_J(\rho)
 \leq\frac{d^{2|J|}\epsilon^2}{1-r}\rho_r
\]
proves the first estimate. Replacing $(K-\id)$ by $(K-\id)^*$ proves the second estimate. The remaining estimates follow similarly.

Define the map
\[
 S_K(Y)=K^*YK^{-*}.
\]
For every $u>0$, the Neumann series and
\eqref{ft:regularized-multipliers} give
\begin{equation}
 \norm{S_K-\idmap}_{M_u(\rho_r,\sigma_r)\to M_u(\rho_r,\sigma_r)}
 \leq4\delta,
 \qquad
 \norm{S_K^{-1}-\idmap}_{M_u(\rho_r,\sigma_r)\to M_u(\rho_r,\sigma_r)}
 \leq6\delta.
 \label{ft:similarity-smallness}
\end{equation}

Fix $u$, set $Z:=Z_u(\rho_r,\sigma_r)$ and $z_{I\cup J}:=\id_{I\cup J}\otimes Z_{u,(I\cup J)^c}(\rho_r,\sigma_r)$. $z_{I\cup J}$ is the orthogonal projection of $Z$ onto $\mathcal M_{(I\cup J)^c}$. Both $\mathcal M_{I^c}$ and
$K^*\mathcal M_{I^c}K^{-*}=S_K(\mathcal M_{I^c})$ contain $\mathcal M_{(I\cup J)^c}$, and $S_K$ fixes $z_{I\cup J}$. Set
\[
 d_I(u)=\dist_{M_u}(Z,\mathcal M_{I^c})^2,\qquad
 g_I(u)=\dist_{M_u}(Z,K^*\mathcal M_{I^c}K^{-*})^2.
\]
Let $z_I=\id_I\otimes Z_{u,I^c}$ and let $v_I$ be the projection of $Z$ onto $K^*\mathcal M_{I^c}K^{-*}$. Nested projections give
\[
 \max\bigl\{\sqrt{d_I(u)},\sqrt{g_I(u)},
       \norm{z_I-z_{I\cup J}}_{M_u},\norm{v_I-z_{I\cup J}}_{M_u}\bigr\}
 \leq\norm{Z-z_{I\cup J}}_{M_u}.
\]
Using $S_K(z_I)$ and $S_K^{-1}(v_I)$ for the triangle inequality and applying \eqref{ft:similarity-smallness},
\[
 \sqrt{g_I(u)}\leq\sqrt{d_I(u)}+4\delta\norm{Z-z_{I\cup J}}_{M_u},
 \qquad
 \sqrt{d_I(u)}\leq\sqrt{g_I(u)}+6\delta\norm{Z-z_{I\cup J}}_{M_u}.
\]
Hence
\[
 \abs{g_I(u)-d_I(u)}
 =\bigl(\sqrt{g_I(u)}+\sqrt{d_I(u)}\bigr)
   \abs{\sqrt{g_I(u)}-\sqrt{d_I(u)}}
 \leq12\delta\norm{Z-z_{I\cup J}}_{M_u}^2.
\]
Integrating yields
\begin{equation}
 \abs{\Gamma_I(K;\rho_r,\sigma_r)-\Delta_I(\rho_r\mid \sigma_r)}
 \leq12\delta\Delta_{I\cup J}(\rho_r\mid \sigma_r).
 \label{ft:regularized-algebra-change}
\end{equation}

Note that
$K^*K-\id=(K-\id)+(K-\id)^*+(K-\id)^*(K-\id)$. Then
\[
 \norm{K^*K-\id}\leq2\epsilon+\epsilon^2\leq3\epsilon,
 \qquad
 \norm{\sigma_r^{-1/2}(K^*K-\id)\sigma_r^{1/2}}
 \leq2\delta+\delta^2\leq3\delta.
\]
Thus, as $\epsilon\le \frac{d^{|J|}\epsilon}{\sqrt{1-r}}=\delta$,
\[
 \eta_r=
 \norm{K^*K-\id}+\norm{\sigma_r^{-1/2}(K^*K-\id)\sigma_r^{1/2}}
 \leq6\delta\leq\frac18.
\]
Apply \eqref{ft:similarity-root} with the support $J$ and subset $I$, 
\[
 \abs{
 \sqrt{\Delta_I(\rho_r^K\mid \sigma_r^K)}
 -\sqrt{\Gamma_I(K;\rho_r,\sigma_r)}
 }
 \leq3\eta_{r}\sqrt{\Gamma_I(K;\rho_r,\sigma_r)}
 +2\eta_{r}\sqrt{\Delta_J(\rho_r\mid \sigma_r)} \leq 5\eta_r\sqrt{\Delta_{I\cup J}(\rho_r\mid \sigma_r)}.
\]
Here we used $\Delta_J(\rho_r\mid \sigma_r)\leq\Delta_{I\cup J}(\rho_r\mid \sigma_r)$, and 
$\Gamma_I(K;\rho_r,\sigma_r)\leq\Delta_{I\cup J}(\rho_r\mid \sigma_r)$ by the algebra $\mathcal M_{(I\cup J)^c}\subseteq\mathcal M_{I^c}\cap K^*\mathcal M_{I^c}K^{-*}$. The difference of squares is at most
\[\abs{
 \Delta_I(\rho_r^K\mid \sigma_r^K)
 -\Gamma_I(K;\rho_r,\sigma_r)
 }
 \leq5\eta_r(2+5\eta_r)\Delta_{I\cup J}(\rho_r\mid \sigma_r)
 \leq14\eta_r\Delta_{I\cup J}(\rho_r\mid \sigma_r).
\]
Combining this with \eqref{ft:regularized-algebra-change},
$\eta_r\leq6\delta$, and \eqref{ft:initial-noise-cost}, gives
\begin{equation}
 \abs{\Delta_I(\rho_r^K\mid \sigma_r^K)-\Delta_I(\rho_r\mid \sigma_r)}
 \leq
 \frac{100d^{|J|}\epsilon}{\sqrt{1-r}}\Delta_{I\cup J}(\rho\mid \sigma).
 \label{ft:regularized-crossing-cost}
\end{equation}
\medskip
\noindent\textbf{Step 3. Noise-order comparison at the original residual.}
We prove, for every $0<r<1$, independently of the reference floor,
\begin{equation}
 \abs{\Delta_I(\rho_r^K\mid \sigma_r^K)
       -\Delta_I(\mathcal D_{J,r}\rho^K\mid \mathcal D_{J,r}\sigma^K)}
 \leq80d^{|J|}\epsilon\Delta_{I\cup J}(\rho\mid \sigma).
 \label{ft:noise-order-cost}
\end{equation}

Define completely positive maps
\[
 \Phi_0=\Ad_K\mathcal D_{J,r},\qquad
 \Phi_1=\mathcal D_{J,r}\Ad_K,\qquad
 \Phi_s=(1-s)\Phi_0+s\Phi_1,\qquad 0\leq s\leq1,
\]
where $\Ad_K(X)=KXK^*$. These maps need not preserve trace. Let
\[
 \rho^{(s)}=\Phi_s(\rho),\qquad \sigma^{(s)}=\Phi_s(\sigma),
 \qquad M_{s,u}=M_u(\rho^{(s)},\sigma^{(s)}),
 \qquad Z_{s,u}=Z_u(\rho^{(s)},\sigma^{(s)}).
\]

At the original pair, let
\[
 M=M_u(\rho,\sigma),\qquad Z=Z_u(\rho,\sigma),
\]
let $z_{I\cup J}=\id_{I\cup J}\otimes Z_{u,(I\cup J)^c}$ be the original $M$-orthogonal projection onto $\mathcal M_{(I\cup J)^c}$, and put
\[
 W_{I\cup J}=Z-z_{I\cup J},
 \qquad
 S_{I\cup J}=MW_{I\cup J}=\rho-\sigma-\rho z_{I\cup J}-uz_{I\cup J}\sigma.
\]
Every Kraus matrix of $\Phi_s$ is supported on $J$ and hence commutes with $z_{I\cup J}$. Thus $\Phi_s$ is a bimodule map for multiplication by $z_{I\cup J}$, and
\begin{equation}
 Y_s:=Z_{s,u}-z_{I\cup J}=M_{s,u}^{-1}\Phi_s(S_{I\cup J}).
 \label{ft:original-residual-source}
\end{equation}
The fixed $z_{I\cup J}$ is a comparison point from the original pair; it is not identified with any transformed reduced score.

Let $\Phi$ be completely positive with $\Phi^*(\id)\leq b\id$, $b>0$, and put
$\rho'=\Phi(\rho)>0$, $\sigma'=\Phi(\sigma)>0$, and $M'=M_u(\rho',\sigma')$.
The Schwarz inequality for the subunital completely positive map $\Phi^*/b$ gives
\begin{align*}
 \norm{\Phi^*(X)}_M^2
 &\leq b\Tr\rho\Phi^*(XX^*)+bu\Tr\sigma\Phi^*(X^*X)\\
 &=b\norm X_{M'}^2.
\end{align*}
Since $MY=\rho Y+uY\sigma$, duality yields
\begin{equation}
 \begin{aligned}
 \norm{(M')^{-1/2}\bigl[\Phi(\rho Y)+u\Phi(Y\sigma)\bigr]}_{\mathrm{HS}}
 &=\sup_{\norm X_{M'}\leq1}\abs{\ip{\Phi^*(X)}{Y}_M}\\
 &\leq\sqrt b\,\norm Y_M.
 \end{aligned}
 \label{ft:cp-source-cauchy}
\end{equation}

Since $\mathcal D_{J,r}$ is trace preserving, its adjoint is unital. Together with $K^*K\leq k^2\id$, this gives $\Phi_s^*(\id)\leq k^2\id$. Applying
\eqref{ft:cp-source-cauchy} to \eqref{ft:original-residual-source} gives
\begin{equation}
 \norm{Y_s}_{M_{s,u}}\leq k\norm{W_{I\cup J}}_M.
 \label{ft:interpolated-residual}
\end{equation}

Both branches have local lower bounds:
\[
 \rho^{(s)}\geq (1-r)c\Omega_J(\rho),\qquad \sigma^{(s)}\geq (1-r)c\Omega_J(\sigma).
\]
Indeed,
\[
 \Phi_0\rho=rK\rho K^*+(1-r)K\Omega_J(\rho)K^*,
 \qquad
 \Phi_1\rho=rK\rho K^*+(1-r)\Omega_J(K\rho K^*),
\]
and each second term is at least $(1-r)c\Omega_J(\rho)$, using the local effect estimate above for $\Phi_1$. The reference argument is identical.

The map derivative is
\[
 \dot\Phi=(1-r)(\Omega_J\Ad_K-\Ad_K\Omega_J).
\]
The matrices $\Omega_J(K\rho K^*)$ and $K\Omega_J(\rho)K^*$ both lie between $c\Omega_J(\rho)$ and $k^2\Omega_J(\rho)$. Their difference is therefore bounded between
$\pm(k^2-c)\Omega_J(\rho)=\pm4\epsilon \Omega_J(\rho)$. The same argument applies to $\sigma$. Hence
\[
 -4(1-r)\epsilon \Omega_J(\rho)\leq\dot \rho^{(s)}\leq4(1-r)\epsilon \Omega_J(\rho),
 \qquad
 -4(1-r)\epsilon \Omega_J(\sigma)\leq\dot \sigma^{(s)}\leq4(1-r)\epsilon \Omega_J(\sigma).
\]
Since $c\geq49/64$, these imply
\begin{equation}
 -\zeta \rho^{(s)}\leq\dot \rho^{(s)}\leq\zeta \rho^{(s)},
 \qquad
 -\zeta \sigma^{(s)}\leq\dot \sigma^{(s)}\leq\zeta \sigma^{(s)},
 \qquad
 \zeta=\frac{4\epsilon}{c}\leq6\epsilon.
 \label{ft:interpolated-metric-order}
\end{equation}
In particular,
$\abs{\ip{T}{\dot M_{s,u}T}_{\mathrm{HS}}}\leq\zeta\norm T_{M_{s,u}}^2$.

The canonical source derivative has the bound
\begin{equation}
 \norm{M_{s,u}^{-1/2}\dot\Phi(S_{I\cup J})}_{\mathrm{HS}}
 \leq6\epsilon d^{|J|}\sqrt{1-r}\,\norm{W_{I\cup J}}_M.
 \label{ft:noise-commutator-source}
\end{equation}
To prove it, choose the $d^{2|J|}$ Weyl unitaries on $\cH_J$. In a basis indexed modulo $d^{|J|}$, phase averaging removes off-diagonal matrix units and shift averaging makes the remaining diagonal constant. Thus
\[
 \Omega_J(X)=\sum_{\ell=1}^{d^{2|J|}}W_\ell XW_\ell^*,
 \qquad
 W_\ell=V_\ell/d^{|J|},
 \qquad
 \norm{W_\ell}=1/d^{|J|}.
\]
Set
\[
 B_\ell=W_\ell K,\qquad
 C_\ell=KW_\ell,\qquad
 E_\ell=B_\ell-C_\ell.
\]
Then
\[
 \norm{B_\ell},\norm{C_\ell}\leq k/d^{|J|},
 \qquad
 \norm{E_\ell}\leq2\epsilon/d^{|J|},
\]
and
\begin{equation}
 \dot\Phi(X)
 =
 (1-r)\sum_\ell
 \bigl(E_\ell XB_\ell^*+C_\ell XE_\ell^*\bigr).
 \label{ft:noise-commutator-kraus}
\end{equation}

For physical $J$-supported matrices $A_0,B_0$ and any test matrix $T$,
\begin{align}
 \abs{\Tr T^*A_0\rho W_{I\cup J}B_0^*}
 &\leq
 \frac{d^{|J|}\norm{A_0}\norm{B_0}}{\sqrt{(1-r)c}}\,
 \norm{W_{I\cup J}}_{\rho,\mathrm{left}}
 \norm{T}_{\rho^{(s)},\mathrm{left}},\label{ft:left-source-test}\\
 \abs{\Tr T^*A_0W_{I\cup J}\sigma B_0^*}
 &\leq
 \frac{d^{|J|}\norm{A_0}\norm{B_0}}{\sqrt{(1-r)c}}\,
 \norm{W_{I\cup J}}_{\sigma,\mathrm{right}}
 \norm{T}_{\sigma^{(s)},\mathrm{right}}.\label{ft:right-source-test}
\end{align}
Here
$\norm Y_{\rho,\mathrm{left}}=\norm{\rho^{1/2}Y}_{\mathrm{HS}}$ and
$\norm Y_{\sigma,\mathrm{right}}=\norm{Y\sigma^{1/2}}_{\mathrm{HS}}$.
For \eqref{ft:left-source-test}, pair $\rho^{1/2}A_0^*TB_0$ with
$\rho^{1/2}W_{I\cup J}$ and use
\[
 A_0\rho A_0^*
 \leq d^{2|J|}\norm{A_0}^2\Omega_J(\rho)
 \leq\frac{d^{2|J|}\norm{A_0}^2}{(1-r)c}\rho^{(s)}.
\]
For \eqref{ft:right-source-test}, pair with $W_{I\cup J}\sigma^{1/2}$, bound left multiplication by $A_0$, and use
\[
 B_0\sigma B_0^*
 \leq d^{2|J|}\norm{B_0}^2\Omega_J(\sigma)
 \leq\frac{d^{2|J|}\norm{B_0}^2}{(1-r)c}\sigma^{(s)}.
\]

The sum of the norm products in
\eqref{ft:noise-commutator-kraus} is at most $4\epsilon k$. Applying
\eqref{ft:left-source-test} and \eqref{ft:right-source-test} to
$S_{I\cup J}=\rho W_{I\cup J}+uW_{I\cup J}\sigma$, then using two-term Cauchy--Schwarz, bounds the tested source by
\[
 \frac{4\epsilon kd^{|J|}\sqrt{1-r}}{\sqrt c}
 \norm{W_{I\cup J}}_M\norm T_{M_{s,u}}.
\]
Since $k\leq9/8$ and $\sqrt c\geq7/8$, this coefficient is at most
$6\epsilon d^{|J|}\sqrt{1-r}$. Duality proves
\eqref{ft:noise-commutator-source}.

Differentiate the Sylvester equation for $Z_{s,u}$. Bimodularity with the fixed $z_{I\cup J}$ yields
\[
 \dot Z_{s,u}
 =
 M_{s,u}^{-1}
 \bigl[\dot\Phi(S_{I\cup J})-\dot \rho^{(s)}Y_s-uY_s\dot \sigma^{(s)}\bigr].
\]
The relative order bounds
\eqref{ft:interpolated-metric-order} imply that the dual norm of the final two terms is at most
$\zeta\norm{Y_s}_{M_{s,u}}$. Combining
\eqref{ft:interpolated-residual} and
\eqref{ft:noise-commutator-source}, we obtain
\[
 \norm{\dot Z_{s,u}}_{M_{s,u}}
 \leq
 [6\epsilon d^{|J|}\sqrt{1-r}+6\epsilon k]\norm{W_{I\cup J}}_M
 \leq13\epsilon d^{|J|}\norm{W_{I\cup J}}_M.
\]

Let
\[
 d_I(s,u)=\dist_{M_{s,u}}(Z_{s,u},\mathcal M_{I^c})^2,
 \qquad
 R_s=Z_{s,u}-\operatorname{proj}_{\mathcal M_{I^c}}^{M_{s,u}}Z_{s,u}.
\]
The metric Gram matrix of the fixed subspace is positive and differentiable, so its projection is differentiable. Orthogonality eliminates the projection derivative and gives
\[
 \dot d_I(s,u)
 =
 2\Re\ip{R_s}{\dot Z_{s,u}}_{M_{s,u}}
 +\ip{R_s}{\dot M_{s,u}R_s}_{\mathrm{HS}}.
\]
Since $z_{I\cup J}\in\mathcal M_{I^c}$,
$\norm{R_s}_{M_{s,u}}\leq\norm{Y_s}_{M_{s,u}}$. Therefore
\[
 \abs{\dot d_I(s,u)}
 \leq[26kd^{|J|}+6k^2]\epsilon\,\norm{W_{I\cup J}}_M^2
 \leq40d^{|J|}\epsilon\,\mathcal Q_{u,I\cup J}(\rho\mid\sigma).
\]
The dominating density is from the original pair.

Integrate first in $s$ and then in $u$. The positive entropy weight, \eqref{ft:homogeneous-conditional-integral}, and
\eqref{noise:eq-conditional-integral} give the unnormalized bound
\begin{equation}
 \abs{\Delta_I(\rho^{(0)}\mid \sigma^{(0)})-\Delta_I(\rho^{(1)}\mid \sigma^{(1)})}
 \leq40d^{|J|}\epsilon\Delta_{I\cup J}(\rho\mid \sigma).
 \label{ft:unnormalized-noise-order}
\end{equation}
No dependence on $r$ remains.

We keep the masses at the original pair and at both interpolation endpoints:
\[
 t=\Tr K\rho K^*,\qquad z=\Tr K\sigma K^*,\qquad
 t_j=\Tr \rho^{(j)},\qquad z_j=\Tr \sigma^{(j)},\qquad j=0,1.
\]
Explicitly,
\begin{align*}
 t_0&=\Tr K\rho_rK^*=rt+(1-r)\Tr K\Omega_J(\rho)K^*,&
 t_1&=\Tr K\rho K^*=t,\\
 z_0&=\Tr K\sigma_rK^*=rz+(1-r)\Tr K\Omega_J(\sigma)K^*,&
 z_1&=\Tr K\sigma K^*=z.
\end{align*}
All four masses lie in $[c,k^2]$. The input mass difference is
\[
 t_0-t_1
 =
 (1-r)\Tr \rho\bigl[\Omega_J(K^*K)-K^*K\bigr],
\]
so
\[
 \abs{t_0-t_1}
 \leq2(1-r)\norm{K^*K-\id}\leq5(1-r)\epsilon.
\]
By \eqref{ft:interpolated-residual},
\[
 \Delta_I(\rho^{(j)}\mid \sigma^{(j)})\leq k^2\Delta_{I\cup J}(\rho\mid \sigma),
 \qquad j=0,1.
\]
By \eqref{sc:independent-scaling}, independently normalizing endpoint $j$ divides its homogeneous deficit by $t_j$, while $z_j$ cancels.
It follows from \eqref{ft:unnormalized-noise-order} that
\[
 \abs{\frac{\Delta_I(\rho^{(0)}\mid \sigma^{(0)})}{t_0}
      -\frac{\Delta_I(\rho^{(1)}\mid \sigma^{(1)})}{t_1}}
 \leq
 \left(\frac{40d^{|J|}}{c}+\frac{5(1-r)k^2}{c^2}\right)
 \epsilon\Delta_{I\cup J}(\rho\mid \sigma)
 \leq80d^{|J|}\epsilon\Delta_{I\cup J}(\rho\mid \sigma).
\]
The normalized $\Phi_0$ pair is exactly $(\rho_r^K,\sigma_r^K)$. Since $\mathcal D_{J,r}$ preserves trace, the normalized $\Phi_1$ pair is exactly
$(\mathcal D_{J,r}\rho^K,\mathcal D_{J,r}\sigma^K)$. This proves
\eqref{ft:noise-order-cost} with the two factor orders and both normalizations explicit.

\medskip
\noindent\textbf{Step 4. Combining the estimates.}
The triangle inequality, together with the bounds
\eqref{ft:initial-noise-cost},
\eqref{ft:regularized-crossing-cost},
\eqref{ft:noise-order-cost}, and
\eqref{ft:final-noise-cost}
gives
\begin{align}
 \abs{\Delta_I(\rho\mid \sigma)-\Delta_I(\rho^K\mid \sigma^K)}\le &  
 \abs{\Delta_I(\rho\mid \sigma)-\Delta_I(\rho_r\mid \sigma_r)}\nonumber \\ &+\abs{\Delta_I(\rho_r\mid \sigma_r)-\Delta_I(\rho_r^K\mid \sigma_r^K)}\nonumber\\ &+
 \abs{\Delta_I(\rho_r^K\mid \sigma_r^K)-\Delta_I(\mathcal D_{J,r}\rho^K\mid \mathcal D_{J,r}\sigma^K)}\nonumber\\ &
 + \abs{\Delta_I(\mathcal D_{J,r}\rho^K\mid \mathcal D_{J,r}\sigma^K)-\Delta_I(\rho^K\mid \sigma^K)}\nonumber\\
 \le &  \Big(\omega_m(1-r)
  +\kappa(K)^2\omega_{m/2}(1-r)
  +\frac{100d^{|J|}\epsilon}{\sqrt{1-r}}
  +80d^{|J|}\epsilon\Big)\Delta_{I\cup J}(\rho\mid \sigma)
 \label{ft:four}
\end{align}
where the last inequality combines the terms of the union conditional relative entropy $\Delta_{I\cup J}$.
It remains to bound this coefficient.

\medskip
\noindent\textbf{Step 5. Coefficient Reduction.}

For $\epsilon>0$, define
\[
 \epsilon_0
 =
 \min\left(\frac18,\frac{m^{3/2}}{d^{|J|}},
                 \frac{48^{-3/2}}{d^{|J|}}\right).
\]
For $\epsilon\leq\epsilon_0$, take
\[
 r=1-(d^{|J|}\epsilon)^{2/3}.
\]
Then $1-r\leq m$, $1-r\leq1/48$, and
$\delta=d^{|J|}\epsilon/\sqrt{1-r}=1-r\leq1/48$, so the noise-dependent bound applies. Using
$\kappa(K)^2\leq2$ and $m\leq1$, the two endpoint noise terms satisfy
\[
 \omega_m(1-r)+\kappa(K)^2\omega_{m/2}(1-r)
 \leq\frac{5(1-r)}{m}\left[1+\log(1+1/(1-r))\right].
\]
The other terms are at most $100(1-r)+80d^{|J|}\epsilon\leq180(1-r)$. Hence the coefficient in
\eqref{ft:four} is at most
\[
 \frac{185(1-r)}{m}\left[1+\log(1+1/(1-r))\right].
\]
For $0\leq r<1$,
\[
 1+\log(1+1/(1-r))\leq2+\log(1/(1-r)),
 \qquad
 (1-r)^{1/4}\bigl[2+\log(1/(1-r))\bigr]\leq4.
\]
Thus the coefficient is at most
\[
\frac{185(1-r)}{m}\left[1+\log(1+1/(1-r))\right]\le  \frac{740d^{|J|/2}}{m}\sqrt\epsilon
 \leq\frac{1000d^{|J|/2}}{m}\sqrt\epsilon.
\]
If $\epsilon_0<\epsilon\leq1/8$, monotonicity of conditional relative entropy and
\eqref{ft:integrated-local-filter} give
\[
 \Delta_I(\rho\mid \sigma)\leq\Delta_{I\cup J}(\rho\mid \sigma),
 \qquad
 \Delta_I(\rho^K\mid \sigma^K)
 \leq 2\Delta_{I\cup J}(\rho\mid \sigma).
\]
Their absolute difference is at most $3\Delta_{I\cup J}(\rho\mid \sigma)$. Also
\[
 \frac1{\sqrt{\epsilon_0}}
 =
 \max\left(\sqrt8,d^{|J|/2}m^{-3/4},
                  d^{|J|/2}\,48^{3/4}\right).
\]
Since $d^{|J|}\geq1$, $m\leq1$, and $48^{3/4}<19$,
\[
 \frac3{\sqrt{\epsilon_0}}
 \leq\frac{57d^{|J|/2}}m
 \leq\frac{1000d^{|J|/2}}m.
\]
This proves \eqref{ft:union-bound} on the remaining range.
Approximation by faithful states and continuity extend the estimate
to every state $\rho\geq0$.

All local maps act trivially on the reference system $R$.
The estimates use only the displayed local parameters and extend
with constants independent of $\dim R$.

\section{Block entropy refinement}
\label{sec:tech-ch_block_refinement}\label{br:section}

\subsection{Two-set estimate from reference floors}

Use the entropy, reset, and weighted-norm conventions of
Appendices~\ref{noise:section} and~\ref{sc:section}.
Throughout this subsection, $\rho\geq0$ and $\sigma>0$ are
states on $\cH_\Lambda$.

\begin{lem}\label{br:repair}
Suppose $\sigma\geq m\Omega_J(\sigma)$, where $0<m\leq1$.
There is a positive matrix $\widetilde\rho\geq\rho$ whose
normalization $\omega=\widetilde\rho/\Tr\widetilde\rho$ satisfies
\begin{equation}
 \omega\geq\frac m2\Omega_J(\omega),\qquad
 0\leq\Tr\widetilde\rho-1
 \leq\frac{\Delta_J(\rho\mid\sigma)}{1-m/2},
 \label{br:eq:normalized-repair}
\end{equation}
and
\begin{equation}
 D(\rho\Vert\omega)\leq\log\Tr\widetilde\rho
 \leq\frac{\Delta_J(\rho\mid\sigma)}{1-m/2}.
 \label{br:eq:repair-entropy}
\end{equation}
\end{lem}

\begin{proof}
Put $k=m/2$ and $\eta=(k\Omega_J(\rho)-\rho)_+$.
The endpoint density in Lemma~\ref{noise:conditional-measure}
and the monotonicity of $G(s)/s$ stated after its proof give
\[
 \Delta_J(\rho\mid\sigma)
 \geq\int_k^m\frac{G(s)}s\,ds
 \geq\frac{m-k}{k}G(k)=\Tr\eta.
\]
Define the explicit repair
\[
 \widetilde\rho
 =\rho+\eta+\frac{k}{1-k}\Omega_J(\eta).
\]
Since $\Omega_J^2=\Omega_J$,
\[
 \widetilde\rho-k\Omega_J(\widetilde\rho)
 =\eta-(k\Omega_J(\rho)-\rho)\geq0.
\]
Moreover,
\[
 \Tr\widetilde\rho-1=\frac{\Tr\eta}{1-k}
 \leq\frac{\Delta_J(\rho\mid\sigma)}{1-m/2}.
\]
Normalization proves~\eqref{br:eq:normalized-repair}.
Operator monotonicity of the logarithm on
$\supp\widetilde\rho$ gives
$D(\rho\Vert\omega)\leq\log\Tr\widetilde\rho$.
Together with $\log t\leq t-1$, this
proves~\eqref{br:eq:repair-entropy}.
\end{proof}

\begin{lem}
Let $I,J\subseteq\Lambda$ be nonempty, possibly overlapping, and
suppose
\[
 \rho\geq k\Omega_J(\rho),\qquad
 \sigma\geq k\Omega_J(\sigma),\qquad 0<k\leq1.
\]
Then
\begin{equation}
 \Delta_{I\cup J}(\rho\mid\sigma)
 \leq\frac{d^{2|J|}}k
 \bigl[\Delta_I(\rho\mid\sigma)+\Delta_J(\rho\mid\sigma)\bigr].
 \label{br:eq:input-union}
\end{equation}
\end{lem}

\begin{proof}
For $u>0$, use $M_u,Z_u$ from Appendix~\ref{noise:section} and put
\[
 M_{0,u}=L_{\Omega_J(\rho)}+uR_{\Omega_J(\sigma)}.
\]
The assumed floors and the physical ceiling
\eqref{noise:eq-physical-ceiling} give
\[
 kM_{0,u}\leq M_u\leq d^{2|J|}M_{0,u}.
\]
In the $M_{0,u}$ metric, $\Omega_J$ is the orthogonal projection
onto $\operatorname{Ran}\Omega_J$.
It preserves $\operatorname{Ran}\Omega_I$ and therefore commutes
with the $M_{0,u}$-orthogonal projection onto this subspace.
Their product projects onto
$\operatorname{Ran}\Omega_{I\cup J}$, so
\begin{equation}
 \operatorname{dist}_{M_{0,u}}
 (Z,\operatorname{Ran}\Omega_{I\cup J})^2
 \leq\operatorname{dist}_{M_{0,u}}
 (Z,\operatorname{Ran}\Omega_I)^2
 +\operatorname{dist}_{M_{0,u}}
 (Z,\operatorname{Ran}\Omega_J)^2.
\end{equation}
Here distance is measured in the indicated weighted norm.
The metric orders transfer this inequality to the $M_u$ metric
with factor $d^{2|J|}/k$.
Taking $Z=Z_u$ and integrating with weight $u/(1+u)^2$,
Lemma~\ref{noise:canonical} proves~\eqref{br:eq:input-union}.
\end{proof}

Whenever $\sigma\geq m_V\Omega_V(\sigma)$ with $0<m_V\leq1$,
tensor-product logarithms give
\[
 \Delta_V(\rho\mid\sigma)
 =D(\rho\Vert\Omega_V(\rho))
 -\Tr\rho[\log\sigma-\log\Omega_V(\sigma)].
\]
The physical ceiling and logarithmic monotonicity on
$\cH_V\otimes\supp\rho_{V^c}$ imply
$D(\rho\Vert\Omega_V(\rho))\leq2|V|\log d$.
Consequently,
\begin{equation}
 0\leq\Delta_V(\rho\mid\sigma)\leq2|V|\log d-\log m_V.
 \label{br:eq:conditional-cap}
\end{equation}

\begin{lem}
If $\rho\leq c\omega$ for a state $\omega$ and $c\geq2$, then
for every $V\subseteq\Lambda$,
\begin{equation}
 \Delta_V(\rho\mid\sigma)
 \leq c\Delta_V(\omega\mid\sigma)+5D(\rho\Vert\omega).
 \label{br:eq:conditional-triangle}
\end{equation}
\end{lem}

\begin{proof}
First suppose $\rho,\omega>0$ and put
$\tau=(c\omega-\rho)/(c-1)$.
For $0\leq t<1$,
\[
 \omega+t(\rho-\omega)\leq c\omega,\qquad
 (1-t)\omega+t\tau\geq(1-t)\omega.
\]
The affine entropy integral and BKM comparison
\eqref{sc:bkm-quadratic-order} give
\[
 D(\rho\Vert\omega)
 \geq\frac{\norm{\rho-\omega}_{\Gamma_\omega^{-1}}^2}{2c},
\]
and
\begin{equation}
 D(\tau\Vert\omega)
 \leq\frac{\norm{\rho-\omega}_{\Gamma_\omega^{-1}}^2}{(c-1)^2}
 \leq\frac{2c}{(c-1)^2}D(\rho\Vert\omega).
 \label{br:eq:complement-entropy-bound}
\end{equation}
For the second bound, the affine integral has weight $1-t$,
which cancels the factor $(1-t)^{-1}$ from the BKM comparison;
the endpoint follows by continuity.

Expanding $\rho+(c-1)\tau=c\omega$ at the full and marginal
levels, then dropping the nonnegative
$\Delta_V(\tau\mid\sigma)$ and marginal relative entropies, gives
\[
 \Delta_V(\rho\mid\sigma)
 \leq c\Delta_V(\omega\mid\sigma)
 +D(\rho\Vert\omega)+(c-1)D(\tau\Vert\omega).
\]
Equation~\eqref{br:eq:complement-entropy-bound} and
$1+2c/(c-1)\leq5$ prove~\eqref{br:eq:conditional-triangle}.
For singular inputs, apply the bound to
$(1-\epsilon)\rho+\epsilon\sigma$ and
$(1-\epsilon)\omega+\epsilon\sigma$.
Joint convexity of $D$ and continuity of the conditional
relative entropies against $\sigma>0$ permit
$\epsilon\downarrow0$.
\end{proof}

\begin{lem}
\label{br:reference-union}
Let $I,J\subseteq\Lambda$ be nonempty, possibly overlapping, with
\[
 \sigma\geq m_I\Omega_I(\sigma),\qquad
 \sigma\geq m_J\Omega_J(\sigma),\qquad 0<m_I,m_J\leq1.
\]
Then
\begin{equation}
 \Delta_{I\cup J}(\rho\mid\sigma)
 \leq C_{I,J}
 \bigl[\Delta_I(\rho\mid\sigma)+\Delta_J(\rho\mid\sigma)\bigr],
 \label{br:eq:reference-union}
\end{equation}
where
\[
 C_{I,J}=\frac{4d^{2|J|}}{m_J}
 \left[1+\frac{2(|I|+|J|)\log d-\log(m_Im_J)}{1-m_J/2}\right]
 +\frac5{1-m_J/2}.
\]
\end{lem}

\begin{proof}
Lemma~\ref{br:repair} gives $\widetilde\rho\geq\rho$ and
$\omega=\widetilde\rho/t$, where $t=\Tr\widetilde\rho$, with
\[
 \omega\geq\frac{m_J}2\Omega_J(\omega),\qquad
 t-1,\ D(\rho\Vert\omega)
 \leq\frac{\Delta_J(\rho\mid\sigma)}{1-m_J/2}.
\]
For $t>1$, write
$\omega=t^{-1}\rho+(1-t^{-1})(\widetilde\rho-\rho)/(t-1)$.
Conditional convexity and~\eqref{br:eq:conditional-cap} yield
\begin{equation}
 t\Delta_V(\omega\mid\sigma)
 \leq\Delta_V(\rho\mid\sigma)
 +(t-1)(2|V|\log d-\log m_V),\qquad V=I,J.
 \label{br:eq:repaired-charges}
\end{equation}
For $t=1$, the same statement holds with equality.

Apply~\eqref{br:eq:conditional-triangle} with $c=2t$, and then
\eqref{br:eq:input-union} to $(\omega,\sigma)$ with $k=m_J/2$.
This gives
\[
 \Delta_{I\cup J}(\rho\mid\sigma)
 \leq\frac{4td^{2|J|}}{m_J}
 [\Delta_I(\omega\mid\sigma)+\Delta_J(\omega\mid\sigma)]
 +\frac5{1-m_J/2}\Delta_J(\rho\mid\sigma).
\]
Substituting~\eqref{br:eq:repaired-charges} and the bound on $t-1$
gives coefficient $4d^{2|J|}/m_J$ for
$\Delta_I(\rho\mid\sigma)$ and coefficient $C_{I,J}$ for
$\Delta_J(\rho\mid\sigma)$.
Since $C_{I,J}\geq4d^{2|J|}/m_J$, this
proves~\eqref{br:eq:reference-union}.
\end{proof}

\subsection{Fixed-size Gibbs refinement}

Let $\sigma$ be the Gibbs state of the model in
Section~\ref{main:sec-spin-chains}.
For an integer $\ell\geq1$, set
\[
 C_\ell=\left[
 4d^2c_{\beta/2}^{\,2}
 \left(1+\frac{\ell(2\log d+2\log c_{\beta/2})}
 {1-c_{\beta/2}^{-2}/2}\right)
 +\frac5{1-c_{\beta/2}^{-2}/2}
 \right]^{\ell-1}.
\]

\begin{prop}
\label{prop:fixed-refinement}
For every state $\rho\geq0$ and every nonempty physical set $V$
with $|V|\leq\ell$,
\begin{equation}
 \Delta_V(\rho\mid\sigma)
 \leq C_\ell\sum_{i\in V}\Delta_{\{i\}}(\rho\mid\sigma).
 \label{br:eq:fixed-refinement}
\end{equation}
The set $V$ may be disconnected.
The constant depends only on $\beta,d,r_0,J_0,\ell$.
\end{prop}

\begin{proof}
Lemma~\ref{gl:conditional-index} gives
\[
 \sigma\geq c_{\beta/2}^{-2|U|}\Omega_U(\sigma)
\]
for every physical subset $U$.
Enumerate $V=\{i_1,\ldots,i_s\}$ and put
$V_r=\{i_1,\ldots,i_r\}$.
The case $s=1$ is immediate.
For $r\geq2$, apply Lemma~\ref{br:reference-union} with
$I=V_{r-1}$, $J=\{i_r\}$, and
\[
 m_I=c_{\beta/2}^{-2(r-1)},\qquad m_J=c_{\beta/2}^{-2}.
\]
Here $|J|=1$ and
\[
 2(|I|+|J|)\log d-\log(m_Im_J)
 =r(2\log d+2\log c_{\beta/2})
 \leq\ell(2\log d+2\log c_{\beta/2}).
\]
Thus each union coefficient is at most the bracket defining
$C_\ell$.
Iterating $s-1\leq\ell-1$ times proves
\eqref{br:eq:fixed-refinement}, since that bracket is at least one.
No connectedness assumption is used.
\end{proof}

\noindent\textbf{Finite reference systems.}
All results remain valid after adjoining a finite reference $R$,
retained under every physical partial trace and reset.
For the Gibbs refinement, use the comparison state
$\eta\otimes\sigma$, where $\eta>0$ is any state on $R$.
The Gibbs orders tensor with $\eta$, and the repair, projection,
and entropy arguments retain their constants.
For a joint input $\rho$, take $\eta=\rho_R$ on
$\supp\rho_R\otimes\cH_\Lambda$.
The conclusions include singular inputs and are independent of
the reference dimension and spectrum.

\section{Interval entropy recursions}
\label{sec:tech-ch_embedded_recursion}
\label{er:section}

\subsection{Recursion from a filtered product marginal}

Recursive interval decompositions for conditional relative entropy
were used for commuting Gibbs states
in~\cite[Section~4]{BCLPR21}.
We use the conditional relative entropy of
Appendix~\ref{noise:section} and the normalized filter notation
of Appendix~\ref{sc:section}.
For an invertible filter $K$, write
\[
 \kappa(K)=\norm K\norm{K^{-1}}.
\]
Recall that erasure monotonicity~\cite{Lindblad75} gives
\[
 0\leq\Delta_I(\rho\mid\sigma)
 \leq\Delta_J(\rho\mid\sigma)\leq D(\rho\Vert\sigma),
 \qquad I\subseteq J.
\]

Decompose the physical system into tensor factors
$E_L,U,V,W,E_R$, allowing one-dimensional factors.
Let $\sigma>0$ be a state whose marginal satisfies
\[
 \sigma_{V^c}=\frac{K\omega K^*}{\Tr K\omega K^*},
 \qquad \omega=\omega_{E_LU}\otimes\omega_{WE_R},
\]
where both factors of $\omega$ are strictly positive states.
Assume that $K$ is invertible and supported on $U_0W_0$,
where $U_0\subseteq U$ and $W_0\subseteq W$ are tensor subfactors.

\begin{lem}
\label{er:two-sided-filter}
Under these assumptions, every state $\rho>0$ satisfies
\begin{equation}
 \kappa(K)^{-2}\Delta_{UVW}(\rho\mid\sigma)
 \leq\Delta_{UVW_0}(\rho\mid\sigma)
 +\Delta_{U_0VW}(\rho\mid\sigma).
 \label{er:two-sided-recursion}
\end{equation}
\end{lem}

\begin{proof}
Put
\[
 \widetilde\rho=(\rho_{V^c})^{K^{-1}},
 \qquad m=\Tr K\widetilde\rho K^*.
\]
Then
\[
 \rho_{V^c}=\frac{K\widetilde\rho K^*}{m},
 \qquad \sigma_{V^c}=\omega^K.
\]

Since $\omega=\omega_{E_LU}\otimes\omega_{WE_R}$,
the comparison-state terms cancel, and strong
subadditivity~\cite{LR73}, as used
in~\cite[Proposition~1]{BCLPR21}, gives
\[
\begin{aligned}
 &\Delta_U(\widetilde\rho\mid\omega)
 +\Delta_W(\widetilde\rho\mid\omega)
 -\Delta_{UW}(\widetilde\rho\mid\omega)\\
 &\qquad=S(\widetilde\rho_{E_LUE_R})
 +S(\widetilde\rho_{E_LWE_R})
 -S(\widetilde\rho_{E_LE_R})-S(\widetilde\rho)\geq0,
\end{aligned}
\]
where $S$ denotes the von Neumann entropy.
Erasure monotonicity therefore yields
\[
 \Delta_{UW}(\widetilde\rho\mid\omega)
 \leq\Delta_{UW_0}(\widetilde\rho\mid\omega)
 +\Delta_{U_0W}(\widetilde\rho\mid\omega).
\]

Each region $J\in\{UW,UW_0,U_0W\}$ contains the support of $K$.
The detailed comparison~\eqref{sc:normalized-filter-detailed}
gives
\[
 \norm{K^{-1}}^{-2}\Delta_J(\widetilde\rho\mid\omega)
 \leq m\Delta_J(\rho_{V^c}\mid\sigma_{V^c})
 \leq\norm K^2\Delta_J(\widetilde\rho\mid\omega).
\]
Use the right inequality for the parent and the left inequality
for the two children.
Combining with the preceding estimate and cancelling $m>0$ gives
\begin{equation}
 \kappa(K)^{-2}\Delta_{UW}(\rho_{V^c}\mid\sigma_{V^c})
 \leq\Delta_{UW_0}(\rho_{V^c}\mid\sigma_{V^c})
 +\Delta_{U_0W}(\rho_{V^c}\mid\sigma_{V^c}).
\end{equation}

Finally, the chain identity
\[
 \Delta_{V\cup J}(\rho\mid\sigma)
 =\Delta_V(\rho\mid\sigma)
 +\Delta_J(\rho_{V^c}\mid\sigma_{V^c}),
 \qquad J\subseteq U\cup W,
\]
applies to the parent and both children.
Each child contributes one copy of
$\Delta_V(\rho\mid\sigma)\geq0$, while the parent coefficient
satisfies $\kappa(K)^{-2}\leq1$.
This proves~\eqref{er:two-sided-recursion}.
\end{proof}

\subsection{Recursion with exterior errors}

Let $\sigma$ be the Gibbs state of the model in
Section~\ref{main:sec-spin-chains} on a finite path.
Partition the path into consecutive intervals $E_L,U,V,W,E_R$,
allowing empty exteriors, and set
\[
 b=|V|,\qquad I=UVW,\qquad
 s=\min\{|U|,|W|\},\qquad m=\lfloor\sqrt b\rfloor\leq s.
\]
Let $U_0$ be the last $m$ sites of $U$, and $W_0$ the first
$m$ sites of $W$. Define
\[
 I_-=UVW_0,\qquad I_+=U_0VW.
\]
All lengths count path sites.
The interval $I^{[r]}$ consists of the sites within distance $r$
of $I$ in the original path.
The decomposition is illustrated schematically in
Figure~\ref{fig:buffered-recursion}.

\begin{figure}[htbp]
\centering
\[
 E_L\ \big|\
 \underbrace{U\setminus U_0\ \big|\ U_0}_{U}
 \ \big|\ V\ \big|\
 \underbrace{W_0\ \big|\ W\setminus W_0}_{W}
 \ \big|\ E_R.
\]
\caption{The children $I_-=UVW_0$ and $I_+=U_0VW$ overlap on
$U_0VW_0$, containing $b+2m$ sites. Lengths are schematic.}
\label{fig:buffered-recursion}
\end{figure}

\begin{prop}
\label{er:original-charged}
There are model-only constants $c_{\mathrm{err}}\geq1$,
$c_{\mathrm{dec}}>0$ and an integer threshold $b_*\geq1$
such that every partition above with $b\geq b_*$ satisfies
\begin{equation}
 \bigl(1-c_{\mathrm{err}}\e^{-c_{\mathrm{dec}}\sqrt b}\bigr)
 \Delta_I(\rho\mid\sigma)
 \leq\Delta_{I_-}(\rho\mid\sigma)
 +\Delta_{I_+}(\rho\mid\sigma)
 +6\sum_{r>s}q_r\Delta_{I^{[r]}}(\rho\mid\sigma)
\end{equation}
for every state $\rho>0$.
The nonnegative coefficients $q_r$ depend only on the model
and $r$, and satisfy
\begin{equation}
 \sum_{r\geq m+1}q_r
 \leq c_{\mathrm{err}}\e^{-c_{\mathrm{dec}}\sqrt b},
 \label{er:q-tail}
\end{equation}
\begin{equation}
 q_r\leq
 \frac{c_{\mathrm{err}}^{\,r+1}}
 {((\lfloor r/r_0\rfloor+1)!)^{1/2}},
 \qquad r\geq m+1.
 \label{er:q-envelope}
\end{equation}
The threshold is chosen so that
\begin{equation}
 c_{\mathrm{err}}\e^{-c_{\mathrm{dec}}\sqrt{b_*}}\leq\frac12.
 \label{er:effective-loss}
\end{equation}
\end{prop}

\begin{proof}
\noindent\textbf{Step 1.}
Take
\[
 b_*\geq\max\{b_{\mathrm{full}},n_{\mathrm{full}}^2\},
\]
so that $b\geq b_*$ implies $m\geq n_{\mathrm{full}}$.
Apply Theorem~\ref{gl:full-reference} to the partition
$E_LU,V,WE_R$.
For $n\geq m$, its filters satisfy
\begin{equation}
 \cN_\sigma(C_{\mathrm{prod}}-C_n)
 \leq f_{\mathrm{full}}(n),\qquad
 \cN_\sigma(C_n),\ \cN_\sigma(C_n^{-1})
 \leq1+\frac1{16}+\frac18<2.
\end{equation}
Here $\cN_\sigma$ is evaluated at the original full Gibbs state.
Each $C_n$ is supported on $J_n$, and $C_n=C_{\mathrm{prod}}$
for all sufficiently large $n$.
Since $m\leq s$, we have $J_m=U_0\cup W_0$.

Put $\omega=\sigma_{E_LU}\otimes\sigma_{WE_R}$.
Exact productization~\eqref{gl:exact-productization} gives
\[
 \Tr_V(C_{\mathrm{prod}}^{-1}\sigma C_{\mathrm{prod}}^{-*})
 =\omega.
\]
Its full trace is therefore one.
Define the separately normalized pair
\[
 (\rho^m,\sigma^m)
 =(\rho^{C_mC_{\mathrm{prod}}^{-1}},
   \sigma^{C_mC_{\mathrm{prod}}^{-1}}).
\]
Since $C_m$ acts outside $V$, partial trace and normalization give
\[
 (\sigma^m)_{V^c}
 =\frac{C_m\omega C_m^*}{\Tr C_m\omega C_m^*}.
\]
The support of $C_m$ lies in $U_0W_0$, so
Lemma~\ref{er:two-sided-filter} yields
\begin{equation}
 \kappa(C_m)^{-2}\Delta_I(\rho^m\mid\sigma^m)
 \leq\Delta_{I_-}(\rho^m\mid\sigma^m)
 +\Delta_{I_+}(\rho^m\mid\sigma^m).
\end{equation}

The smallness and approximation bounds in
Theorem~\ref{gl:full-reference} imply
\[
 \norm{C_m-\id}
 \leq e_m:=c_{\mathrm{loc}}\e^{-c_{\mathrm{full}}\sqrt b}
 +f_{\mathrm{full}}(m)\leq\frac3{16}.
\]
Consequently,
\[
 \kappa(C_m)^{-2}
 \geq\left(\frac{1-e_m}{1+e_m}\right)^2,\qquad
 1-\kappa(C_m)^{-2}
 \leq\frac{4e_m}{(1+e_m)^2}\leq4e_m.
\]

\noindent\textbf{Step 2.}
For $n\geq m$, set
\[
 K_n=C_mC_n^{-1},\qquad
 (\rho_n,\sigma_n)=(\rho^{K_n},\sigma^{K_n}).
\]
Then $K_n$ is supported on $J_n$, $K_m=\id$, and
\[
 K_n-\id=(C_m-C_n)C_n^{-1},\qquad
 K_n^{-1}-\id=(C_n-C_m)C_m^{-1}.
\]
By Step~1 and submultiplicativity,
\[
 \cN_\sigma(K_n-\id)+\cN_\sigma(K_n^{-1}-\id)
 \leq4[f_{\mathrm{full}}(m)+f_{\mathrm{full}}(n)]
 \leq u_m,
\]
where $u_m=8\sum_{k\geq m}f_{\mathrm{full}}(k)$.
Since $u_m$ decreases to zero, increase $b_*$ so that
$u_{\lfloor\sqrt{b_*}\rfloor}\leq1/8$.
Thus, for $b\geq b_*$,
\[
 \kappa(K_n)^2
 \leq\left(\frac{1+u_m}{1-u_m}\right)^2
 \leq\left(\frac97\right)^2<2.
\]

The modular part of the same estimate gives
\[
 (1-u_m)^2\sigma
 \leq K_n\sigma K_n^*
 \leq(1+u_m)^2\sigma.
\]
Taking traces yields
\[
 (1-u_m)^2\leq\Tr K_n\sigma K_n^*\leq(1+u_m)^2.
\]
After normalization, the same ratio therefore gives
\[
 \frac12\sigma\leq\sigma_n\leq2\sigma.
\]
Lemma~\ref{gl:conditional-index} and positivity of $\Omega_J$
imply, for every physical subset $J$,
\begin{equation}
 \sigma_n\geq\frac12\sigma
 \geq\frac12c_{\beta/2}^{-2|J|}\Omega_J(\sigma)
 \geq\frac14c_{\beta/2}^{-2|J|}\Omega_J(\sigma_n).
\end{equation}

For $J\supseteq J_n$, the input normalization satisfies
\[
 \norm{K_n^{-1}}^{-2}
 \leq\Tr K_n\rho K_n^*
 \leq\norm{K_n}^2.
\]
Dividing~\eqref{sc:normalized-filter-detailed} by this trace
gives comparison factors $\kappa(K_n)^{\pm2}$. Hence
\begin{equation}
 \frac12\Delta_J(\rho\mid\sigma)
 \leq\Delta_J(\rho_n\mid\sigma_n)
 \leq2\Delta_J(\rho\mid\sigma).
\end{equation}

\noindent\textbf{Step 3.}
For $r\geq m+1$, the increment $K_rK_{r-1}^{-1}$ is invertible,
supported on $J_r$, and satisfies
\[
 K_rK_{r-1}^{-1}-\id
 =C_mC_r^{-1}(C_{r-1}-C_r)C_m^{-1}.
\]
Put $\epsilon_r=\norm{K_rK_{r-1}^{-1}-\id}$.
The three factors outside $C_{r-1}-C_r$ have norm at most $2$.
Hence
\begin{equation}
 \epsilon_r
 \leq8[f_{\mathrm{full}}(r-1)+f_{\mathrm{full}}(r)]
 \leq8(1+c_{\mathrm{loc}})f_{\mathrm{full}}(r-1),
\end{equation}
where $f_{\mathrm{full}}(r)
\leq c_{\mathrm{loc}}f_{\mathrm{full}}(r-1)$.
The first upper bound is at most $u_m\leq1/8$.

The intermediate trace normalizations cancel separately
for the input and comparison state, giving
\[
 (\rho_{r-1}^{K_rK_{r-1}^{-1}},
  \sigma_{r-1}^{K_rK_{r-1}^{-1}})
 =(\rho_r,\sigma_r).
\]
Since $C_n=C_{\mathrm{prod}}$ for all sufficiently large $n$,
this sequence is eventually equal to $(\rho^m,\sigma^m)$.

Step~2 supplies the local order bound
\[
 \sigma_{r-1}\geq
 \frac14c_{\beta/2}^{-2|J_r|}\Omega_{J_r}(\sigma_{r-1}).
\]
Using $|J_r|\leq4r$, the coefficient in
Theorem~\ref{ft:union-transport} before the factor
$\sqrt{\epsilon_r}$ is bounded by
\[
 4000(\sqrt d\,c_{\beta/2}^{\,2})^{|J_r|}
 \leq4000(d^2c_{\beta/2}^{\,8})^r.
\]
For $r\geq n_{\mathrm{full}}+1$, choose
\[
 q_r=4000\sqrt{8(1+c_{\mathrm{loc}})}
 (d^2c_{\beta/2}^{\,8})^r
 \sqrt{f_{\mathrm{full}}(r-1)}.
\]
Theorem~\ref{ft:union-transport} therefore bounds each increment
by $q_r$ times the conditional relative entropy on the union
of the target region and $J_r$.

For $J\in\{I,I_-,I_+\}$, this union contains $J_{r-1}$,
so Step~2 compares its conditional relative entropy with
that of the original pair with factor $2$.
Moreover, $J_r\subseteq I$ when $r\leq s$, while
$J\cup J_r\subseteq I^{[r]}$ when $r>s$.
Erasure monotonicity consequently gives
\[
 \abs{\Delta_J(\rho_r\mid\sigma_r)
 -\Delta_J(\rho_{r-1}\mid\sigma_{r-1})}
 \leq
 \begin{cases}
 2q_r\Delta_I(\rho\mid\sigma),&m<r\leq s,\\
 2q_r\Delta_{I^{[r]}}(\rho\mid\sigma),&r>s.
 \end{cases}
\]

\noindent\textbf{Step 4.}
Sum the finitely many nonzero increments in Step~3 for
the parent and both children, and substitute into Step~1.
The parent error has weight $\kappa(C_m)^{-2}$,
while each child error has weight one. Since
\[
 2\bigl(\kappa(C_m)^{-2}+2\bigr)\leq6,
\]
rearranging gives
\[
 \left[
 \kappa(C_m)^{-2}-6\sum_{r=m+1}^{s}q_r
 \right]\Delta_I(\rho\mid\sigma)
 \leq\Delta_{I_-}(\rho\mid\sigma)
 +\Delta_{I_+}(\rho\mid\sigma)
 +6\sum_{r>s}q_r\Delta_{I^{[r]}}(\rho\mid\sigma).
\]

The factorial indices at $r-1$ and $r$ differ by at most one.
Using $\sqrt{r+1}\leq2^r$, the definition of $q_r$ yields
\[
 q_r\leq
 \frac{4000\sqrt{8(1+c_{\mathrm{loc}})}
 (2d^2c_{\beta/2}^{\,8}\sqrt{c_{\mathrm{loc}}})^r}
 {((\lfloor r/r_0\rfloor+1)!)^{1/2}}.
\]
The factorial bounds ensure that the weighted series below
are finite.
Since $m\geq\sqrt b-1$ and $m+1>\sqrt b$,
\[
 f_{\mathrm{full}}(m)\leq\e^{1-\sqrt b}
 \sum_{n\geq n_{\mathrm{full}}}\e^nf_{\mathrm{full}}(n),
\]
\[
 \sum_{r\geq m+1}q_r\leq\e^{-\sqrt b}
 \sum_{r\geq n_{\mathrm{full}}+1}\e^rq_r.
\]

Take $c_{\mathrm{dec}}=\min\{1,c_{\mathrm{full}}\}$ and choose
\[
 c_{\mathrm{err}}\geq
 \max\left\{
 4000\sqrt{8(1+c_{\mathrm{loc}})},
 2d^2c_{\beta/2}^{\,8}\sqrt{c_{\mathrm{loc}}}
 \right\},
\]
\[
 c_{\mathrm{err}}\geq4c_{\mathrm{loc}}
 +4\e\sum_{n\geq n_{\mathrm{full}}}\e^nf_{\mathrm{full}}(n)
 +6\sum_{r\geq n_{\mathrm{full}}+1}\e^rq_r.
\]
These choices give~\eqref{er:q-tail} and~\eqref{er:q-envelope}.
Together with Step~1, they also give
\[
 \kappa(C_m)^{-2}-6\sum_{r=m+1}^{s}q_r
 \geq1-4e_m-6\sum_{r\geq m+1}q_r
 \geq1-c_{\mathrm{err}}\e^{-c_{\mathrm{dec}}\sqrt b}.
\]
Finally, increase $b_*$, preserving the conditions in
Steps~1 and~2, so that~\eqref{er:effective-loss} holds.
Substitution into the first inequality proves the proposition.
\end{proof}

\subsection{Unified path and cycle recursion}

On a path, let $\pi$ be the identity map.
On a cycle labeled by $\Lambda=\{0,\ldots,N-1\}$, set
\[
 \pi(i)=\min(i,N-i).
\]
For either geometry, define
\[
 \Lambda_\pi=\pi(\Lambda),\qquad G_k=\pi^{-1}(\{k\}),\qquad
 \delta_I=\Delta_{\pi^{-1}(I)}(\rho\mid\sigma).
\]
Lengths and enlargements $I^{[r]}$ below are measured on
the index path $\Lambda_\pi$, with clipping at its endpoints.
Each group contains at most two physical sites.
Since $\pi$ changes by at most one along each physical edge,
every connected physical set has an interval image.

By the reflection regrouping of Appendix~\ref{gl:section},
the padded cycle is a path model with parameters
$(\beta,d^2,r_0,2J_0)$.
Tensoring the input and Gibbs state with the same padding state
leaves the conditional relative entropies unchanged.
We choose common constants $c_{\mathrm{err}},c_{\mathrm{dec}}$
and thresholds in Proposition~\ref{er:original-charged}
for the original and grouped models.

Let $n_0\geq1$ be an integer threshold to be chosen in the proof.
For $n\geq n_0$, define
\[
 b_n=\lceil\sqrt n\rceil,\qquad
 m_n=\lfloor\sqrt{b_n}\rfloor,\qquad
 \mathcal K_n=
 \{\lceil n/4\rceil+m_n,\ldots,\lfloor n/2\rfloor\}.
\]
For $I=[u,u+n-1]\subseteq\Lambda_\pi$ and $k\in\mathcal K_n$, put
\[
 I_-=[u,u+k+b_n+m_n-1],\qquad
 I_+=[u+k-m_n,u+n-1].
\]

\begin{thm}
\label{sf:recursion-interface}
There is a model-only integer threshold $n_0$ such that,
for every state $\rho>0$, every $n\geq n_0$,
every interval $I\subseteq\Lambda_\pi$ of length $n$,
and every $k\in\mathcal K_n$,
\begin{equation}
 \delta_I\leq\frac{\delta_{I_-}+\delta_{I_+}}{1-e(n)}
 +\sum_{r\geq\lceil n/4\rceil}a_r\delta_{I^{[r]}},
 \label{sf:eq:embedded-recursion}
\end{equation}
where
\begin{equation}
 0\leq e(n)=c_{\mathrm{err}}\e^{-c_{\mathrm{dec}}n^{1/4}}
 \leq\frac12,\qquad
 a_r=\frac{12c_{\mathrm{err}}^{\,r+1}}
 {((\lfloor r/r_0\rfloor+1)!)^{1/2}}.
 \label{sf:eq:recursion-envelopes}
\end{equation}
The children satisfy
\[
 I_-\cup I_+=I,\qquad I_-,I_+\subseteq I,\qquad
 |I_-|,|I_+|\leq\frac{3n}{4}.
\]
If $k$ is chosen uniformly from $\mathcal K_n$, then,
for every $x\in I$,
\[
 \Pr\{x\in I_-\cap I_+\}\leq\frac{10}{\sqrt n}.
\]
\end{thm}

\begin{proof}
Take $n_0\geq\max\{256,b_*^2\}$.
The bounds $b_n\leq\sqrt n+1$ and $m_n\leq n^{1/4}+1$
imply, for $n\geq n_0$,
\[
 b_n+m_n\leq n/4,\qquad
 b_n+2m_n\leq2\sqrt n,\qquad m_n+1\leq n/20.
\]
Also,
\[
 b_n\geq b_*,\qquad
 e(n)\leq c_{\mathrm{err}}\e^{-c_{\mathrm{dec}}\sqrt{b_*}}
 \leq\frac12.
\]
Split $I=UVW$ into lengths $k,b_n,n-k-b_n$. Then
\[
 s=\min\{k,n-k-b_n\}\geq n/4,\qquad m_n\leq s.
\]
Apply Proposition~\ref{er:original-charged} on the index path.
Since $\sqrt{b_n}\geq n^{1/4}$, it gives
\[
 (1-e(n))\delta_I
 \leq\delta_{I_-}+\delta_{I_+}
 +6\sum_{r>s}q_r\delta_{I^{[r]}}.
\]
Now $6/(1-e(n))\leq12$, $12q_r\leq a_r$,
and $r>s$ implies $r\geq\lceil n/4\rceil$.
Division by $1-e(n)$ proves~\eqref{sf:eq:embedded-recursion}.
For cycles, the preceding regrouping identifies the indexed
erasures with their physical preimages under $\pi$,
so the same argument applies.

The definitions give $I_-\cup I_+=I$ with both children
contained in $I$, and
\[
 |I_-|=k+b_n+m_n\leq3n/4,\qquad
 |I_+|=n-k+m_n\leq3n/4.
\]
Moreover,
\[
 |\mathcal K_n|
 =\lfloor n/2\rfloor-\lceil n/4\rceil-m_n+1
 \geq n/4-m_n-1\geq n/5.
\]
A fixed $x$ lies in the overlap for at most $b_n+2m_n$ cuts.
Hence
\[
 \Pr\{x\in I_-\cap I_+\}
 \leq\frac{b_n+2m_n}{|\mathcal K_n|}
 \leq\frac{2\sqrt n}{n/5}=\frac{10}{\sqrt n}.
\]
\end{proof}

\noindent\textbf{Finite reference systems.}
All results remain valid with a finite reference $R$ retained
by every physical erasure.
Replace the comparison state $\sigma$ by $\eta\otimes\sigma$,
where $\eta>0$ is any fixed state on $R$.
By~\eqref{sc:reference-cancellation}, the conditional relative
entropies are independent of $\eta$, even when filtering
changes the input marginal on $R$.
In Lemma~\ref{er:two-sided-filter}, strong subadditivity
conditions on $RE_LE_R$.
The comparisons in Appendices~\ref{sc:section}
and~\ref{ft:section} retain their constants, and
\[
 \cN_{\eta\otimes\sigma}(\id_R\otimes X)=\cN_\sigma(X).
\]

For a joint input $\rho\geq0$, take $\eta=\rho_R$ and work on
$\supp\rho_R\otimes\cH_\Lambda$, where
$\rho_R\otimes\sigma>0$.
Approximation by
$(1-\epsilon)\rho+\epsilon(\rho_R\otimes\sigma)$,
followed by continuity, extends the conclusions to singular
inputs.
In particular, the unified recursion uses
\[
 \delta_I=
 \Delta_{\pi^{-1}(I)}(\rho\mid\rho_R\otimes\sigma)
\]
with the same constants, independently of the reference
dimension and spectrum.

\section{Spatial entropy factorization}
\label{sec:tech-ch_spatial_factorization}\label{sf:preliminaries}

\subsection{Fixed-window entropy factorization}

Let $\sigma$ be the physical Gibbs state.
Use the index path $\Lambda_\pi$, the group map $\pi$, and the recursion
of Theorem~\ref{sf:recursion-interface}, with
$\delta_I=\Delta_{\pi^{-1}(I)}(\rho\mid\sigma)$.

\begin{prop}
\label{prop:regional-charges}
There exist an integer $L\geq n_0$ and a constant $C\geq1$,
depending only on $\beta,d,r_0,J_0$, with the following properties.
For $j\in\Lambda_\pi$, set
\[
 J_j=\Lambda_\pi\cap\{j,\ldots,j+L-1\},\qquad W_j=\pi^{-1}(J_j).
\]
For every state $\rho\geq0$,
\begin{equation}
 D(\rho\Vert\sigma)\leq C\sum_j\Delta_{W_j}(\rho\mid\sigma).
 \label{sf:eq:fixed-cover}
\end{equation}
For every integer $\ell\geq0$,
\begin{equation}
 \sum_i\Delta_{S_{i,\ell}}(\rho\mid\sigma)
 \leq C(4r_0\ell+2)\sum_j\Delta_{W_j}(\rho\mid\sigma).
 \label{sf:eq:summed-ball-charges}
\end{equation}
Each window contains at most $L$ physical sites on paths and
at most $2L$ on cycles. Every physical site belongs to at most
$L$ indexed windows.
\end{prop}

\begin{proof}
Fix an integer $L\geq n_0$, to be chosen below, and put $q=4/3$.
Using the explicit expression for $e$ in
\eqref{sf:eq:recursion-envelopes}, define
\[
 c_L=\prod_{t\geq0}
 \frac{1+10/(\sqrt L\,q^{t/2})}{1-e(Lq^t)}.
\]
The stretched-exponential decay of $e$ gives
$\sum_{t\geq0}e(Lq^t)\to0$ as $L\to\infty$. Hence
\[
 0\leq\log c_L
 \leq\frac{10}{\sqrt L(1-q^{-1/2})}
 +2\sum_{t\geq0}e(Lq^t)\longrightarrow0.
\]

Start each root interval $I$ with weight one.
At every node $P$ with $|P|>L$, choose an admissible cut
uniformly and independently, assign the child weights
\[
 w_{P_-}=w_{P_+}=\frac{w_P}{1-e(|P|)},
\]
and stop when the interval length is at most $L$.
Write $\mathbb E_I$ for expectation over the cuts.
Repeated intervals are counted with their tree occurrences.

Each branch visits a band $Lq^t<|P|\leq Lq^{t+1}$ at most once,
since each child has length at most $3|P|/4$.
The averaged overlap bound in Theorem~\ref{sf:recursion-interface},
together with the monotonicity of $e$, therefore gives
\[
 \mathbb E_I\sum_{T\text{ leaf}}w_T\mathbf1_{\{x\in T\}}
 \leq c_L\mathbf1_{\{x\in I\}}.
\]
Nodes in one band have disjoint descendant trees, and each node
containing $x$ has a descendant leaf containing $x$ of at least
its weight. Consequently,
\[
 \mathbb E_I
 \sum_{\substack{P\text{ internal}\\Lq^t<|P|\leq Lq^{t+1}}}
 w_P\mathbf1_{\{x\in P\}}
 \leq c_L\mathbf1_{\{x\in I\}}.
\]

All interval sums below range over nonempty intervals of $\Lambda_\pi$.
For the fixed input, define
\[
 \mathfrak M=
 \max\left\{\sum_I\lambda_I\delta_I:
 \lambda_I\geq0,\quad
 \sum_{I\ni x}\lambda_I\leq1
 \ \text{for every }x\in\Lambda_\pi\right\}.
\]
Since there are finitely many intervals and the constraints imply
$0\leq\lambda_I\leq1$, the feasible set is compact and the maximum
is finite.

Fix any feasible root weights $(\lambda_I)$.
Expanding their averaged recursion trees gives leaf contributions
and enlarged-interval errors.
Every leaf starting at $j$ is contained in $J_j$, so its
conditional relative entropy is at most
$\Delta_{W_j}(\rho\mid\sigma)$.
The leaf coverage bound and feasibility give
\[
 \sum_I\lambda_I\mathbb E_I
 \sum_{\substack{T\text{ leaf}\\\min T=j}}w_T
 \leq c_L\sum_{I\ni j}\lambda_I\leq c_L.
\]
Consequently,
\[
 \sum_I\lambda_I\delta_I
 \leq c_L\sum_j\Delta_{W_j}(\rho\mid\sigma)
 +\sum_Jh_J\delta_J,
\]
where
\[
 h_J=\sum_I\lambda_I\mathbb E_I
 \sum_{P\text{ internal}}w_P
 \sum_{r\geq\lceil|P|/4\rceil}
 a_r\mathbf1_{\{J=P^{[r]}\}}.
\]
These nonnegative coefficients are finite because the trees
are finite and $\sum_ra_r<\infty$.

A contributing pair $(P,r)$ in band $t$ satisfies
$Lq^t<|P|\leq4r$.
If $x\in P^{[r]}$, then $|x-y|\leq5r$ for every $y\in P$,
and hence
\[
 \mathbf1_{\{x\in P^{[r]}\}}
 \leq\frac1{Lq^t}
 \sum_{y\in P}\mathbf1_{\{|x-y|\leq5r\}}.
\]
Applying the band coverage bound at each $y$, summing against
the feasible root weights, and counting at most $10r+1$
such indices, including near the endpoints, gives
\[
 \sum_I\lambda_I\mathbb E_I
 \sum_{\substack{P\text{ internal}\\
 Lq^t<|P|\leq Lq^{t+1}\\|P|\leq4r}}
 w_P\mathbf1_{\{x\in P^{[r]}\}}
 \leq\frac{c_L(10r+1)}{Lq^t}
 \leq\frac{11c_Lr}{Lq^t}.
\]
Summing over $r$ and $t$, using $\sum_{t\geq0}q^{-t}=4$, yields
\[
 \sum_{J\ni x}h_J\leq\kappa_L,\qquad
 \kappa_L:=\frac{44c_L}{L}
 \sum_{r\geq\lceil L/4\rceil}r\,a_r.
\]
Since $c_L\to1$ and $a_r$ satisfies the factorial bound in
\eqref{sf:eq:recursion-envelopes}, we have $\kappa_L\to0$.
Choose a model-dependent integer $L\geq n_0$ such that
$\kappa_L\leq1/2$.

For $\kappa_L>0$, the weights $h_J/\kappa_L$ are feasible
for $\mathfrak M$, so
\[
 \sum_Jh_J\delta_J\leq\kappa_L\mathfrak M.
\]
This is also immediate when $\kappa_L=0$.
Taking the maximum over the feasible root weights gives
\[
 \mathfrak M\leq
 c_L\sum_j\Delta_{W_j}(\rho\mid\sigma)+\kappa_L\mathfrak M.
\]
Absorbing the last term, we obtain
\[
 \mathfrak M\leq C\sum_j\Delta_{W_j}(\rho\mid\sigma),
 \qquad C:=\frac{c_L}{1-\kappa_L}\leq2c_L.
\]

The full index interval is a feasible single root, so
\[
 D(\rho\Vert\sigma)=\delta_{\Lambda_\pi}
 \leq\mathfrak M\leq C\sum_j\Delta_{W_j}(\rho\mid\sigma).
\]
This proves~\eqref{sf:eq:fixed-cover}.

For either geometry, each ball has an interval image
$\pi(S_{i,\ell})$ and satisfies
$S_{i,\ell}\subseteq\pi^{-1}(\pi(S_{i,\ell}))$.
For every group index $x$,
\[
 \#\{i:x\in\pi(S_{i,\ell})\}
 =\left|\bigcup_{v\in G_x}S_{v,\ell}\right|
 \leq|G_x|(2r_0\ell+1)\leq4r_0\ell+2.
\]
Counting repeated image intervals with multiplicity,
this root family divided by $4r_0\ell+2$ is feasible
for $\mathfrak M$.
Monotonicity of conditional relative entropy gives
\[
 \sum_i\Delta_{S_{i,\ell}}(\rho\mid\sigma)
 \leq\sum_i\delta_{\pi(S_{i,\ell})}
 \leq(4r_0\ell+2)\mathfrak M
 \leq C(4r_0\ell+2)\sum_j\Delta_{W_j}(\rho\mid\sigma).
\]
This proves~\eqref{sf:eq:summed-ball-charges}.

Finally, $|J_j|\leq L$, and each group contains one physical
site on a path and at most two on a cycle.
Thus $|W_j|\leq L$ on paths and $|W_j|\leq2L$ on cycles.
If a physical site $i$ belongs to $W_j$, then
\[
 \pi(i)-L+1\leq j\leq\pi(i).
\]
There are at most $L$ such indices $j$, proving the stated
window multiplicity.
\end{proof}

\subsection{Proof of Theorem~\ref{main:thm-static}}
\label{sf:static-proof}

\begin{proof}
The windows in Proposition~\ref{prop:regional-charges}
have at most $2L$ physical sites, and each site belongs
to at most $L$ windows.
Proposition~\ref{prop:fixed-refinement} therefore gives
\[
 \sum_j\Delta_{W_j}(\rho\mid\sigma)
 \leq C_{2L}\sum_j\sum_{i\in W_j}\Delta_{\{i\}}(\rho\mid\sigma)
 \leq C_{2L}L\sum_i\Delta_{\{i\}}(\rho\mid\sigma).
\]
Combining this with the two bounds of
Proposition~\ref{prop:regional-charges} yields
\[
 D(\rho\Vert\sigma)
 \leq CC_{2L}L\sum_i\Delta_{\{i\}}(\rho\mid\sigma),
\]
and, for every integer $\ell\geq0$,
\[
 \sum_i\Delta_{S_{i,\ell}}(\rho\mid\sigma)
 \leq CC_{2L}L(4r_0\ell+2)
 \sum_i\Delta_{\{i\}}(\rho\mid\sigma).
\]
Since $4r_0\ell+2\leq(4r_0+2)(1+\ell)$, we may take
\[
 C_s=CC_{2L}L,\qquad C_r=(4r_0+2)C_s.
\]
Both constants depend only on $\beta,d,r_0,J_0$.
\end{proof}

\noindent\textbf{Finite reference systems.}
For any finite reference system $R$ and joint state $\rho$,
use the pair $(\rho,\rho_R\otimes\sigma)$ on
$\supp\rho_R\otimes\cH_\Lambda$, retaining $R$ under all
physical partial traces.
The recursion of Appendix~\ref{sec:tech-ch_embedded_recursion}
and the refinement of Proposition~\ref{prop:fixed-refinement}
hold with the same constants, while the remaining arguments
use only physical-region monotonicity and scalar covering
estimates. Moreover,
\[
 \Delta_\Lambda(\rho\mid\rho_R\otimes\sigma)
 =D(\rho\Vert\rho_R\otimes\sigma).
\]
Thus Proposition~\ref{prop:regional-charges} and
Theorem~\ref{main:thm-static} hold for arbitrary joint states,
including singular inputs, with constants independent of
the reference dimension and the spectrum of $\rho_R$.

\section{Arithmetic energy and entropy production}\label{sec:tech-ch_actual_rows}
\label{ar:introduction}

\subsection{Arithmetic energy and onsite control}
\label{noise:arithmetic-section}

We use the reset maps, operators $M_u$, scores $Z_u$, and weighted
norms of Appendix~\ref{noise:section}.

For states $\rho\geq0$, $\sigma>0$ and a Hermitian physical operator $X$, define the arithmetic energy
\begin{equation}
 E_X(\rho\mid\sigma)=\int_0^\infty\frac{u}{(1+u)^2}\norm{[X,Z_u]}_{M_u}^2\,du\in[0,\infty].
\end{equation}
The auxiliary dephasing generator associated with $X$ is
\[
 \widetilde{\cL}_X(Y):=XYX-\tfrac12(X^2Y+YX^2)=-\tfrac12[X,[X,Y]].
\]
It equals its trace dual, $\widetilde{\cL}_{X,*}=\widetilde{\cL}_X$, so the same formula acts on observables and states.

Assume now that $\rho,\sigma>0$. The two-weight channel identity of~\cite[Section~3.2, (3.3)]{GG26Entropy} has the following infinitesimal form: for Hermitian variations $A,B$, differentiation of the source and inverse metric gives
\begin{equation}\label{noise:eq-moving-resolvent}
 d\mathcal Q_u(\rho\mid\sigma)[A,B]
 =2\Re\Tr[(A-B)Z_u]-\Tr(AZ_uZ_u^*)-u\Tr(BZ_u^*Z_u).
\end{equation}
The square field of $\widetilde{\cL}_X$ satisfies
\[
 \Gamma_{\widetilde{\cL}_X}(Y):=\widetilde{\cL}_X(Y^*Y)-\widetilde{\cL}_X(Y^*)Y-Y^*\widetilde{\cL}_X(Y)=[X,Y]^*[X,Y].
\]
Set $A=\widetilde{\cL}_X(\rho)$ and $B=\widetilde{\cL}_X(\sigma)$. Using the adjoint score equation $Z_u^*\rho+u\sigma Z_u^*=\rho-\sigma$, trace duality in~\eqref{noise:eq-moving-resolvent} yields
\[
 -d\mathcal Q_u(\rho\mid\sigma)[A,B]
 =\Tr\rho\Gamma_{\widetilde{\cL}_X}(Z_u^*)+u\Tr\sigma\Gamma_{\widetilde{\cL}_X}(Z_u)
 =\norm{[X,Z_u]}_{M_u}^2.
\]

To justify differentiation under the integral, let $\rho_t=\rho+tA$ and $\sigma_t=\sigma+tB$, where $A,B$ are Hermitian and traceless. On a sufficiently small interval about zero, both states have a common positive lower bound. Thus, for constants $c_1,c_2>0$ independent of $u$ and $t$ in this interval,
\[
 \norm{M_u(t)^{-1}}\leq\frac1{c_1(1+u)},\qquad \norm{Z_u(t)}_2\leq\frac{c_2}{1+u}.
\]
The differential identity then bounds the absolute value of the weighted derivative by $c_3u/(1+u)^3$ for some $c_3>0$. Dominated convergence permits differentiation of~\eqref{noise:eq-homogeneous-full-integral}; its linear trace terms have zero derivative. These local constants may depend on the states and tangents and enter no subsequent comparison constant.

Since the derivative depends only on the tangent, any differentiable state paths starting at $(\rho,\sigma)$ with $\rho'(0)=\widetilde{\cL}_X(\rho)$ and $\sigma'(0)=\widetilde{\cL}_X(\sigma)$ satisfy
\begin{equation}\label{noise:eq-moving-pair-entropy}
 -\left.\frac d{dt}D(\rho(t)\Vert\sigma(t))\right|_{t=0}=E_X(\rho\mid\sigma)<\infty.
\end{equation}
Both states evolve here, so no stationarity assumption on $\sigma$ is required. If $X$ is supported on $V$, partial-trace cyclicity gives
\[
 \Tr_V\widetilde{\cL}_X(\rho)=\Tr_V\widetilde{\cL}_X(\sigma)=0.
\]
The marginal relative entropy consequently has zero derivative, and hence
\[
 -\left.\frac d{dt}\Delta_V(\rho(t)\mid\sigma(t))\right|_{t=0}=E_X(\rho\mid\sigma).
\]

At a site $i$ of dimension $d\geq2$, take a Hermitian basis $U_0=\id_i,U_1,\ldots,U_{d^2-1}$ with $\Tr_i(U_aU_b)=d\delta_{ab}$, extended by the identity outside $i$. Rescaling~\eqref{main:eq-proposal-completeness} from $U_a/d$ to $U_a$, and using $\widetilde{\cL}_{U_0}=0$, gives
\[
 \sum_{a>0}\widetilde{\cL}_{U_a}=d^2(\Omega_i-\idmap).
\]
For $\rho(t)=(1-t)\rho+t\Omega_i(\rho)$ and $\sigma(t)=(1-t)\sigma+t\Omega_i(\sigma)$, joint convexity gives
\[
 D(\rho(t)\Vert\sigma(t))\leq D(\rho\Vert\sigma)-t\Delta_i(\rho\mid\sigma),\qquad 0\leq t\leq1,
\]
because $D(\Omega_i(\rho)\Vert\Omega_i(\sigma))=D(\rho_{i^c}\Vert\sigma_{i^c})$. Taking the right derivative at zero and applying~\eqref{noise:eq-moving-pair-entropy}, with linearity in the tangent, yields
\begin{equation}\label{noise:eq-onsite-frame}
 \Delta_i(\rho\mid\sigma)\leq\frac1{d^2}\sum_{a>0}E_{U_a}(\rho\mid\sigma).
\end{equation}

\subsection{Paired-noise stability}

\begin{thm}
\label{noise:arithmetic-stability}
Let $\rho,\sigma>0$ be states on $\cH_\Lambda$, and let $V\subseteq\Lambda$ satisfy $\dim\cH_V=d^{|V|}$. Assume $\sigma\geq m\Omega_V(\sigma)$ for $0<m\leq1$, and let $X=X^*$ be supported on $V$. For $0<a\leq1/2$, set $r=1-a$ and
\[
 \rho_r=r\rho+a\Omega_V(\rho),\qquad \sigma_r=r\sigma+a\Omega_V(\sigma).
\]
There is an absolute constant $c_{\mathrm{st}}\geq1$ such that
\begin{equation}\label{noise:eq-arithmetic-stability}
 E_X(\rho_r\mid\sigma_r)\leq rE_X(\rho\mid\sigma)
 +c_{\mathrm{st}}d^{6|V|}m^{-8}a\norm X^2\Delta_V(\rho\mid\sigma).
\end{equation}
\end{thm}

\begin{proof}
\medskip
\noindent\textbf{Step 1.}
Let $\Phi_t=\e^{t\widetilde{\cL}_X}$, which equals its trace dual. Since $\widetilde{\cL}_X=-\ad_X^2/2$, where $\ad_X(Y)=[X,Y]$, the Gaussian Fourier identity gives
\[
 \Phi_t(Y)=\frac1{\sqrt{2\pi t}}\int_\R\e^{-s^2/(2t)}\e^{\ii sX}Y\e^{-\ii sX}\,ds,\qquad t>0.
\]
Thus $\Phi_t$ is completely positive, trace-preserving and unital. Since $X$ is supported on $V$, the map $\Phi_t$ is a bimodule map over operators on $V^c$ and satisfies
\[
 \Omega_V\Phi_t=\Omega_V=\Phi_t\Omega_V,\qquad \mathcal D_{V,r}\Phi_t=\Phi_t\mathcal D_{V,r}.
\]
Set $\rho(t)=\Phi_t(\rho)$, $\sigma(t)=\Phi_t(\sigma)$, and apply the reset family of Appendix~\ref{noise:section} to these states. Then
\[
 \rho_r(t)=\Phi_t(\rho_r),\qquad \sigma_r(t)=\Phi_t(\sigma_r).
\]
All these states remain strictly positive. Their $V^c$-marginals are fixed, so $Z_{u,V^c}$ is independent of $t$. The reference floor is preserved, and~\eqref{noise:eq-physical-ceiling} gives
\[
 m\Omega_V(\sigma)\leq\sigma(t)\leq d^{2|V|}\Omega_V(\sigma),\qquad
 \rho(t)\leq d^{2|V|}\Omega_V(\rho).
\]

Use the coordinates of Lemma~\ref{noise:conditional-measure} with the time parameter included:
\[
 M_{r,u}(t)=L_{\rho_r(t)}+uR_{\sigma_r(t)},\qquad
 S_u(t)=\rho(t)-\sigma(t)-M_{1,u}(t)(\id_V\otimes Z_{u,V^c}),
\]
\[
 T_u(t)=M_{0,u}^{-1/2}M_{1,u}(t)M_{0,u}^{-1/2},\qquad
 v_u(t)=M_{0,u}^{-1/2}S_u(t).
\]
Here $M_{0,u}=L_{\Omega_V(\rho)}+uR_{\Omega_V(\sigma)}$ is independent of $t$. The reduced score equation gives $\Tr_VS_u(t)=0$, while bimodularity gives $S_u(t)=\Phi_t(S_u(0))$.

The operator $M_{0,u}$ commutes with $\ad_X$. Consequently,
\[
 v_u(t)=\e^{-t\ad_X^2/2}v_u(0),\qquad \norm{v_u(t)}_2\leq\norm{v_u(0)}_2.
\]
Since $\ad_X$ is self-adjoint on Hilbert--Schmidt space and $\norm{\ad_X}\leq2\norm X$,
\[
 \norm{v_u(t)-v_u(0)}_2\leq2t\norm X^2\norm{v_u(0)}_2.
\]
Conjugation by $\e^{\ii sX}$ induces conjugation by $\e^{\ii s\ad_X}$ on the Sylvester operators. The same Gaussian average therefore yields
\[
 \partial_tT_u(t)=-\tfrac12[\ad_X,[\ad_X,T_u(t)]].
\]
The physical ceilings imply $0<T_u(t)\leq d^{2|V|}\idmap$, and hence
\[
 \norm{T_u(t)-T_u(0)}\leq8t\norm X^2d^{2|V|}.
\]

\medskip
\noindent\textbf{Step 2.}
Write $\Delta_1(t)=\Delta_V(\rho(t)\mid\sigma(t))$ and $\Delta_r(t)=\Delta_V(\rho_r(t)\mid\sigma_r(t))$, and let $\mu_t$ be the measure supplied by Lemma~\ref{noise:conditional-measure} for $(\rho(t),\sigma(t))$. Then
\[
 \Delta_r(t)=r^2\int_{(0,d^{2|V|}]}\frac{\mu_t(d\lambda)}{r\lambda+a}.
\]
Its construction and mass bound give
\begin{equation}\label{noise:eq-flow-mass}
 \int_0^\infty\frac{u}{(1+u)^2}\norm{v_u(0)}_2^2\,du
 =\mu_0((0,d^{2|V|}])\leq d^{2|V|}\Delta_1(0).
\end{equation}
Define the noise gap
\begin{equation}
 K_r(t):=\Delta_1(t)-\frac{\Delta_r(t)}r
 =\int_{(0,d^{2|V|}]}\frac{a}{\lambda(r\lambda+a)}\,\mu_t(d\lambda).
\end{equation}
The preserved reference floor permits the endpoint-density formula~\eqref{noise:eq-endpoint-density}:
\begin{equation}
 \mu_t(d\lambda)=G_t(\lambda)\,d\lambda,\qquad
 G_t(\lambda)=\Tr(\lambda\Omega_V(\rho)-\rho(t))_+,\qquad 0<\lambda<m.
\end{equation}
For Hermitian $Y$, self-duality and positivity of $\Phi_s$ give
\[
 \Tr(\Phi_s(Y))_+=\sup_{0\leq P\leq\id}\Tr\Phi_s(P)Y\leq\Tr Y_+.
\]
Since $\Phi_s(\Omega_V(\rho))=\Omega_V(\rho)$, it follows that
\[
 G_{t+s}(\lambda)\leq G_t(\lambda),\qquad s,t\geq0,\quad 0<\lambda<m.
\]

\medskip
\noindent\textbf{Step 3.}
Fix a smooth function $\chi:\R\to[0,1]$ with $\chi(x)=1$ for $x\leq1/4$ and $\chi(x)=0$ for $x\geq1/2$. Define
\[
 F(\lambda)=
 \begin{cases}
 \displaystyle\frac{a[1-\chi(\lambda/m)]}{\lambda(r\lambda+a)},&\lambda>0,\\[4pt]
 0,&\lambda\leq0.
 \end{cases}
\]
Splitting the gap kernel with $\chi(\lambda/m)$, and using the spectral construction of $\mu_t$, gives $K_r=K_{\mathrm{end}}+K_{\mathrm{reg}}$, where
\[
 K_{\mathrm{end}}(t)=\int_0^m\frac{a\chi(\lambda/m)G_t(\lambda)}{\lambda(r\lambda+a)}\,d\lambda,
\]
\[
 K_{\mathrm{reg}}(t)=\int_0^\infty\frac{u}{(1+u)^2}
 \ip{v_u(t)}{F(T_u(t))v_u(t)}_{\mathrm{HS}}\,du.
\]
Both terms are finite and nonnegative. Step~2 gives $K_{\mathrm{end}}(t)\leq K_{\mathrm{end}}(0)$.

It remains to estimate the regular part. Since $r\geq1/2$,
\[
 \left|\frac{d^j}{d\lambda^j}\frac1{\lambda(r\lambda+a)}\right|
 \leq2(j+1)!\lambda^{-j-2},\qquad \lambda>0,\quad 0\leq j\leq3.
\]
The derivatives of the cutoff are supported in $[m/4,m/2]$. The product rule and integration over $[m/4,\infty)$ therefore give, for an absolute constant $c\geq1$,
\begin{equation}
 \norm F_\infty\leq ca m^{-2},\qquad \norm F_1\leq ca m^{-1},\qquad \norm{F'''}_1\leq ca m^{-4}.
\end{equation}
The function $F$ vanishes near zero, and its first three derivatives decay at infinity. Thus three integrations by parts give
\[
 \abs{\widehat F(\xi)}\leq\min\{\norm F_1,|\xi|^{-3}\norm{F'''}_1\}\quad(\xi\ne0),\qquad
 \widehat F(\xi)=\int_\R F(\lambda)\e^{-\ii\xi\lambda}\,d\lambda.
\]
Splitting at $|\xi|=1$, and increasing the absolute constant $c$ if necessary, yields
\[
 \frac1{2\pi}\int_\R|\xi|\abs{\widehat F(\xi)}\,d\xi\leq ca m^{-4}.
\]
Fourier inversion and the Duhamel estimate $\norm{\e^{\ii\xi Y}-\e^{\ii\xi Z}}\leq|\xi|\norm{Y-Z}$ give the dimension-independent functional-calculus bound~\cite[Theorem~1.1.1]{AP16}
\begin{equation}
 \norm{F(Y)-F(Z)}\leq ca m^{-4}\norm{Y-Z}
\end{equation}
for self-adjoint $Y,Z$.

For each $u$, changing the two vector factors costs at most
\[
 2\norm F_\infty\norm{v_u(0)}_2\norm{v_u(t)-v_u(0)}_2,
\]
and changing the operator factor costs at most
\[
 ca m^{-4}\norm{T_u(t)-T_u(0)}\norm{v_u(0)}_2^2.
\]
The motion estimates of Step~1, together with $m\leq1$ and $d^{2|V|}\geq1$, bound their sum by
\[
 12ca\,t\norm X^2d^{2|V|}m^{-4}\norm{v_u(0)}_2^2.
\]
Integration and~\eqref{noise:eq-flow-mass} now give
\[
 \abs{K_{\mathrm{reg}}(t)-K_{\mathrm{reg}}(0)}
 \leq12ca\,t\norm X^2d^{4|V|}m^{-4}\Delta_1(0).
\]
Combining this with the endpoint monotonicity, we obtain
\begin{equation}\label{noise:eq-gap-motion}
 K_r(t)-K_r(0)\leq12ca\,t\norm X^2d^{4|V|}m^{-4}\Delta_1(0).
\end{equation}

The entropy derivative identity~\eqref{noise:eq-moving-pair-entropy} applies to both pairs, whose $V^c$-marginals are fixed. Hence
\[
 \Delta_1'(0)=-E_X(\rho\mid\sigma),\qquad
 \Delta_r'(0)=-E_X(\rho_r\mid\sigma_r),
\]
and therefore
\[
 K_r'(0)=-E_X(\rho\mid\sigma)+\frac1rE_X(\rho_r\mid\sigma_r).
\]
Dividing~\eqref{noise:eq-gap-motion} by $t>0$ and letting $t\downarrow0$ proves~\eqref{noise:eq-arithmetic-stability} with $c_{\mathrm{st}}=12c$, since $r\leq1$ and $d^{4|V|}m^{-4}\leq d^{6|V|}m^{-8}$. The argument also covers $\Delta_1(0)=0$.
\end{proof}

\subsection{Local operator perturbations}

Let $\Lambda$ be a finite physical system with local dimension $d\geq2$.
Fix $i\in\Lambda$ and a Hermitian onsite basis
\[
 U_0=\id_i,\quad U_1,\ldots,U_{d^2-1},\qquad \Tr_i(U_aU_b)=d\delta_{ab},
\]
extended by the identity outside $i$. For $r_0\geq1$, take nested physical sets
\[
 S_0=\{i\}\subseteq S_1\subseteq\cdots\subseteq S_{\ell_{\max}}\subseteq\Lambda,\qquad
 |S_\ell|\leq s_\ell:=1+2r_0\ell,
\]
where $\ell_{\max}\geq0$ is finite. Let $e_{a,\ell}=e_{a,\ell}^*$ be supported on $S_\ell$, and set
\[
 C_a=U_a+\sum_{\ell=0}^{\ell_{\max}}e_{a,\ell},\qquad a>0.
\]

\begin{lem}
\label{noise:local-perturbation}
Fix $d\geq2$, $r_0\geq1$, $c_{\mathrm{ref}}\geq1$, and a nonnegative sequence $(g_\ell)_{\ell\geq0}$ satisfying
\[
 \sum_{\ell\geq0}\e^{\nu\ell}g_\ell<\infty\qquad(\nu>0).
\]
There exist $0<\delta_*\leq1$ and nonnegative coefficients $f_\ell$, depending only on $d,r_0,c_{\mathrm{ref}},(g_\ell)$, such that
\[
 \sum_{\ell\geq0}\e^{\nu\ell}f_\ell<\infty\qquad(\nu>0)
\]
and the following holds. For any system and operator family specified
above, let $\rho,\sigma>0$ be states on $\cH_\Lambda$.
If $0<\delta\leq\delta_*$ and
\[
 \sigma\geq c_{\mathrm{ref}}^{-s_\ell}\Omega_{S_\ell}(\sigma),\qquad
 \norm{e_{a,\ell}}\leq\delta g_\ell,\qquad a>0,\quad 0\leq\ell\leq \ell_{\max},
\]
then, with $\varepsilon_\delta=\delta[1+\log(1/\delta)]$,
\begin{equation}\label{noise:eq-local-perturbation}
 \Delta_i(\rho\mid\sigma)\leq\frac2{d^2}\sum_{a>0}E_{C_a}(\rho\mid\sigma)
 +\varepsilon_\delta\sum_{\ell=0}^{\ell_{\max}}f_\ell\Delta_{S_\ell}(\rho\mid\sigma).
\end{equation}
The threshold and coefficients are uniform in $\ell_{\max}$, the volume,
the state spectra and the normalized onsite basis.
\end{lem}

For the physical Gibbs state $\sigma$, Lemma~\ref{gl:conditional-index}
permits $c_{\mathrm{ref}}=c_{\beta/2}^{\,2}$.

\begin{proof}
\medskip
\noindent\textbf{Step 1.}
Initially let $0<\delta\leq1$. Define
\[
 c_{\mathrm{op}}=\sqrt d+\sum_{\ell\geq0}g_\ell,\qquad
 m_\ell=c_{\mathrm{ref}}^{-s_\ell},\qquad \theta_\ell=2^{-\ell-2},
\]
and choose
\[
 b_\ell=g_\ell+\e^{-(\ell+1)^2},\qquad a_\ell=\frac{\delta b_\ell m_\ell}{2c_{\mathrm{op}}}.
\]
Since $0<b_\ell\leq c_{\mathrm{op}}$, we have $0<a_\ell\leq m_\ell/2\leq1/2$. The Gaussian term ensures positivity even when $g_\ell=0$, and gives
\[
 \frac{g_\ell^2}{b_\ell}\leq b_\ell,\qquad \log(1/b_\ell)\leq(\ell+1)^2.
\]

Set
\[
 C_{a,-1}=U_a,\qquad C_{a,\ell}=U_a+\sum_{j=0}^\ell e_{a,j}.
\]
For $0\leq\ell\leq \ell_{\max}$, these operators are Hermitian, supported on $S_\ell$, and satisfy $\norm{C_{a,\ell}}\leq c_{\mathrm{op}}$, since $\norm{U_a}\leq\sqrt d$. Moreover, $C_{a,\ell_{\max}}=C_a$.

Starting from $(\rho^{(\ell_{\max})},\sigma^{(\ell_{\max})})=(\rho,\sigma)$, apply the resets in descending order:
\[
 \rho^{(\ell-1)}=\mathcal D_{S_\ell,1-a_\ell}(\rho^{(\ell)}),\qquad
 \sigma^{(\ell-1)}=\mathcal D_{S_\ell,1-a_\ell}(\sigma^{(\ell)}),\qquad \ell=\ell_{\max},\ldots,0.
\]
All pairs remain strictly positive. The nesting identity
\[
 \Omega_{S_\ell}\Omega_{S_j}=\Omega_{S_j}\Omega_{S_\ell}=\Omega_{S_{\max\{\ell,j\}}}
\]
implies that every reset preserves all the reference orders:
\[
 \sigma^{(k)}\geq m_\ell\Omega_{S_\ell}(\sigma^{(k)}),\qquad -1\leq k\leq \ell_{\max},\quad 0\leq\ell\leq \ell_{\max}.
\]
The reset at step $\ell$ also supplies the input floor
\[
 \rho^{(\ell-1)}\geq a_\ell\Omega_{S_\ell}(\rho^{(\ell-1)}).
\]

Write
\[
 \Delta_\ell=\Delta_{S_\ell}(\rho\mid\sigma),\qquad
 \widetilde\Delta_\ell=\Delta_{S_\ell}(\rho^{(\ell)}\mid\sigma^{(\ell)}),
\]
and define
\[
 E_\ell=\sum_{a>0}E_{C_{a,\ell}}(\rho^{(\ell)}\mid\sigma^{(\ell)}),\qquad -1\leq\ell\leq \ell_{\max}.
\]
Thus $E_{\ell_{\max}}=\sum_{a>0}E_{C_a}(\rho\mid\sigma)$, while $E_{-1}$ is the sum of the onsite energies at the final pair.

\medskip
\noindent\textbf{Step 2.}
Theorem~\ref{noise:arithmetic-stability}, the inherited reference order and $1-a_\ell\leq1$ give
\[
 \sum_{a>0}E_{C_{a,\ell}}(\rho^{(\ell-1)}\mid\sigma^{(\ell-1)})
 \leq E_\ell+(d^2-1)c_{\mathrm{st}}d^{6s_\ell}m_\ell^{-8}a_\ell c_{\mathrm{op}}^2\widetilde\Delta_\ell.
\]

To control the perturbation energy, let $e=e^*$ be supported on $S_\ell$. The physical ceiling, the local floors and commutation of $e$ with reset states yield
\[
 e\rho^{(\ell-1)}e\leq d^{2s_\ell}a_\ell^{-1}\norm e^2\rho^{(\ell-1)},\qquad
 e\sigma^{(\ell-1)}e\leq d^{2s_\ell}m_\ell^{-1}\norm e^2\sigma^{(\ell-1)}.
\]
Let $M_u,Z_u,Z_{u,S_\ell^c}$ refer to the pair $(\rho^{(\ell-1)},\sigma^{(\ell-1)})$, and put $W_u=Z_u-\id_{S_\ell}\otimes Z_{u,S_\ell^c}$. Since $[e,Z_u]=[e,W_u]$, cyclicity of trace and $a_\ell\leq m_\ell$ bound left and right multiplication by $e$, giving
\[
 \norm{[e,W_u]}_{M_u}^2\leq4d^{2s_\ell}a_\ell^{-1}\norm e^2\norm{W_u}_{M_u}^2.
\]
The conditional integral representation~\eqref{noise:eq-conditional-integral} and Lemma~\ref{noise:entropy-retention} therefore imply
\[
 E_e(\rho^{(\ell-1)}\mid\sigma^{(\ell-1)})
 \leq4d^{2s_\ell}a_\ell^{-1}\norm e^2\widetilde\Delta_\ell.
\]

Apply the quadratic-form inequality to $C_{a,\ell-1}=C_{a,\ell}-e_{a,\ell}$, using the factors $1+\theta_\ell$ and $1+\theta_\ell^{-1}$. Substituting the chosen $a_\ell$, together with $g_\ell^2/b_\ell\leq b_\ell$ and $1+\theta_\ell\leq2$, gives
\[
 E_{\ell-1}\leq(1+\theta_\ell)E_\ell+\delta q_\ell\widetilde\Delta_\ell,
\]
where
\[
 q_\ell=(d^2-1)c_{\mathrm{op}}b_\ell
 \left[c_{\mathrm{st}}d^{6s_\ell}m_\ell^{-7}
 +8(1+\theta_\ell^{-1})d^{2s_\ell}m_\ell^{-1}\right].
\]
Since $\prod_{\ell\geq0}(1+\theta_\ell)\leq\e^{1/2}<2$, finite iteration yields
\begin{equation}\label{ds:finite-descending-energy}
 E_{-1}\leq2E_{\ell_{\max}}+2\delta\sum_{\ell=0}^{\ell_{\max}}q_\ell\widetilde\Delta_\ell.
\end{equation}

The continuity coefficient at step $\ell$ is $\lambda_\ell:=\omega_{m_\ell}(a_\ell)$, with $\omega_m$ as in Appendix~\ref{noise:section}. Its explicit formula gives
\[
 \lambda_\ell\leq\frac{\delta b_\ell}{2c_{\mathrm{op}}}
 \left[1+\log\left(1+\frac{2c_{\mathrm{op}}}{\delta b_\ell}\right)\right]
 \leq\frac{\delta b_\ell}{2c_{\mathrm{op}}}
 \left[1+\log(3c_{\mathrm{op}})+(\ell+1)^2+\log(1/\delta)\right].
\]
Here we used $c_{\mathrm{op}}/(\delta b_\ell)\geq1$ and the Gaussian lower bound on $b_\ell$.

Define the common envelope
\[
 h_\ell=q_\ell+\frac{b_\ell}{2c_{\mathrm{op}}}\left[1+\log(3c_{\mathrm{op}})+(\ell+1)^2\right].
\]
Then
\[
 q_\ell\leq h_\ell,\qquad \lambda_\ell\leq\varepsilon_\delta h_\ell.
\]
The sequence $h_\ell$ is independent of $\delta$ and has every exponential moment: $b_\ell$ has this property, while the remaining factors in $q_\ell$ grow at most exponentially because $s_\ell=1+2r_0\ell$.

Choose $0<\delta_*\leq1$ so that
\[
 \varepsilon_{\delta_*}\sum_{\ell\geq0}h_\ell\leq\tfrac12.
\]
This is possible since the sum is finite and $\varepsilon_\delta\to0$ as $\delta\downarrow0$. As $\varepsilon_\delta$ is nondecreasing on $(0,1]$, the same inequality holds whenever $0<\delta\leq\delta_*$.

\medskip
\noindent\textbf{Step 3.}
For $j\leq\ell$, Corollary~\ref{noise:union-continuity} and $S_j\cup S_\ell=S_\ell$ give
\[
 \abs{\Delta_{S_j}(\rho^{(\ell-1)}\mid\sigma^{(\ell-1)})
 -\Delta_{S_j}(\rho^{(\ell)}\mid\sigma^{(\ell)})}
 \leq\lambda_\ell\widetilde\Delta_\ell.
\]
The pair at index $\ell$ has undergone precisely the resets at indices $k>\ell$. Telescoping therefore yields
\[
 \widetilde\Delta_\ell\leq\Delta_\ell+\sum_{k=\ell+1}^{\ell_{\max}}\lambda_k\widetilde\Delta_k.
\]
Multiplying by $h_\ell$ and exchanging the finite sums,
\[
 \sum_{\ell=0}^{\ell_{\max}}h_\ell\widetilde\Delta_\ell
 \leq\sum_{\ell=0}^{\ell_{\max}}h_\ell\Delta_\ell
 +\sum_{k=0}^{\ell_{\max}}\lambda_k\left(\sum_{\ell<k}h_\ell\right)\widetilde\Delta_k.
\]
The last sum is bounded by
\[
 \varepsilon_\delta\left(\sum_{\ell\geq0}h_\ell\right)
 \sum_{k=0}^{\ell_{\max}}h_k\widetilde\Delta_k
 \leq\tfrac12\sum_{k=0}^{\ell_{\max}}h_k\widetilde\Delta_k.
\]
Absorbing this term gives
\begin{equation}\label{ds:weighted-absorption}
 \sum_{\ell=0}^{\ell_{\max}}h_\ell\widetilde\Delta_\ell\leq2\sum_{\ell=0}^{\ell_{\max}}h_\ell\Delta_\ell.
\end{equation}

For the singleton, telescoping through all resets gives
\[
 \Delta_i(\rho\mid\sigma)\leq\Delta_i(\rho^{(-1)}\mid\sigma^{(-1)})
 +\sum_{\ell=0}^{\ell_{\max}}\lambda_\ell\widetilde\Delta_\ell.
\]
Apply~\eqref{noise:eq-onsite-frame} at the final pair and~\eqref{ds:finite-descending-energy}. Since $\delta\leq\varepsilon_\delta$,
\begin{equation}
 \Delta_i(\rho\mid\sigma)\leq\frac2{d^2}E_{\ell_{\max}}
 +\left(1+\frac2{d^2}\right)\varepsilon_\delta
 \sum_{\ell=0}^{\ell_{\max}}h_\ell\widetilde\Delta_\ell.
\end{equation}
The weighted absorption estimate~\eqref{ds:weighted-absorption} proves~\eqref{noise:eq-local-perturbation} with
\[
 f_\ell=2\left(1+\frac2{d^2}\right)h_\ell.
\]
These coefficients have every exponential moment. Their construction and the choice of $\delta_*$ depend only on $d,r_0,c_{\mathrm{ref}},(g_\ell)$, as required.
\end{proof}

\subsection{CKG entropy production and frequency comparison}

Let $\sigma$ be the physical Gibbs state of Appendix~\ref{gl:section}.
We use the original CKG generators from the main text, with Hermitian
proposals $A_{i,a}=U_{i,a}/d$ and either rate
$\gamma\in\{\gamma_G,\gamma_M\}$. The Gaussian filter is
\[
 f_\beta(t)=\beta^{-1/2}(2/\pi)^{1/4}\e^{-t^2/\beta^2}.
\]
In an energy basis $H\ket m=E_m\ket m$, its Fourier transform gives
\begin{equation}
 q_\beta(x):=\frac1{\sqrt{2\pi}}\int_\R f_\beta(t)\e^{-\ii xt}\,dt
 =\frac{\beta^{1/2}}{(2\pi)^{1/4}}\e^{-\beta^2x^2/4},
\end{equation}
\[
 J_{i,a}(\omega)_{mn}=(A_{i,a})_{mn}q_\beta\bigl(\omega-(E_m-E_n)\bigr).
\]
Gaussian decay and $0<\gamma\leq1$ make the jump integrals finite. We have $\norm{A_{i,a}}\leq d^{-1/2}\leq1$, and the scalar proposal $A_{i,0}=\id_i/d$ contributes zero to $\cL_i$. With energy-width parameter $1/\beta$, \cite[Theorem~I.1 and Lemma~II.1]{CKG23} apply to this normalization, both rates and the coherent correction~\eqref{main:eq-original-correction}. They give
\[
 B_{i,a}=B_{i,a}^*,\qquad \cL_{i,*}(\sigma)=0.
\]

\paragraph{Fourier shells and frequency truncation.}
Let $\ell_{\max}$ be the first index for which $S_{i,\ell_{\max}}=\Lambda$. For each proposal $A_{i,a}$, let $X_{i,a,\ell}(t)$ be its evolution shells from Lemma~\ref{gl:onsite}, and define
\[
 Y_{i,a,\ell}(\omega)=\frac1{\sqrt{2\pi}}\int_\R f_\beta(t)\e^{-\ii\omega t}X_{i,a,\ell}(t)\,dt.
\]
These operators are supported on $S_{i,\ell}$, vanish for $\ell>\ell_{\max}$, and satisfy
\[
 J_{i,a}(\omega)=\sum_{\ell=0}^{\ell_{\max}}Y_{i,a,\ell}(\omega).
\]

Put
\[
 p_\ell=\frac{q_\beta(0)}{\sqrt\pi}
 \frac{(v_0\beta)^\ell\Gamma((\ell+1)/2)}{\ell!},\qquad
 c_{\mathrm{freq}}=2\e\,c_\beta q_\beta(0).
\]
By homogeneity of Lemma~\ref{gl:onsite} and the Gaussian moment formula
\[
 \int_\R|t|^\ell\e^{-t^2/\beta^2}\,dt=\beta^{\ell+1}\Gamma((\ell+1)/2),
\]
we obtain $\norm{Y_{i,a,\ell}(\omega)}\leq d^{-1/2}p_\ell$.

For each restricted-evolution term in the shell, shift the integration contour to $t-\ii\beta\operatorname{sgn}(\omega)$. The integrand is entire, and Gaussian decay makes the vertical-side integrals vanish. The shift contributes $\e^{-\beta|\omega|}$ from the Fourier factor and $\e$ from the Gaussian, while Lemma~\ref{gl:onsite} bounds the imaginary-time evolution by $c_\beta\norm{A_{i,a}}$. Each shell contains at most two such terms, so
\begin{equation}
 \norm{Y_{i,a,\ell}(\omega)}
 \leq d^{-1/2}\min\{p_\ell,c_{\mathrm{freq}}\e^{-\beta|\omega|}\}
 \leq d^{-1/2}\sqrt{p_\ell c_{\mathrm{freq}}}\,\e^{-\beta|\omega|/2}.
\end{equation}
Moreover,
\[
 \frac{p_{\ell+2}}{p_\ell}=\frac{(v_0\beta)^2}{2(\ell+2)}.
\]
Consequently, $p_\ell$ and $\sqrt{p_\ell}$ have every exponential moment.

For a cutoff $\omega_{\mathrm c}>0$, define
\[
 c_f=[\sqrt{2\pi}f_\beta(0)]^{-1},\qquad
 C_{i,a}=dc_f\int_{-\omega_{\mathrm c}}^{\omega_{\mathrm c}}J_{i,a}(\omega)\,d\omega,
\]
\[
 e_{i,a,\ell}=-dc_f\int_{|\omega|>\omega_{\mathrm c}}Y_{i,a,\ell}(\omega)\,d\omega.
\]
Since $Y_{i,a,\ell}(\omega)^*=Y_{i,a,\ell}(-\omega)$, both $C_{i,a}$ and $e_{i,a,\ell}$ are Hermitian, and $e_{i,a,\ell}$ is supported on $S_{i,\ell}$.

The Gaussian transform satisfies $\int_\R q_\beta(\omega-\nu)\,d\omega=c_f^{-1}$ for every $\nu$. Hence
\[
 dc_f\int_\R J_{i,a}(\omega)\,d\omega=dA_{i,a}=U_{i,a},
\]
and the finite shell decomposition yields the exact identity
\begin{equation}
 C_{i,a}-U_{i,a}=\sum_{\ell=0}^{\ell_{\max}}e_{i,a,\ell}.
\end{equation}
Set
\[
 \delta=\e^{-\beta\omega_{\mathrm c}/2},\qquad
 g_\ell=\frac{4c_f\sqrt d}{\beta}\sqrt{p_\ell c_{\mathrm{freq}}}.
\]
Then
\begin{equation}
 \norm{e_{i,a,\ell}}
 \leq c_f\sqrt{dp_\ell c_{\mathrm{freq}}}
 \int_{|\omega|>\omega_{\mathrm c}}\e^{-\beta|\omega|/2}\,d\omega
 =\delta g_\ell,
\end{equation}
\[
 \sum_{\ell\geq0}\e^{\nu\ell}g_\ell<\infty\qquad(\nu>0).
\]
The sequence $g_\ell$ depends only on the model parameters,
independently of the cutoff, volume, site, normalized onsite basis
and choice of rate.

For either rate, define
\[
 c_\gamma(\omega_{\mathrm c}):=d^2c_f^2\int_{-\omega_{\mathrm c}}^{\omega_{\mathrm c}}\gamma(\omega)^{-1}\,d\omega.
\]

\begin{lem}\label{ar:window-control}
For a state $\rho>0$ on $\cH_\Lambda$, define the site entropy
production by
\[
 \cI_i(\rho\mid\sigma):=-\Tr\!\left[\cL_{i,*}(\rho)(\log\rho-\log\sigma)\right].
\]
For every site $i$, cutoff $\omega_{\mathrm c}>0$ and rate $\gamma\in\{\gamma_G,\gamma_M\}$,
\begin{equation}\label{ar:window-production}
 \sum_{a>0}E_{C_{i,a}}(\rho\mid\sigma)\leq c_\gamma(\omega_{\mathrm c})\cI_i(\rho\mid\sigma).
\end{equation}
The coefficient is independent of the input state.
\end{lem}

\begin{proof}
The square field of $\cL_i$ is
\[
 \Gamma_{\cL_i}(Y):=\cL_i(Y^*Y)-\cL_i(Y^*)Y-Y^*\cL_i(Y),
\]
\[
 \Gamma_{\cL_i}(Y)=\sum_{a>0}\int_\R\gamma(\omega)
 [J_{i,a}(\omega),Y]^*[J_{i,a}(\omega),Y]\,d\omega.
\]
The coherent term is a derivation and contributes zero to this expression. Each site generator preserves $\sigma$, so $\Tr\sigma\cL_i(Y)=0$ for every $Y$.

Apply~\eqref{noise:eq-moving-resolvent} with $A=\cL_{i,*}(\rho)$ and $B=0$. Trace duality gives
\[
 -d\mathcal Q_u(\rho\mid\sigma)[A,0]
 =\Tr\rho\cL_i(Z_uZ_u^*)-2\Re\Tr\rho\cL_i(Z_u).
\]
Using $\rho-\sigma=Z_u^*\rho+u\sigma Z_u^*$ in the linear term, together with stationarity and preservation of adjoints, yields
\[
 -d\mathcal Q_u(\rho\mid\sigma)[A,0]
 =\Tr\rho\Gamma_{\cL_i}(Z_u^*)+u\Tr\sigma\Gamma_{\cL_i}(Z_u).
\]
The differentiation argument in Section~\ref{noise:arithmetic-section} applies to this trace-zero tangent. Integrating therefore gives
\begin{equation}
 \cI_i(\rho\mid\sigma)
 =\int_0^\infty\frac{u}{(1+u)^2}
 \left[\Tr\rho\Gamma_{\cL_i}(Z_u^*)+u\Tr\sigma\Gamma_{\cL_i}(Z_u)\right]\,du.
\end{equation}
In particular, $\cI_i(\rho\mid\sigma)\geq0$.

Since $C_{i,a}$ is Hermitian,
\[
 \norm{[C_{i,a},Z_u]}_{M_u}^2
 =\norm{[C_{i,a},Z_u^*]\rho^{1/2}}_2^2
 +u\norm{[C_{i,a},Z_u]\sigma^{1/2}}_2^2.
\]
Apply weighted Cauchy--Schwarz to the window integral defining $C_{i,a}$, separately in these two Hilbert--Schmidt norms. Both rates are positive on the window, so the coefficient is $c_\gamma(\omega_{\mathrm c})$. Summing over $a$ and enlarging the nonnegative frequency integrals to $\R$ gives
\[
 \sum_{a>0}\norm{[C_{i,a},Z_u]}_{M_u}^2
 \leq c_\gamma(\omega_{\mathrm c})
 \left[\Tr\rho\Gamma_{\cL_i}(Z_u^*)+u\Tr\sigma\Gamma_{\cL_i}(Z_u)\right].
\]
Multiplying by $u/(1+u)^2$ and integrating proves~\eqref{ar:window-production}, including finiteness.
\end{proof}

\subsection{Proof of Theorem~\ref{main:prop-dynamic}}\label{ar:dynamic-proof}

\begin{proof}
Fix $\kappa>0$, let $\rho>0$ be a state on $\cH_\Lambda$, and set
$c_{\mathrm{ref}}=c_{\beta/2}^{\,2}$.
Lemma~\ref{gl:conditional-index} gives
\[
 \sigma\geq c_{\mathrm{ref}}^{-|S_{i,\ell}|}\Omega_{S_{i,\ell}}(\sigma)
 \geq c_{\mathrm{ref}}^{-(1+2r_0\ell)}\Omega_{S_{i,\ell}}(\sigma).
\]
The sets $S_{i,\ell}$ are nested, and the preceding construction supplies
\[
 C_{i,a}=U_{i,a}+\sum_{\ell=0}^{\ell_{\max}}e_{i,a,\ell},\qquad
 \norm{e_{i,a,\ell}}\leq\delta g_\ell.
\]
Thus Lemma~\ref{noise:local-perturbation} applies with
$S_\ell=S_{i,\ell}$ and
$(C_a,e_{a,\ell})=(C_{i,a},e_{i,a,\ell})$.

Let $\delta_*,f_\ell$ be the threshold and coefficients provided by that lemma. They depend only on the model parameters, since $c_{\mathrm{ref}}$ and $g_\ell$ do, and are independent of the frequency cutoff. Choose one $\omega_{\mathrm c}>0$ sufficiently large that
\[
 \delta=\e^{-\beta\omega_{\mathrm c}/2}\leq\delta_*,\qquad
 \varepsilon_\delta\sum_{\ell\geq0}f_\ell\e^{r_0\ell}\leq\kappa,\qquad
 \varepsilon_\delta=\delta[1+\log(1/\delta)].
\]
Such a choice exists because $f_\ell$ has every exponential moment and $\varepsilon_\delta\to0$ as $\omega_{\mathrm c}\to\infty$.

Define
\[
 z_\ell=\varepsilon_\delta f_\ell,\qquad
 A=\frac2{d^2}\max\{c_{\gamma_G}(\omega_{\mathrm c}),c_{\gamma_M}(\omega_{\mathrm c})\}.
\]
Lemma~\ref{noise:local-perturbation} and Lemma~\ref{ar:window-control} give
\[
 \Delta_i(\rho\mid\sigma)\leq A\cI_i(\rho\mid\sigma)
 +\sum_{\ell\geq0}z_\ell\Delta_{S_{i,\ell}}(\rho\mid\sigma).
\]
Here the finite error sum was extended by nonnegative terms, and nonnegativity of the site entropy production permits the common coefficient $A$ for both rates. By construction,
\[
 \sum_{\ell\geq0}z_\ell\e^{r_0\ell}\leq\kappa.
\]
All choices depend only on $\beta,d,r_0,J_0,\kappa$, uniformly in
the site and volume, for both chains and cycles.
This proves~\eqref{main:eq-dynamic-bridge}
and~\eqref{main:eq-dynamic-error}.
\end{proof}

\noindent\textbf{Finite reference systems.}
The faithful-state estimates above remain valid for joint states
$\rho,\sigma>0$ on $R\Lambda$, with the same physical order assumptions.
Extend every physical operator and map by the identity on $R$,
and retain $R$ under all physical partial traces.
The dimension factors in the reset ceilings, spectral bounds, and
operator estimates involve only the physical systems.
Thus the comparison constants, perturbation threshold, and coefficient
sequences are independent of the reference dimension and state spectra.

For the Gibbs applications, use the comparison state
$\eta\otimes\sigma$, where $\sigma$ is the physical Gibbs state
and $\eta>0$ is any state on $R$.
The Gibbs orders of Lemma~\ref{gl:conditional-index} tensor with $\eta$,
and $\idmap_R\otimes\cL_{i,*}$ preserves $\eta\otimes\sigma$.
The physical supports and norm bounds of the Fourier shells
are unchanged.
For a faithful joint input $\rho>0$, take $\eta=\rho_R$.
The proof of Theorem~\ref{main:prop-dynamic} then gives the stated
joint-state estimate with the same choices of $A$ and $(z_\ell)$.

\bibliographystyle{references}
\bibliography{references}
\end{document}